# Physical properties and device applications of graphene oxide

Xiaoming Huang[1], Lizhao Liu[2,3]*, Si Zhou[3]*, Jijun Zhao[3]

[1]*School of Ocean Science and Technology, Dalian University of Technology, Panjin 124221, China*

[2]*School of Mathematical and Physical Sciences, Dalian University of Technology, Panjin 124221, China*

[3]*Key Laboratory of Materials Modification by Laser, Ion and Electron Beams (Dalian University of Technology), Ministry of Education, Dalian 116024, China*

Graphene oxide (GO), the functionalized graphene with oxygenated groups (mainly epoxy and hydroxyl), has attracted resurgent interests in the past decade owing to its large surface area, superior physical and chemical properties, and easy composition with other materials via surface functional groups. Usually, GO is used as an important raw material for mass production of graphene via reduction. However, under different conditions, the coverage, types, and arrangements of oxygen-containing groups in GO can be varied, which give rise to excellent and controllable physical properties, such as tunable electronic and mechanical properties depending closely on oxidation degree, suppressed thermal conductivity, optical transparency and fluorescence, and nonlinear optical properties. Based on these outstanding properties, many electronic, optical, optoelectronic, and thermoelectric devices with high performance can be achieved on the basis of GO. Here we present a comprehensive review on recent progress of GO, focusing on the atomic structures, fundamental physical properties, and related device applications, including transparent and flexible conductors, field-effect transistors, electrical and optical sensors, fluorescence quenchers, optical limiters and absorbers, surface enhanced Raman scattering detectors, solar cells, light-emitting diodes, and thermal rectifiers.

---

* Corresponding authors. Email: lizhao_liu@dlut.edu.cn (L. Liu), sizhou@dlut.edu.cn (S. Zhou)

# Contents

## 1 Introduction

Currently, nanotechnology has fomented a revolution in science and technology. The essence behind nanotechnology is to pursue nanomaterials possessing superior physical and chemical properties for device applications, including flexible displays, sensors, and optical limiters.

Graphene oxide (GO), which presents controllable electronic, mechanical, thermal and optical properties depending on the detailed samples, is a promising candidate for numerous nanodevices [1]. In experiments, GO can be obtained by exfoliating graphite oxide into nanosheets through thermal and mechanical methods [2, 3]. Large sized GO sheets with lateral sizes above 10 μm generally require expanded graphite as the starting material [4]. As the bulk counterpart of GO, graphite oxide has a long history of 160 years [5]. Currently, there are three major methods to produce graphite oxide from graphite, i.e., Brodie, Staudenmaier, and Hummers methods, which are different in oxidizing agent [6]. Compared with graphene, GO possesses various oxygen-containing groups (mainly epoxy and hydroxyl groups) on the graphene-based lattice, as demonstrated by spectroscopic and microscopic techniques [7-10]. The existence of oxygenated groups brings GO abundant and attractive properties.

Depending on the details of samples, such as the oxidation degree and thickness [11, 12], GO possesses tunable mechanical properties. The reported Young's modulus and intrinsic strength show a wide range of distributions of 6–42 GPa and 76–293 MPa for GO papers [13] and $250 \pm 150$ GPa and 30−50 GPa for GO monolayer, respectively [11]. Moreover, GO is very flexible under bending. The flexibility of GO can be ~100 times higher than net graphene [14]. Similar to the mechanical properties, electronic properties of GO layers are also tunable by controlling the functional group types and oxidation degree. Typically, GO is insulating with a sizeable band gap due to the large portion of $sp^3$ hybridized carbon atoms functionalized with the oxygenated groups, resulting in a sheet resistance of ~$10^{10}$ Ω/square or higher [15]. However, after reduction, the sheet resistance of reduced graphene oxide (namely, RGO) can be degraded by several orders of magnitude, transforming the material into a semiconductor or even into a graphene-like semimetal [9]. It has been predicted that the band gaps of GO sheets can be easily tailored in a wide range from 0 to 4 eV by varying the coverage, arrangement, and relative ratio of the epoxy and hydroxyl groups [16, 17], as well as external tensile strain [11].

On the other hand, owing to its heterogeneous atomic and electronic structures, GO shows good optical transmittance, as well as intrinsic fluorescence and nonlinear optical properties [18]. The transmittance depends closely on the GO film thickness [19]. Particularly, the intrinsic fluorescence of GO is the most notable difference from graphene since there is no fluorescence in graphene due to absence of an energy gap [20]. Besides, GO is an excellent candidate for thermoelectric materials [21] with a suppressed thermal conductivity compared with that of graphene (> 2000 W/mK at room temperature [22, 23]). The presence of oxygen functional groups significantly reduces the thermal conductivity of GO mainly through the oxygen defect scattering [24]. In addition to the outstanding physical properties of GO itself, the oxygenated groups on graphene basal plane can also largely expand the physical and chemical properties of GO by further decoration with other materials like ions and polymers to form a variety of GO-based composites [25-28].

Utilizing the excellent physical properties of GO and its composites, numerous GO-based devices have been developed, as illustrated in Fig. 1. For example, the tunable electronic properties combined with mechanical flexibility and optical transparency enable GO a good choice for transparent and flexible film conductors to replace the commercially standard materials of indium tin oxide (ITO) and fluorine tin oxide (FTO) [29, 30]. Besides, the oxygen-containing groups in GO can significantly improve the bipolar field emission performance of graphene through doping effects. Therefore, GO is a promising candidate material for field-effect transistors (FETs) [8, 31]. Meanwhile, change of conductance after molecular adsorption endows GO with electrical sensing for small amount and high selectivity detections [32-34]. In addition to the electrical devices, GO is also useful for fabricating optical devices, such as fluorescence quenching and sensing devices, surface enhanced Raman scattering (SERS), nonlinear-optical-based optical limiters and saturable absorbers [18, 35, 36]. The intrinsic fluorescence affords GO ability of photodetection and photoimaging in various fields, especially for biotechnology [37-40]. Moreover, the superior electronic and optical properties of GO can be utilized in optoelectronics, including organic photovoltaic (OPV), and light-emitting diode (LED) devices [41-43].

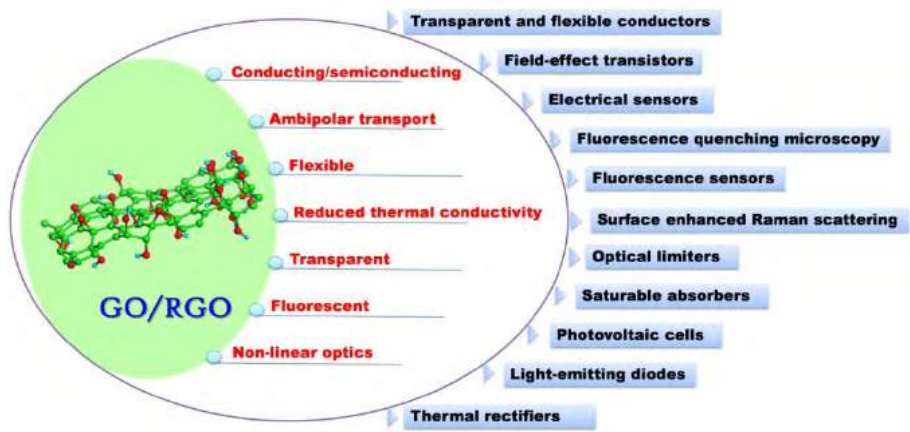

**Fig. 1** Physical properties and their relevant device applications of GO and RGO materials.

Due to the superior properties and promising applications of GO, numerous efforts have been devoted to exploring the fundamental structures and properties of GO sheets and to designing and fabricating the GO-based materials and devices. A fast growing year-by-year trend of the relevant publications is displayed in Fig. 2. As a result, more than 300 excellent review articles on GO-based materials have been published in recent years. However, most of them address GO/RGO focusing on the fabrication and reduction [3, 44-48], structure and characterization [9, 49-55], surface activity and catalysis [56-60], chemical functionalization and compositing [25, 28, 61-64], energy storage and conversion [65-70], environmental monitoring and treatment [10, 71-76], and biological and medical sciences [37, 40, 77-85]. Only few review articles have paid attention to the physical properties of GO as well as the device applications from the physical point of view. Particularly, most of them emphasize one or two aspects of the physical properties and related applications of GO. For example, Gao *et al*. [49] and Krishnan *et al*. [86] reviewed the stability of GO, including the stable structures respect to oxygen coverage, thermal stability, and flammability. Young *et al*. [87] overviewed the mechanical properties of GO and the GO-based composites. The electric properties of GO and related devices were reviewed by several groups [8, 30, 88, 89]. The optical properties and optoelectronic applications were also reviewed by several review articles as mentioned above [18, 35, 41]. Since researches in GO have boosted a lot in recent few years (see Fig. 2), it is imminent to have a comprehensive review to cover the recent progress on all the main physical properties of GO including mechanical, electronic, optical and thermal properties, as well as their related device applications.

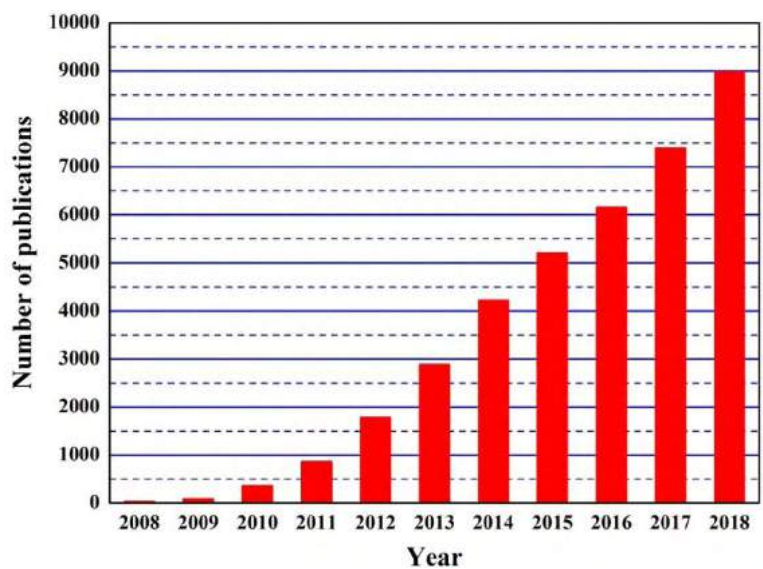

**Fig. 2** Trend of SCI-indexed publications concerning graphene oxide in last decade from 2008 to 2018 (searched from Web of Science using the title words: *graphene oxide*).

This review article is organized in the following sequence. We start with a comprehensive introduction of the stability and mechanical properties of GO in Section 2, including the thermodynamic and kinetic stabilities (Section 2.1), in-plane mechanics (Section 2.2), and out-of-plane mechanics (Section 2.3). Band structures (Section 3.1) and electronic transport properties (Section 3.2) are focused and then transparent and flexible devices (Section 3.3) and FETs (Section 3.4) are further discussed. Afterwards, thermal properties and thermoelectric devices of GO are described in Section 4.1 and Section 4.2 separately. Besides, nonlinear optics including the saturable absorbers, reverse saturable absorption, two-photon absorption and optical limiting are discussed from Sections 5.1 to 5.4 respectively. Finally, we summarize the essential physics and related applications of GO and give an outlook of the future research of GO in Section 6.

## 2. Stability and mechanical properties of GO

### 2.1 Thermodynamic and kinetic stabilities

Similar to graphene (monolayer of graphite), GO can be considered as a monolayer of graphite oxide. Graphite oxide has a long history and its fabrication can be traced back to more than one and a half centuries ago. Early in 1859, by treating graphite with the mixture of $KClO_3$ and $HNO_3$ at 60 °C for three to four days, Brodie [5] showed the product was composed of carbon, hydrogen, and oxygen, i.e. graphite oxide. Later, Staudenmaier [90] improved Brodie's approach by adding the chlorate (such as $KClO_3$ or $NaClO_3$) in multiple aliquots during the reaction, as well as adding the concentrated $H_2SO_4$ to increase the acidity of the mixture. Then, Hummers and Offeman [91] developed an alternative oxidation method by reacting graphite with a mixture of $KMnO_4$ and concentrated $H_2SO_4$, which also achieved similar levels of oxidation. Those three methods are major methods to produce graphite oxide from graphite. Then GO can be obtained by exfoliating graphite oxide into nanosheets through mechanical and thermal methods. Mechanical exfoliation is usually carried out by stirring or sonicating graphite oxide in agitated solutions, which is convenient but obtain small GO sheets due to cracking. An improved method to obtain large GO sheets is to use thermally or microwave expanded graphite as the starting material [4]. For the thermal exfoliation of graphite oxide, the functional groups such as epoxy and hydroxyl will decompose into gases and diffuse along the lateral direction during heating. The exfoliation can only occur when the decomposition rate of functional groups surpasses the diffusion rate of evolved gases. Generally, a minimum temperature of 550 °C was suggested to be necessary for the successful exfoliation at atmospheric pressure and the product is RGO [92].

Although GO is a long history material, the structure of GO remains an elusive problem due to its complicated stoichiometry, which highly relies on the synthesis process [6, 44]. Various oxygen-containing functional groups can exist in GO, including hydroxyl, epoxy, lactol, carboxyl, carbonyl, ketone, as identified by spectroscopic methods [10, 93]. It's generally accepted that GO mainly bears hydroxyl and epoxy groups on its basal plane and other oxygen-containing functional groups prefer the defective sites and edges of GO. On the other hand, structural features such as holes, graphitic and highly oxidized regions, wrinkles, folds and non-hexagon topological defects are also demonstrated by microscopic techniques [53, 94].

Beyond the structural characterization, how about the arrangement of functional groups in GO from the structural stability point of view attracts much theoretical attention. Several low-energy ordered phases are predicted for GO based on first-principles calculations [16, 95-97]. Studies show that epoxy and hydroxyl groups favor aggregation on the graphene basal plane [98-100]. The lowest-energy motifs consisting of dimeric, trimeric, and teramaric epoxy and hydroxyl groups on graphene have been identified (Fig. 3a), and they present energies up to 1 eV per oxygen lower than that of isolated units on graphene. Based on these motifs, Yan *et al.* [16, 98] predicted the lowest-energy crystalline phases for graphene fully covered by pure epoxy or hydroxyl groups, with stoichiometric formulae of $C_4O_2$ and $C_4(OH)_4$, respectively (Fig. 3b). For mixed epoxy-hydroxyl functionalization at full coverage, they found that three ordered structures with epoxy to hydroxyl ratios of 1:4, 1:2 and 1:1, respectively, are more stable than the two pure phases of $C_4O_2$ and $C_4(OH)_4$. In the case of intermediate coverage, it is energetically favorable for oxygenated groups to form specific chain-like structures with $sp^2$ carbon regions in between. These partially oxidized models are nearly as stable as the two pure phases, as illustrated in Fig. 3c.

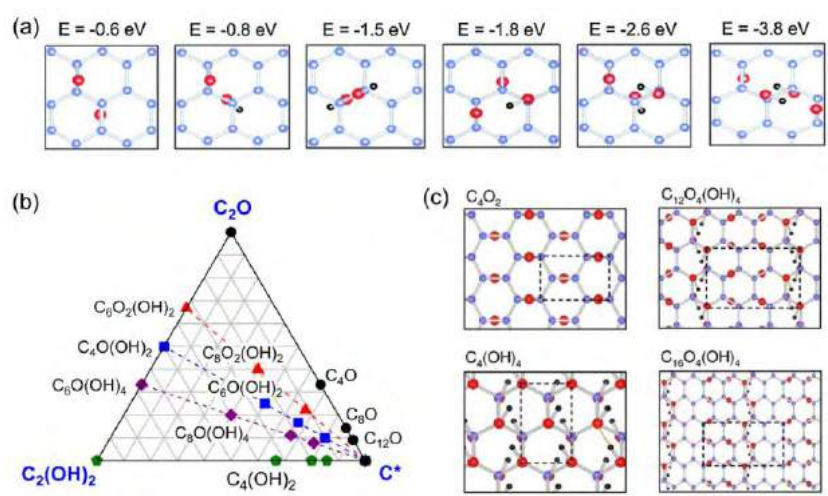

**Fig. 3** (a) Atomic configurations of dimeric and trimeric oxygen functional groups on graphene, and their corresponding association energies. Carbon, oxygen and hydroxyl atoms are represented in blue, red and black, respectively. The association energies are referred to the energy of isolated oxygen units on graphene. (b) Ternary diagram of ordered phases of GO, consisting of different fractions of $sp^2$ C, epoxy and hydroxyl pair. The dashed lines indicate GO phases with the same relative fractions of epoxy and hydroxyl pairs. (c) A few ordered phases of GO. Reproduced from Ref. [98].

Similarly, Wang *et al*. [100, 101] proposed a series of structural motifs for GO materials with epoxies only, hydroxyls only, and co-existence of epoxy and hydroxyl groups. First, several stable zero-dimensional (0D) motifs are constructed by appropriately placing epoxy and hydroxyl groups. It is found that the 0D motifs prefer to form one-dimensional (1D) chains, while these 1D chains tend to bind with each other via strain compensation, leading to various two-dimensional (2D) structures of GO with different C:O:H ratios. The crystalline structures of GO, although possessing very low energies, are difficult to achieve in experiment. Possible explanations include the entropy effect at finite temperature, the kinetic barriers for oxygen species to diffuse on graphene, and the lattice mismatch between GO and the substrate upon phase transitions.

However, in experiment, despite of the low-energy ordered GO, amorphous GO samples are often obtained. Based on the experimental observations, great attentions have been drawn to amorphous structures of GO. Using the 0D structural motifs proposed by Wang *et al*. [100, 101],

Liu *et al.* [17] constructed some disordered models of GO. They found the ones consisting of locally ordered motifs of oxygen functional groups have lower energies (Fig. 4a), as a compromise of the formation of hydrogen bonds, the existence of dangling bonds, and the retention of the bonds. These amorphous systems are as stable as the ordered phase at lower coverage less than 5%. At the same coverage rate, the stability of the amorphous system increases with the fractions of hydroxyl groups. Nguyen *et al.* [102] considered epoxy-functionalized graphene systems, where epoxy groups locally aggregate and exhibit long-range disorder. The system with an O:C ratio of 0.25 is lower in energy by up to 0.5 eV per oxygen compared to that of uniformly oxidized graphene system at the same oxidation level (Fig. 4b). They suggested that formation of highly oxidized islands is favored during the oxidation of graphene.

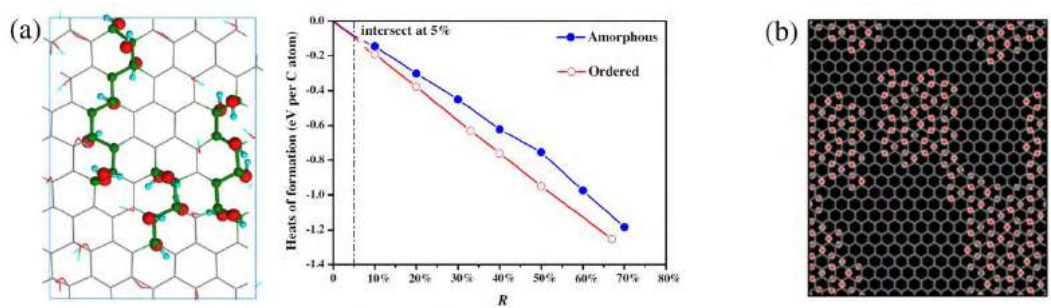

**Fig. 4** (a) Heats of formation of amorphous and ordered GO structures as a function of oxygen coverage rate. An example of the amorphous structure of GO used for the calculation is shown on the left, where carbon, oxygen and hydrogen are represented in green, red and cyan, respectively. Reproduced from Ref. [17]. (b) A disordered model structure of GO containing epoxy groups only. It shows superior stability compared to the ordered and homogeneously functionalized GO at the same oxygen content. Reproduced from Ref. [102].

The tendency of agglomeration of oxygenated groups indicates strong interactions between oxygen species on the graphene basal plane [102-104]. According to DFT calculations, as two oxygen functional groups approach each other along the graphene hexagonal lattice, the energy of the system exhibits an oscillatory variation [105, 106] (Fig. 5a and b). This is related to the charge density profile of the oxidized carbon surface, and the chemisorption of oxygen favors the C–C bond having the highest charge density. For GO consisting of epoxy groups aggregated on the carbon sheet, the binding energy of an additional epoxy group increases with the size of oxygen cluster, and reaches as large as 3 eV [105] (Fig. 5c). Moreover, the stability of the system increases with the size of oxygen agglomeration, and approaches to that of the ordered pure epoxy phase at full coverage [107] (Fig. 5d). All these results support the oxygen domain formation on graphene, and coincide well with the inhomogeneous structure of GO observed in TEM and STM experiments [94, 108, 109].

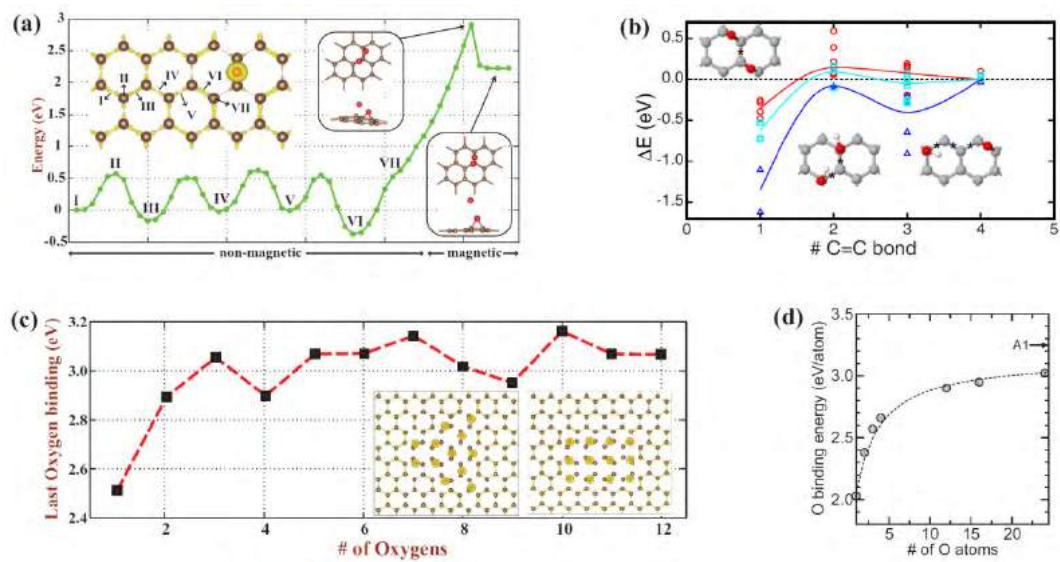

**Fig. 5** (a) Variation of the energy of two epoxy groups approaching each other on one side of graphene. The positions of epoxy groups corresponding to each energy value are labeled in the inset structures. Reproduced from Ref. [105]. (b) The association energy of two epoxy groups (red), two hydroxyl groups (blue), and one epoxy and one hydroxyl groups (cyan) on graphene, as a function of the number of C–C bonds in-between the two oxygen species. The association energy is referred to the energy of isolated oxygen units on graphene. Reproduced from Ref. [106]. (c) Variation of the binding energy of the last oxygen species to the oxygen agglomeration on graphene. Insets show the structures of GO with aggregated epoxy groups on one side of graphene. Reproduced from Ref. [105]. (d) The binding energy as a function of the number of oxygen atoms for a GO model with epoxy groups forming locally ordered structure on graphene. Reproduced from Ref. [107].

So far, many theoretical efforts have been devoted to predicting the realistic models of GO having inhomogeneous phases. Zhou *et al*. [106] explored the interactions between two oxygenated groups in various configurations by DFT calculations. Based on these interactions, they simulated the reconstruction of epoxy and hydroxyl groups on graphene basal plane using a lattice-model

Monte Carlo method. The GO systems in equilibrium present non-uniform oxidation at nanoscale consisting of highly oxidized domains surrounded by the pristine graphene regions. For mixed epoxy-hydroxyl functionalization, phase segregation occurs within the highly oxidized domains, forming regions rich in epoxy and hydroxyl groups, respectively, with maximized boundaries in-between these two phases. Alternatively, Yang *et al*. [110] simulated the evolution of graphene oxidation based on the reaction constant rates of various oxidation loci, which is described by DFT calculations and the transition state theory (Fig. 6a). They observed oxygen nucleation during the oxidation process, and obtained inhomogeneous GO structures similar to the Zhou's finding [106]. Hence, they attributed these results to the high correlation between oxidation loci, and suggested that the correlation length can be as long as 5 nm. Luo *et al*. [111] further studied the GO edge structure with carboxyl and carbonyl groups and proposed a possible more realistic GO model, as shown in Fig. 6b. Through steric energy analysis, they pointed out that the carboxyl prefers to locate in the armchair orientation, while the carbonyl prefers the zigzag orientation. Then the vacant edge carbon atoms are occupied by hydroxyls. However, for the epoxy and hydroxyl groups in graphene basal plane, Luo *et al*. predicted a different pattern from the ordered chain-like structure by Yan [16] and Wang [101]. They suggested that para-substituted epoxies and para-substituted hydroxyls in trans-configuration shown in Fig. 6b is 1.31 eV lower in binding energy than that of the chain-like structure. In Luo's model, the C–C bond connecting the hydroxyl has a length 1.42–1.44 Å. But in the chain-like model, this bond has a length of 1.54 Å. The less stretched C–C bonds by epoxy groups contribute to the lower energy of this trans-configuration model.

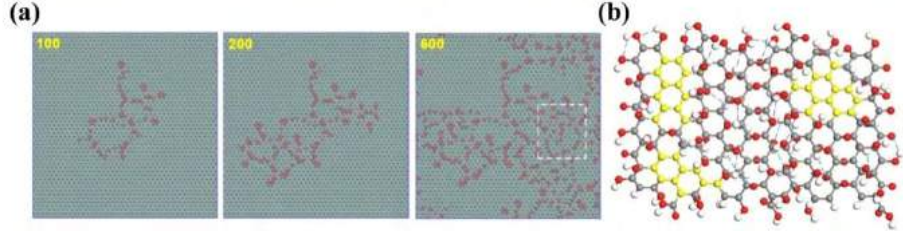

**Fig. 6** (a) Oxidation evolution from pristine graphene. The numbers indicate the oxidation times, wherein each oxidation means the oxidation reaction on one pair of carbon atoms at the oxidation loci. The unoxidized and oxidized carbon atoms are represented in green and red bonds, respectively. Reproduced from Ref. [110]. (b) Luo's thermodynamic GO structure model with carbon in grey, oxygen in red, hydrogen in white, graphene regions in yellow, and hydrogen bonds in blue dashed lines. Reproduced from Ref. [111].

In short, various model structures have been proposed to explain the chemical and structural features of GO revealed by experimental characterizations. These models are important starting points for further exploring and tailoring the physical properties of GO with technological applications.

GO is not the native oxide of graphene, and harsh treatment is involved in the synthesis process. Hence the stability of GO is an interesting issue for both fundamental science and technological applications [86]. Previous experiments showed that GO starts to lose oxygen at temperature above 100 ºC [45, 112], and a more recent study observed spontaneous reduction in the oxidized epitaxial graphene films at room temperature [113]. These results indicate the volatility of the oxygen adatoms in GO [114-116], and thus great efforts have been devoted to exploring the thermochemical stability of this material [117-120].

The thermodynamic stability of GO is usually characterized by the Gibbs free energy of formation defined as [106]:

$$\Delta G = G(C_l O_m H_n) - lG(C) - m\mu_O - n\mu_H, \tag{1}$$

where $G(C_l O_m H_n)$ is the Gibbs free energy of GO consisting of $l$ carbon, $m$ oxygen and $n$ hydrogen atoms within unit cell; $G(C)$ is the Gibbs free energy of pristine graphene per C atom; $\mu_O$ and $\mu_H$ are the chemical potentials of O and H in the form of gaseous $O_2$ and $H_2$ (or $H_2O$), respectively. Based on the 2D structural models of GO, Wang *et al*. [100] obtained the energy diagram of GO as a function of the atomic chemical potentials using DFT calculations (Fig. 7, left panel). They found that only four fully-oxidized ordered phases of GO can be thermodynamically stable, including the pure epoxy and hydroxyl phases, and two mixed epoxy-hydroxyl phases with the epoxy to hydroxyl ratios of 1:1 and 1:2, respectively. The conditions for the existence of these four GO structures are actually stringent and require an extreme O-rich and H-poor environment.

Zhou *et al.* [106] investigated the thermodynamic stability of non-uniformly oxidized graphene systems with amorphous structures at room temperature and standard pressure ($P = 1$ bar) (Fig. 7, right panel). They showed that the Gibbs free energy for the formation of GO decreases as the O:C ratio increases and approaches to zero at full oxidation, reaching equilibrium with gaseous $O_2$ and $H_2O$. The partially oxidized graphene systems, although thermodynamically unstable in the air, have $\Delta G$ very close to that of the fully oxidized phases due to the aggregation of oxygenated groups. For graphene with mixed epoxy-hydroxyl functionalization, the stability increases with the hydroxyl content at a fixed O:C ratio.

**Fig. 7** Left panel: total energy stability of GO in ordered phases with respect to the chemical potentials of oxygen and hydrogen. Insets show the structures of corresponding GO phases. Reproduced from Ref. [100]. Right panel: Gibbs free energy of formation of GO with epoxy functionalization (a), hydroxyl functionalization (b), and mixed epoxy/hydroxyl functionalization (c), from MC simulations. The black symbols in (a) and (b) are the energy values by DFT calculations. Selected model structures of GO obtained by the MC simulations are shown next to the curves. Red and yellow spheres represent epoxy groups on top and on bottom of graphene, respectively. Blue and cyan spheres represent hydroxyl groups on top and on bottom of graphene, respectively. In panel (c), two O:C ratios of 0.3 (magenta) and 0.4 (cyan) are considered. Reproduced from Ref. [106].

The experimentally fabricated GO, usually presenting partial oxygen coverages, are considered as kinetically constrained metastable phases. Lu *et al*. [121] showed that the diffusion barrier for a hydroxyl or epoxy group to migrate on graphene surface can be as large as 3 eV, when it is surrounded by other oxygenated groups, compared to those of 0.3 eV and 0.7 eV for an isolated hydroxyl and an isolated epoxy group, respectively. Therefore the transition of GO between two local energy minima involves large energy barriers and is kinetically hindered. Zhou *et al*. [106] suggested that the desorption of oxygenated groups as gaseous molecules is energetically unfavorable and involves large energy barriers within highly oxidized domains, due to the strong attractions between the oxygen species. Hence the agglomeration of oxygen functional groups can enhance the kinetic stability of GO.

On the other side, the chemical and structural evolution of GO upon thermal treatments has been intensively studied from both experimental and theoretical aspects [122, 123]. Bagri *et al*. [124, 125] investigated the thermal reduction of GO based on molecular dynamic simulations (Fig. 8a-d). They found that hydroxyl groups desorb at temperature below ~1000 K through the formation of water without altering the graphene sheet. Isolated epoxy groups are very stable and remain on the carbon layer up to 1000 K. Upon thermal annealing, carbonyl and ether groups are generated from two neighboring oxygenated groups. They are thermodynamically very stable and hinder complete reduction of GO to graphene. At temperature above 1500 K, CO and $CO_2$ molecules are formed from epoxy groups in close proximity to the other saturated $sp^3$ C bonds, leading to the creation of vacancies on the carbon basal plane. The transition states and corresponding energy barriers for the formation of a carbonyl pair from a pair of epoxies and a phenol-carbonyl pair from hydroxyl and epoxy groups are presented in Fig. 8e. Moreover, it was suggested that annealing GO or RGO in the hydrogen atmosphere can heal the broken C–C bonds associated with carbonyl and ether species, and improve the reduction efficiency [126].

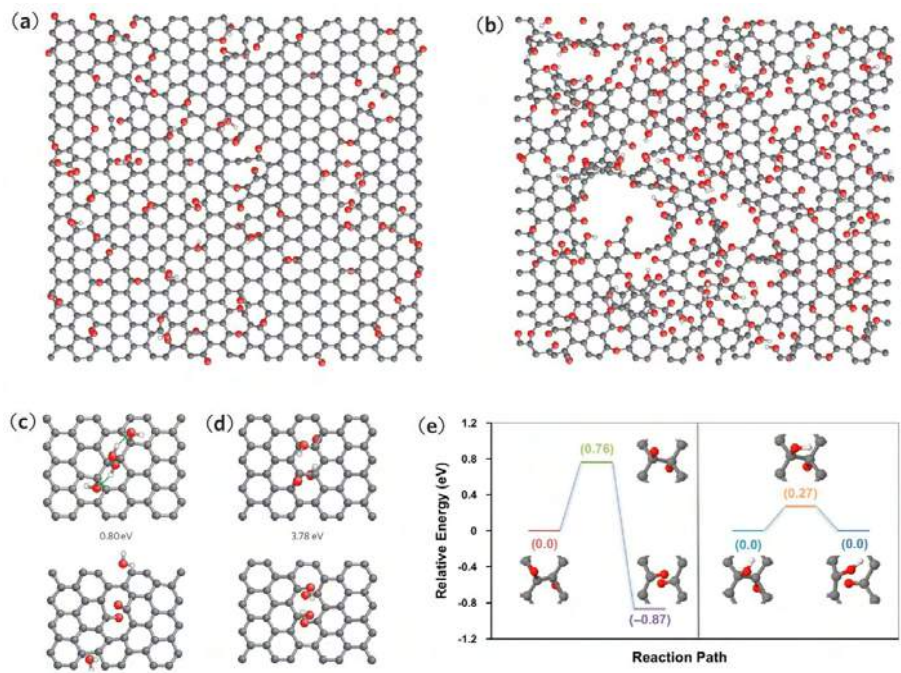

**Fig. 8** (a), (b) The structures of RGO by thermal annealing at 1500 K, with initial oxygen contents of 20% (a) and 30% (b), respectively, and the epoxy to hydroxyl ratio is 3:2. (c), (d) The initial configurations of epoxy and hydroxyl groups on graphene (top), leading to the formation of a carbonyl pair (c, bottom), and phenol groups (d, bottom), respectively. The numbers indicate the formation energies of the reactions in eV. (e) The reaction pathways for two epoxy groups to the formation of a carbonyl pair (left), and for one hydroxyl and one epoxy group to the formation of a phenol group (right). Reproduced from Ref. [124].

Larciprete *et al.* [127] explored the thermal evolution of oxidized graphene monolayer on Ir(111) in a temperature range of 350–550 K. They found that the reduction mechanism of oxidized graphene is closely related to the oxidation level of the material. At low oxygen coverage of 0.03 per monolayer, thermal treatment leads to desorption of the oxygenated groups, mainly through the cycloaddition reaction to the formation of $O_2$ molecules (Fig. 9a). At high oxygen coverage of 0.25 per monolayer, reduction is accompanied by the loss of carbon. CO and $CO_2$ molecules are formed through the precursors of lactone-ether pairs, which are originated from the diffusion and aggregation of epoxy groups (Fig. 9b). DFT calculations showed that both decomposition pathways are exothermic with activation energies about 1 eV.

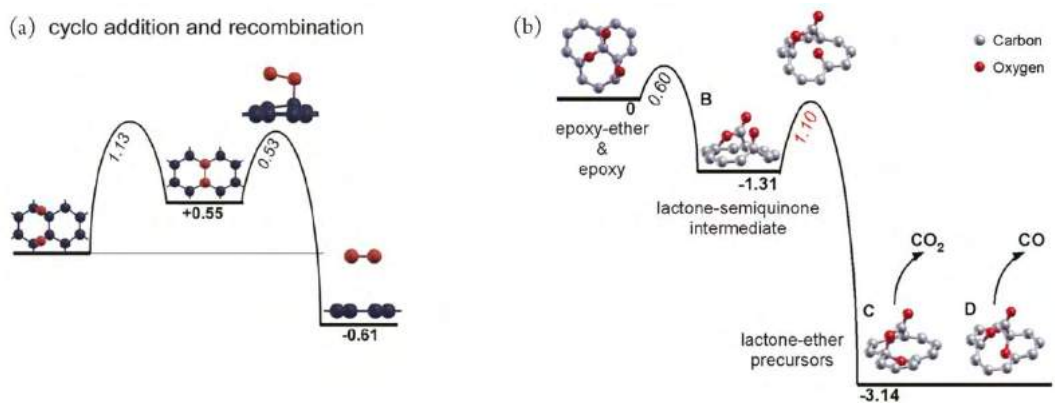

**Fig. 9** (a) Reaction pathway for two epoxy groups on graphene to the formation of a molecular oxygen. (b) Reaction pathway for three epoxy groups on graphene to the formation of the lactone-ether surface precursors and $CO/CO_2$ molecules. The numbers indicate the reaction barriers and the relative energy levels in eV. Reproduced from Ref. [127].

Kim *et al.* [113] reported that the oxidized epitaxial graphene film on SiC substrate is unstable at room temperature. The O:C ratio of the material drops by 0.06 in an one-month period, accompanied by chemical and structural modifications. Infrared radiation (IR) measurement further showed that hydrogen species exist in the as-synthesized GO, forming C–H bonds. According to DFT calculations, the hydrogen species prefer to react with epoxy and hydroxyl groups to form hydroxyl and water, and hence they are the origin of the metastability of the GO films. Further computational studies suggested that the aging process of GO leads to the aggregation of oxygenated groups on the carbon sheets, forming highly oxidized domains in the aged GO film [51]. Recently, Kumar *et al.* [128] applied a mild annealing procedure at 50–80 ℃ to manipulate the structure of GO. Under thermal treatment, the as-synthesized GO suspension with a mixed $sp^2$–$sp^3$ hybridized phase is transformed into the distinct oxidized and graphitic phases, directly revealed by the Auger electron spectroscopy images. The domain sizes are of the nanoscale, resulting in the bandgap opening observed from the electronic and optical measurements.

To briefly summarize, GO prefer the structure with epoxy and hydroxyl groups orderly aligned on its basal plane without considering the temperature and pressure. Considering the experimental condition, highly oxidized domains surrounded by graphene regions due to large migrating barrier between the oxidized domains, making GO inhomogeneous structure. Moreover, GO is only thermodynamically stable under certain conditions since high temperature (>1000 K) will decompose the oxygen-containing functional groups, which should be a major concern for its applications. Moreover, the thermochemical stability of GO can be utilized to tailor the structure and physical properties of this material. It creates many opportunities to fabricate new forms of GO for various applications.

## 2.2 In-plane mechanics

It is known that graphene has extraordinary mechanical properties owing to its unique 2D honeycomb structure and very strong C–C bonds [129]. The $sp^2$ carbon atoms form tight $\sigma$ bonds with three neighbors, resulting in a robust network. Experiments show that graphene with an in-plane Young's modulus E of 1 TPa and an intrinsic strength $\tau_c$ of 130 GPa [130], is the strongest material ever studied. As for GO, oxygen functionalization transforms some $sp^2$ carbon atoms into the tetrahedral $sp^3$ hybridization and thus weakens the in-plane mechanical strength of the lattice structure [131, 132]. Studies showed that the mechanical properties of GO are affected not only by the oxidation degree and atomic structure, but also the stacking geometry, water content, and intercalated molecules or ions [8, 133, 134]. Therefore, GO exhibits tunable mechanical properties and can be used for nanomechanical resonators, flexible electronics, and strain-engineering of the electronic band structure.

To date, a number of experimental and theoretical studies have been performed to explore the mechanical behaviors of GO/RGO [135-137]. It was demonstrated that the oxidation degree is a key factor to affect the mechanical strength of GO. Employing DFT calculations, Liu *et al*. [11] showed that as the oxidation degree increases from 0 to 0.7, the in-plane E of GO monolayer under tensile strain drops continuously due to weakening of $sp^2$ C–C bonds. As shown in Fig. 10a, the E degrades from 495 GPa (scaled value of graphene) to 294 GPa for the amorphous GO as the oxidation degree increases from 0 to 0.7, and degrades from 495 GPa (scaled value of graphene) to 388 GPa for the ordered GO as the oxidation degree increases from 0 to 0.67, respectively. Moreover, the theoretically predicted E of GO was further verified by experimental measurements. For example, for the GO monolayer with an oxidation degree of 0.2, the measured in-plane E was $384 \pm 31$ GPa obtained from the atomic force microscopy (AFM) nanoindentation by Cao *et al*. [138], close to the theoretical value of 395 GPa by Liu *et al*. [11]. Then, the GO monolayer with an oxidation degree of 0.7 has a measured E of $256.4 \pm 28.2$ GPa [139], agreeing well with the DFT value of 293 GPa [11]. Similar to the in-plane E, the $\tau_c$ of GO also drops continuously from 48 GPa (scaled value of graphene) to 28 GPa as the oxidation degree increases up to 0.7 [11]. The tendency of decreased in-plane E and $\tau_c$ with increasing oxidation degree was also reproduced by the density functional-based tight-binding (DFTB) calculations from Espinosa's group [139, 140], as shown in Fig. 10b and c. Besides, the fracture strain also slightly decreases with the oxidation degree [11]. In contrast to the prominent role of oxidation degree, the relative fractions of hydroxyl and epoxy groups have less effect on the mechanical strength of GO [11, 140], as shown in Fig. 10d, but affect the facture behavior a lot. It was found that hydroxyl groups cause GO to be brittle, while epoxy groups make GO ductile induced by an epoxy-to-ether transformation [140].

Particularly, even considering each functional group separately, the mechanical strength of GO also decreases with increasing coverage of the functional groups [131, 141, 142]. Zheng *et al*. [131] showed that a small coverage rate of 0.075 for hydroxyl significantly reduces the shear modulus of graphene to 8 GPa, decreased by about 50% compared to that of graphene, while the in-plane E is reduced by 20% from 1050 to 864 GPa. Zandiatashbar *et al*. [141] studied the in-plane E and $\tau_c$ of defective graphene containing only epoxy groups and only hydroxyl groups individually. As the coverage of epoxy increases from 0 to 0.25, the E of pristine graphene drops from 435 to 174 N/m and the $\tau_c$ drops from 50 to 20 N/m separately. While, compared with the epoxy groups, hydroxyl alone in graphene shows a weak effect on both properties. As the coverage of hydroxyl increases from 0 to 0.175, the E of pristine graphene is slightly decreased within a descent ratio of 10%, and the $\tau_c$ drops within a ratio of 30%.

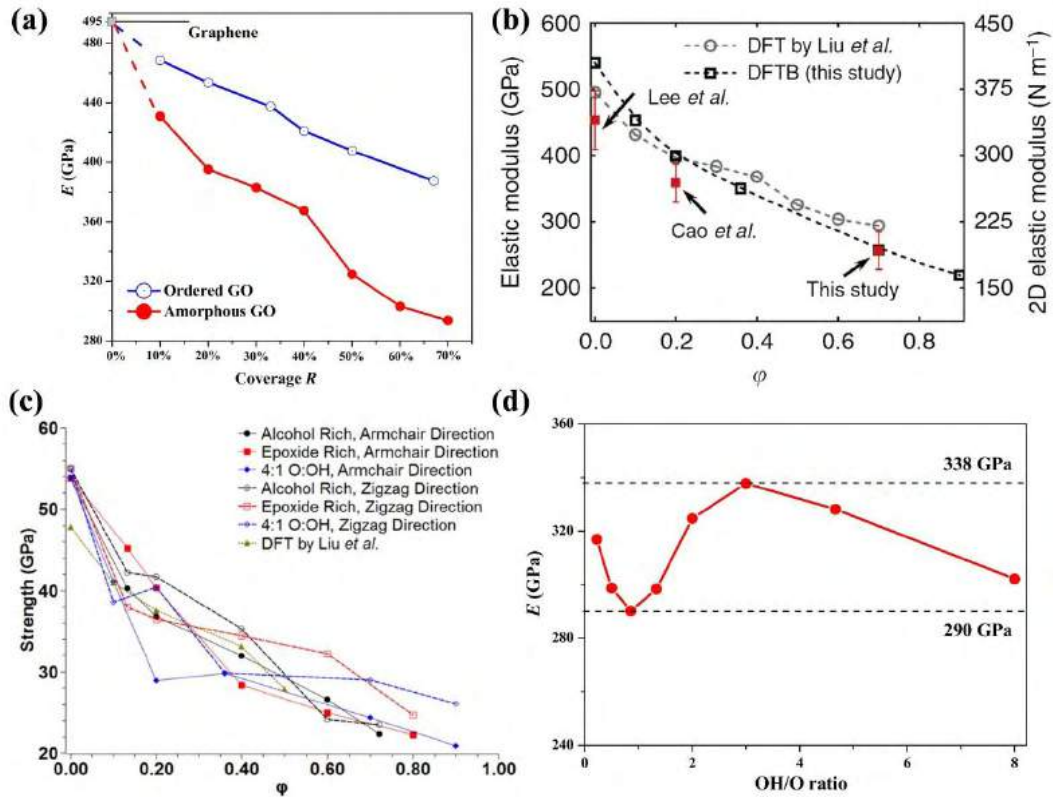

**Fig. 10** (a) Young's modulus of ordered and amorphous GO as a function of the oxygen coverage *R* for a fixed hydroxyl to epoxy ratio of 2:1. (d) Young's modulus of amorphous GO as a function of the hydroxyl to epoxy ratio for a fixed coverage *R* of 50%. Reproduced from Ref. [11]. (b) Comparison of Young's modulus at different oxidation degree φ for GO predicted by DFT [11] and DFTB [139] and the experimental results for graphene (φ = 0) , GO with φ = 0.2 [138], and GO with φ = 0.7 [139]. Reproduced from Ref. [139]. (c) Intrinsic strength for GO with varying the oxidation degree φ. Reproduced from Ref. [140].

Another main factor to affect the mechanical properties of GO is the sample detail, such as the degree of structural order, defects, thickness, and stacking orientation. For example, RGO exhibits lower mechanical strength than that of graphene due to the presence of defects and oxygen residues [143]. Various RGO samples show a standard deviation of Young's modulus up to 150 GPa, indicating the impact of different microscopic structures on the mechanical performance of this material. Liu *et al*. [11] pointed out that with the same chemical composition, GO in ordered structural phases exhibits larger E than the amorphous GO, as indicated in Fig. 10a. Paci et al. [117] suggested that defects in GO can significantly degrade the mechanical performance of graphene. For the defective GO models, the one with a hole defect (missing a carbon hexagon) has the lowest E and $\tau_c$, followed by the heated pristine GO with regard to the GO with an epoxy line defect. Zandiatashbar et al. [141] studied the effects of vacancies on the stiffness and strength of GO. They found that with increasing the vacancy concentration, both the E and $\tau_c$ drop more remarkably than those of increasing the epoxy or hydroxyl concentration. The AFM measurement by Suk et al. [144] indicates that the E of GO is closely related to the number of layers. The effective E for monolayer GO is 207.6 ± 23.4 GPa, but slightly increases to 223.9 ± 17.7 GPa for two-layer GO and 229.5 ± 27.0 GPa for three-layer GO. Similarly, Cui et al. [145] also studied the effect of layer number on the mechanical strength of GO by means of fracture force normalized by the number of layers. As shown in Fig. 11a, monolayer and bilayer GO show comparable normalized fracture force. The normalized fracture force of bilayer GO is slightly higher than the monolayer GO indicating a higher strength in bilayer GO, which agrees well with Suk's results [144]. However, when the number of layers increases to 5 and further to 12, the normalized fracture force shows a sharp deterioration of ~60%. Meanwhile, the stacking angle also affects the mechanical performance of GO, which has a more significant effect on the fracture pathways rather than the strength of GO [145]. Molecular simulation (MD) indicates that the bilayer GO stacking at 0° presents a common crack behaviour where the top and bottom layers break in a scattered fashion with a shared fracture path, as shown in Fig. 11b. However, the one at 21.8° mainly shows an individual layer cracking pattern with a decoupled and uncorrelated fracture pathway, as shown in Fig. 11c. While, for the GO with the same layer number, the stacking angle only has a little influence on the mechanical strength. For example, the bilayer GO stacking at 0° shows slightly higher ultimate tensile strength and toughness than those of 21.8° in average, but the statistically differences in ultimate tensile strength and toughness are within the error bars, as shown in Fig. 11d. Besides, the mechanical performance of GO also depends on the tensile direction. Peng and De [146] suggested that the monolayer GO shows a larger E of 323 N/m under the biaxial tensile strain than that (~270 N/m) under the uniaxial strain. Under the uniaxial strain, the GO has the E of 272 N/m, $\tau_c$ of 13.3 N/m and fracture strain of 0.08 under armchair direction, all of which are lower than those (275 N/m, 25.5 N/m, and 0.26) under zigzag direction, especially the $\tau_c$ and fracture strain. Through MD simulations, Wang et al. [12] also indicated that E and $\tau_c$ of GO are slightly larger along the zigzag direction than those along the armchair direction by 10−15%. On the other hand, some works have reported very low E of 34−77 GPa for multilayer GO [12], especially 12−42 GPa for GO papers [147, 148], which may be due to the wrinkles in GO.

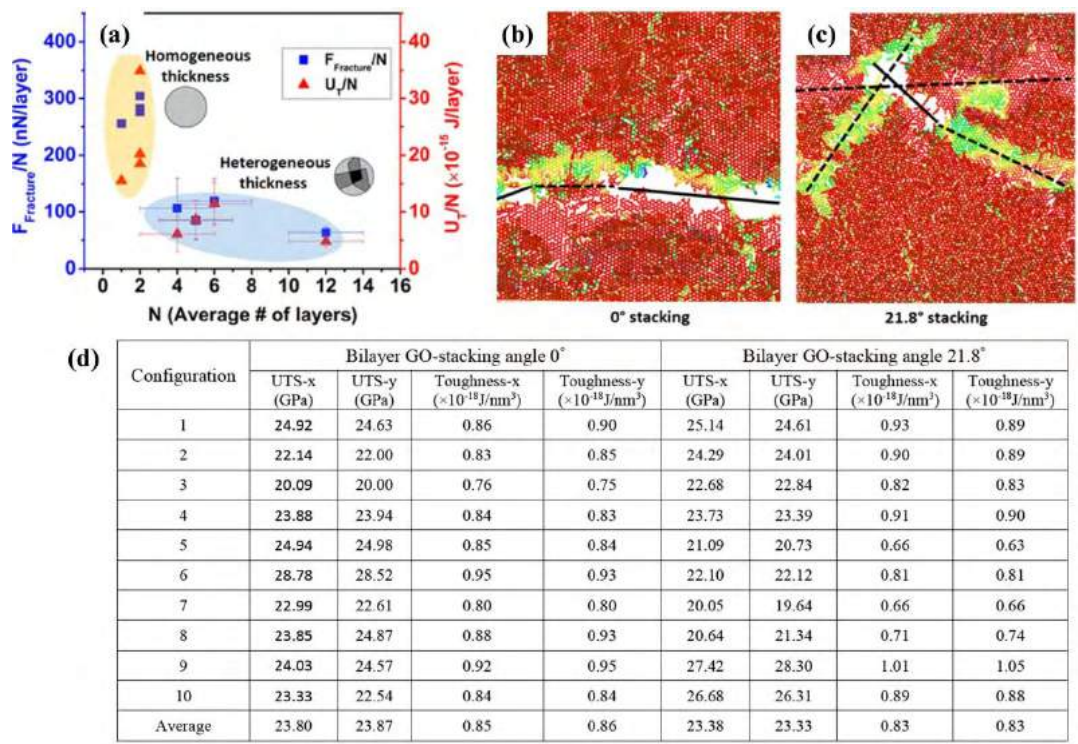

| (d) Configuration | Bilayer GO-stacking angle 0° | | | | Bilayer GO-stacking angle 21.8° | | | |
|---|---|---|---|---|---|---|---|---|
| | UTS-x (GPa) | UTS-y (GPa) | Toughness-x (×10⁻¹⁸J/mm³) | Toughness-y (×10⁻¹⁸J/mm³) | UTS-x (GPa) | UTS-y (GPa) | Toughness-x (×10⁻¹⁸J/mm³) | Toughness-y (×10⁻¹⁸J/mm³) |
| 1 | 24.92 | 24.63 | 0.86 | 0.90 | 25.14 | 24.61 | 0.93 | 0.89 |
| 2 | 22.14 | 22.00 | 0.83 | 0.85 | 24.29 | 24.01 | 0.90 | 0.89 |
| 3 | 20.09 | 20.00 | 0.76 | 0.75 | 22.68 | 22.84 | 0.82 | 0.83 |
| 4 | 23.88 | 23.94 | 0.84 | 0.83 | 23.73 | 23.39 | 0.91 | 0.90 |
| 5 | 24.94 | 24.98 | 0.85 | 0.84 | 21.09 | 20.73 | 0.66 | 0.63 |
| 6 | 28.78 | 28.52 | 0.95 | 0.93 | 22.10 | 22.12 | 0.81 | 0.81 |
| 7 | 22.99 | 22.61 | 0.80 | 0.80 | 20.05 | 19.64 | 0.66 | 0.66 |
| 8 | 23.85 | 24.87 | 0.88 | 0.93 | 20.64 | 21.34 | 0.71 | 0.74 |
| 9 | 24.03 | 24.57 | 0.92 | 0.95 | 27.42 | 28.30 | 1.01 | 1.05 |
| 10 | 23.33 | 22.54 | 0.84 | 0.84 | 26.68 | 26.31 | 0.89 | 0.88 |
| Average | 23.80 | 23.87 | 0.85 | 0.86 | 23.38 | 23.33 | 0.83 | 0.83 |

**Fig. 11** (a) The influence of number of layers on the normalized fracture force ($F_{Fracture}/N$) and normalized toughness ($U_T/N$); (b) MD biaxial tension results for bilayer GO stacking at 0° and 21.8° where UTS is the ultimate tensile strength; (c) and (d) are the fracture behaviours for the bilayer GO stacking at 0° and 21.8° respectively. Reproduced from Ref. [145].

Due to the existence of water in realistic GO materials, the effect of water content on the mechanical strength of GO is also an interesting question. In experiment, Dikin *et al.* [147] showed that for GO papers with increasing water content, the E decreases from 25 to 17 GPa. Gao *et al.* [13] also reported that hydrated GO film has E of 6.5 GPa, lower than the value (E = 10.5 GPa) of as-prepared GO sample, but almost keeps $\tau_c$ (~63 MPa) of the as-prepared GO. They suggested that the intercalated water molecules weaken and toughen the GO layers through reorientation of hydrogen bond. Compton *et al.* [149] showed that the E of GO paper increases from 20 GPa to a maximum value of 40 GPa as the water content increases from 0 to 7 wt.%, and then drops to 34

GPa as the water content further goes up to 15 wt.% (Fig. 12a). MD simulations of multilayer GO models yield similar tendency of E versus water content, where E achieves a maximum of 25 GPa at a water concentration of 5%. The optimal water content for the peak modulus is attributed to the saturation of hydrogen bonds in GO formed by oxygen functional groups and water molecules. Beyond this optimal value, intercalating additional water in-between GO sheets swells the structure and facilitates the lateral sliding of GO layers, and hence degrades the mechanical strength of GO. Riobóo *et al*. [150] pointed out that the in-plane E of GO thin film increases from 256 GPa at room temperature to 582 GPa at 200 °C due to heating-induced water elimination and interlayer distance shrinking. Moreover, the shear strength of GO also highly replies on the water content. As shown in Fig. 12b, interfacial shear strength for GO−GO interfaces strengthens steadily from 28 to 63 GPa with increasing the water content from 13 to 24 wt%, i.e., relative humidity from 13% to 61%, demonstrated by the AFM measurement by Soler-Crespo *et al*. [151]. Further MD simulations [151] indicated the hardening mechanism for interfacial shear strength is ascribed to the formation of strongly interacting hydrogen-bond networks between two GO surfaces.

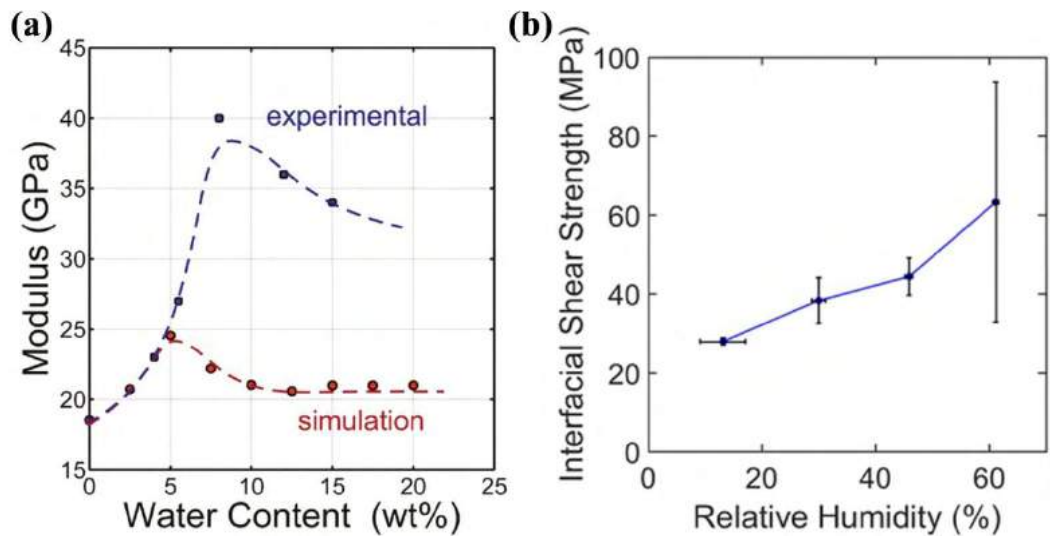

**Fig. 12** (a) In-plane Young's modulus of GO paper as a function of the water content from experiment and simulation. Reproduced from Ref. [149]. (b) Interfacial shear strength for GO−GO interfaces as a function of relative humidity. Reproduced from Ref. [151].

Similar to the water effect, the mechanical properties of GO can be also effectively enhanced by intercalation of organic molecules and metal ions into the gallery of GO layers, as shown in Table

1. The inserted molecules and ions bridge the adjacent GO sheets and help resist to the tensile force [152, 153]. For instance, GO papers modified by $Mg^{2+}$ and $Ca^{2+}$ ions with concentrations below 1 wt.% have E and $\tau_c$ significantly enhanced by 10−200% and 50%, respectively, compared to those of the unmodified GO [154]. GO functionalized by polyvinyl alcohol (PVA) chains presents superior E and $\tau_c$ values, which are increased by up to 60% and 400%, respectively [149].

Another way to improve the mechanical strength is to crosslink the GO sheets covalently with functional polymers. GO papers intercalated by glutaraldehyde (GA) molecules show enhanced E and $\tau_c$ by 190% and 60%, respectively, due to increased interlayer adhesion caused by the GA molecules [13]. Similarly, GO crosslinked papers by polyetherimide (PEI) have E and $\tau_c$ increased by up to 174% and 54%, respectively [155].

**Table 1** Young's modulus E and tensile strength $\tau_c$ of GO modified by various ions and molecules, compared with the values ($E_0$, $\tau_{c0}$) of GO before modification from experimental studies.

| Decorator | E (GPa) | $E_0$ (GPa) | $\tau_c$ (MPa) | $\tau_{c0}$ (MPa) | Ref. |
|---|---|---|---|---|---|
| $Mg^{2+}$ | 15−28 | 6−26 | 81−88 | 82 | [154] |
| $Ca^{2+}$ | 10−28 | 6−26 | 75−126 | 82 | [154] |
| glutaraldehyde | 26−35 | 9−13 | 101 | 64 | [13] |
| polyvinyl alcohol | 30−60 | 25−40 | -- | -- | [149] |
| polyvinyl alcohol | 8 | 5 | 125 | 25 | [133] |
| polyetherimide | 41−85 | 31 | 114−179 | 116 | [155] |

To summarize, GO exhibits large stiffness and breaking strength, owing to its unique intralayer structure. The mechanical properties of GO can be effectively tailored by the oxygen content, atomic structure, hydration level, and intercalted molecules or ions. Therefore, GO is a very promising material for nanomechanical systems, as well as many other related applications [29, 88, 156].

## 2.3 Out-of-plane mechanics

In addition to the in-plane performance, out-of-plane mechanical property of GO is also interesting. Compared with the large in-plane Young's modulus, the out-of-plane elastic modulus of GO is quite lower and closely related to the interlayer environment. Medhakar *et al.* [134] calculated the perpendicular Young's modulus ($E_\perp$) of hydrated GO using MD simulations. They showed that $E_\perp$ is in the range of 3.6–18.6 GPa and dominated by water-mediated interlayer hydrogen bonds in the system (Fig. 13a). At the same hydration level, higher oxygen content leads to larger $E_\perp$. For an O:C ratio of 0.4, $E_\perp$ reaches the maximum value at a water content of 8.6 wt.%. Riobóo et al. [150] demonstrated that water elimination under heating will increase the $E_\perp$ of GO few-layer film. The GO thin film has an $E_\perp$ of 118 GPa at room temperature, which increases to 190 GPa at 100 ºC due to partial elimination of embedded water layers and further reaches 535 GPa at 200 ºC after almost completely removing of water. Gao et al. [157] also investigated $E_\perp$ of few-layer GO films by using experiment and DFT calculations. It was found that the $E_\perp$ of GO increases with increasing the amount of water trapped in between the layers and reaches the maximum when a full monolayer is produced. Further forming the second water layer will lead to a modulus decrease. In particular, in ambient humidity, the conventional GO has the maximum $E_\perp$ of 35 ± 10 GPa, but the epitaxial graphene oxide (EGO) shows almost invariable modulus of ~22 GPa. To gain deep insight into the perpendicular elasticity of GO, DFT calculations were performed. As shown in Fig. 13b, for the conventional GO with fully oxidized hydroxyl groups, the $E_\perp$ increases with the amount of intercalated $H_2O$ molecules and reaches a maximum value of about 31 GPa at 25% $H_2O$, agreeing well with the experimental data that the $E_\perp$ reaches the maximum of 35 GPa at the relative humidity (RH) of 25%, as shown in Fig. 13c. The curves shown in Fig. 13b and c can be understood by the following picture. At low RH, water molecules swell the GO structure, increasing the interlayer distance from 3.4 Å to about 6.2 Å, but leaving the interlayer space mainly empty and therefore producing a soft structure with a low perpendicular elastic modulus. As the RH increases, the interlayer space is gradually filled without changing interlayer distance significantly, leading to increase of the interlayer modulus and reaching the maximum at a RH of 25%, where a full water monolayer is formed. Above 25% water, the perpendicular elastic modulus decreases because a second water layer starts to form, further swelling and softening the GO structure. Following this picture, large interlayer perpendicular elasticity in conventional GO compared with EGO at RH = 25% can be also explained. Since the EGO is not a porous structure, and water intercalation is minimal and independent of humidity, its $E_\perp$ remains constant of ~22 GPa for all RHs. However, the conventional GO is a porous structure and the $E_\perp$ will change with the humidity and reach a large value of ~35 GPa at RH = 25%.

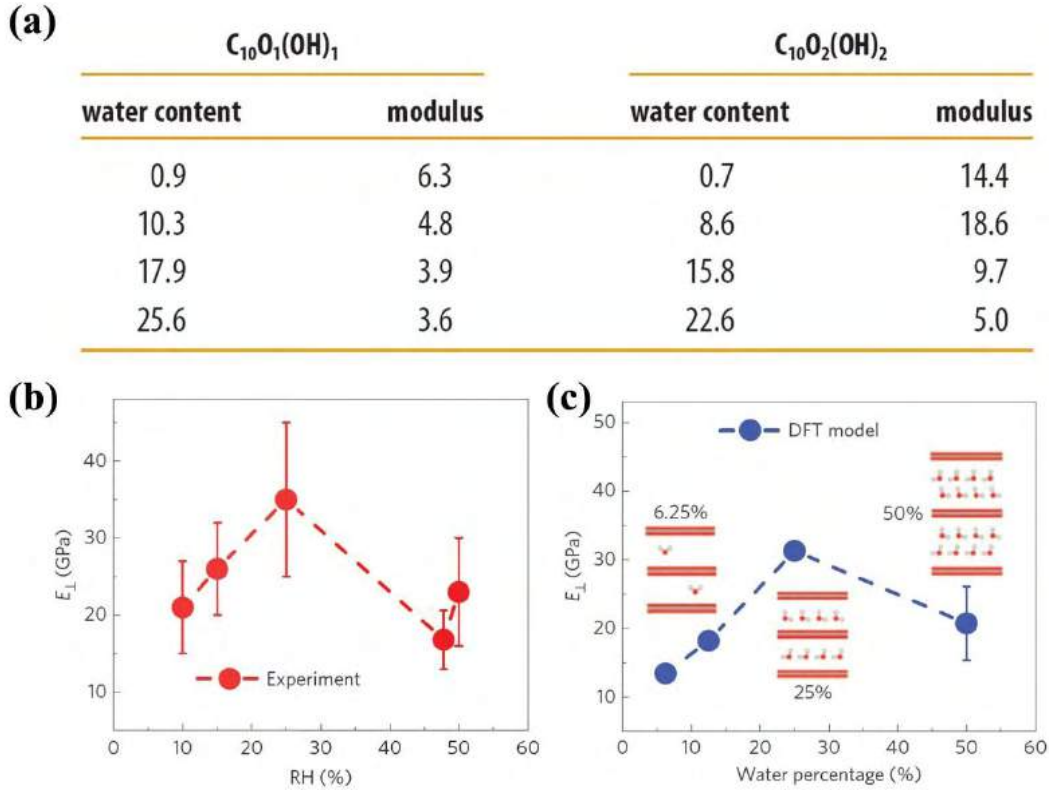

**Fig. 13** (a) Out-of-plane Young's modulus of GO models with various water contents from MD simulations. Reproduced from Ref. [134]. (b) Out-of-plane Young's modulus of GO as a function of the water content from DFT calculations. Insets are the schematic models of GO showing how the water molecules fill the interlayer spacing. (c) Out-of-plane Young's modulus of GO as a function of the relative humidity from experiment. Reproduced from Ref. [157].

On the other hand, oxygen functional groups and non-hexagonal structure defects in GO make it flexible. Bending experiments for several samples of GO papers by Dikin *et al.* [147] showed that GO papers can sustain more bending deformation (1.1 ± 0.1%) than uniaxial tension (0.6%). As shown in Fig. 14b–d, a strip of GO paper is bent and then compressed until a kink (or more than one kink) is formed. The positive (or negative) normal strain $\varepsilon_x$ at the outer (or inner) bar surface is $|\varepsilon_x| = 0.5t / r$, where $t$ is the thickness of the GO paper and $r$ is the curvature radius of the formed kink under compression. The linear fitting of experimental points, as displayed in the red line of Fig. 14a, gives the average normal strain value $\varepsilon_x$ of 1.1 ± 0.1%, which is significantly larger than the uniaxial tensile strain (0.6%). The uniaxial tension and bending deformation are schematized in Fig. 14e, along with the breaking of interactions between the interlamellar water molecules (blue) during deformation. The good bendability of GO papers was also demonstrated by Lee *et al*. [158], as well as the polydopamine-embedded RGO papers. Besides, the 3D carbon films from self-assembly of GO platelets through a breath-figure method are also mechanically flexible, along with tunable open porous morphologies [159]. For the RGO assembly film with a curvature radius of about 2 mm, even the pores are slightly elongated, the macropore morphology is well-maintained. After releasing the strain, the film is restored to its original planar geometry and the macropores also recover their initial shapes, indicating high flexibility of the film.

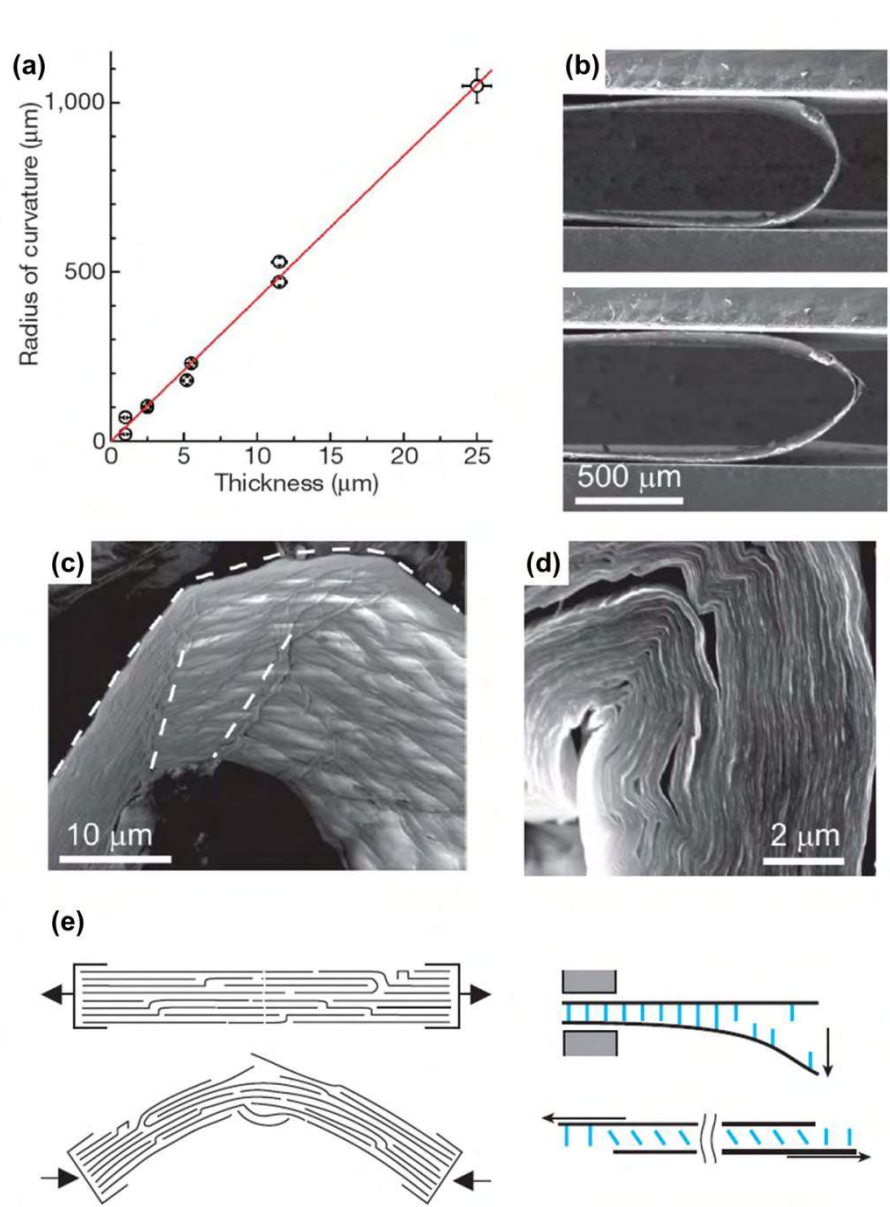

**Fig. 14** Bending experiments for GO paper samples with different thicknesses: (a) the curvature radius at which the GO strip loses mechanical stability during bending, where the red line is a linear

fit of experimental points; (b) two SEM images of a strip cut from a 5.2-mm-thick GO papers during compressing, where the upper image is taken immediately before and the lower one after the sample buckled; (c) an SEM image of a 1-mm-thick curved GO paper strip with two major buckling creases (dotted white lines); (d) a high-resolution SEM image of an 11-mm-thick buckled strip; (e) schematics of the uniaxial in-plane load-to-fracture and a bending-to-buckling test, where the right parts show the broken of interactions between the interlamellar water molecules (blue), which hold neighboring GO sheets together. Reproduced from Ref. [147].

Schniepp *et al*. [160] studied the bending properties of RGO using atomic force microscopy (AFM) complemented with computational modeling. They demonstrated that RGO can be folded easily and reaches a stable folded state, most likely through attractive van der Waals (vdW) forces. Figure 15 shows the detailed folding process by AFM from the completely flat conformation to a stable conformation. Moreover, the RGO can be folded back and forth at exactly the same position multiple times, suggesting its good flexibility. Huang *et al*. [161] also demonstrated that the RGO is ultraflexible which can be single folded, double folded and crumpled, but springs back to its original shape without yielding or plastic deformation after removing the applied stress.

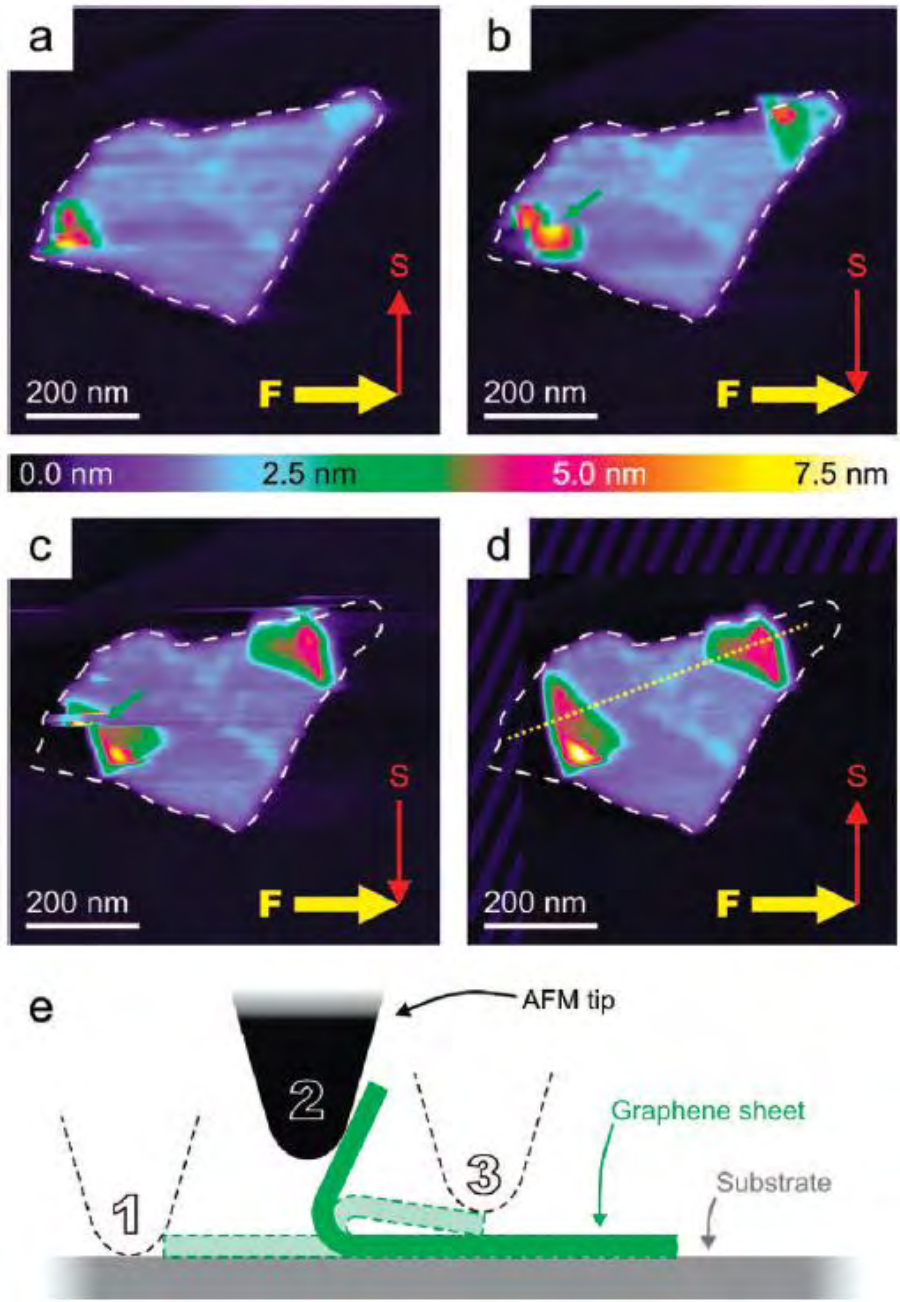

**Fig. 15** Folding the RGO by contact mode AFM: (a) the completely flat conformation; (b) the partially folded state with top right and bottom left edges folded; (c) folding progresses and (d) the stable folded conformation. The hatched area was not part of the actual scan but added later to make comparison with (a)–(c) easier. The green arrows in (b) and (c) indicate lines in which folding events occurred. (e) Illustration of the folding process: (1) the tip first lifts a part of the sheet (green) off the substrate (gray). (2) The sheet is bent until it touches itself (3). Reproduced from Ref. [160].

Due to existence of defects and functional groups, the GO shows intrinsic wrinkles. DFT

simulations [160] indicated that folding of RGO sheets with pre-existing kink lines is energetically favored over folding of unwrinkled sheets (graphene). Especially, in the stable folded conformation, the kink line is located in the apex of wrinkle. Therefore, wrinkles play important roles on the out-of-plane mechanical properties of GO. The out-of-plane wrinkles of GO can be created by shear stress [131] and Langmuir−Blodgett compression [162]. For the graphene functionalized with 7.5% hydroxyl groups, it starts to wrinkle at a critical shear strain of 4.5%, as well as 4% for graphene functionalized with 7.5% carboxyl groups. Both of the critical shear strains are lower than that of pristine graphene (6%), as shown in Fig. 16a. Figure 16b shows a typical wrinkling of a pristine graphene. Beyond the critical strain, a sudden drop of shear stress occurs and the average perpendicular displacement increases greatly, indicating further wrinkling of functionalized graphene. Besides, the wrinkles prefer to occur along the direction about 30º inclined to the shear direction. The degree of wrinkling at the edge is higher than that in the center.

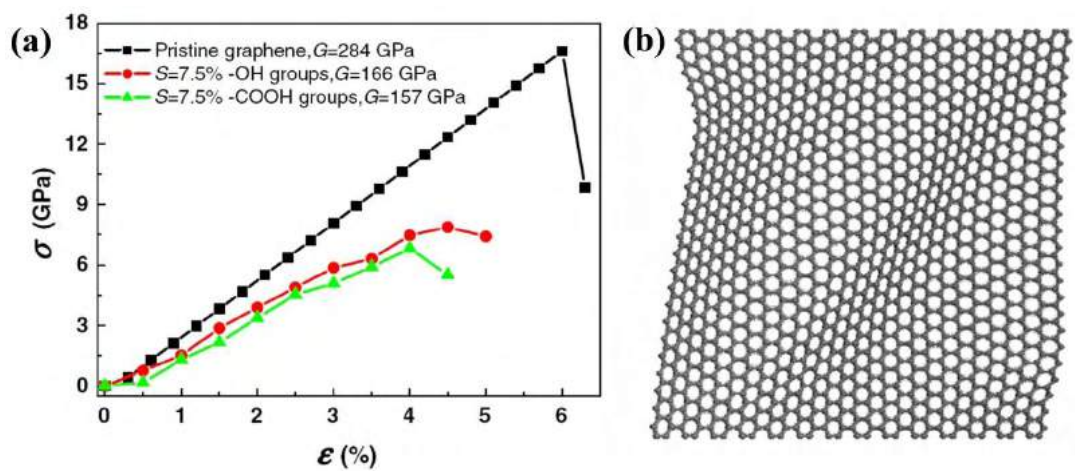

**Fig. 16** (a) Shear stress–shear strain curves for pristine and functionalized graphene; (b) Onset of wrinkling of a pristine armchair graphene at a critical strain of 6%. Reproduced from Ref. [131].

Directly measuring the bending modulus of GO monolayer was achieved in 2016 by Poulin *et al*. [14] through characterizing the flattening of thermal undulations in response to shear forces. As schematically plotted in Fig. 17a, under shear forces, the undulation of amplitude $h(r)$ of GO flakes is flattened, leading to decrease of the average spacing distance from $d_0$ to $d$. Then, from the following equation, the bending modulus ($\xi$) of GO monolayer can be determined:

$$\frac{d_0 - d}{d_0} = \frac{kT}{24\pi\xi}\ln(1 + \frac{R^6\eta^2\dot{\gamma}^2}{24\pi^4\xi^2}),\qquad(2)$$

where $kT$ is the thermal energy, $R$ is the GO flake diameter, and $\eta$ is the water viscosity, respectively. As a result, the GO monolayer has a bending modulus of ~1 $kT$, fitted from the curves of normalized variation of the spacing as a function of shear rate presented in Fig. 17b. Compared with the bending modulus (40–80 $kT$) of graphene [14], GO monolayer shows much lower bending modulus, indicating its superflexibility.

Afterwards, Tardani *et al*. [163] obtained the bending stiffness of GO by characterizing the wavelength of periodic winkles in GO flakes. The aligned wrinkles were produced through an evaporative lithography method to treat the GO concentrated dispersions. It was found that the bending stiffness ($B = E_s\lambda^3$ with $E_s$ being the Young's modulus of substrate and $\lambda$ being the wavelength of the wrinkles) increases with the GO concentration exponentially, as shown in Fig. 17c. The extraordinary low bending stiffness in the order of $10^{-9}$ N·m indicates the ultrahigh flexibility of GO flakes.

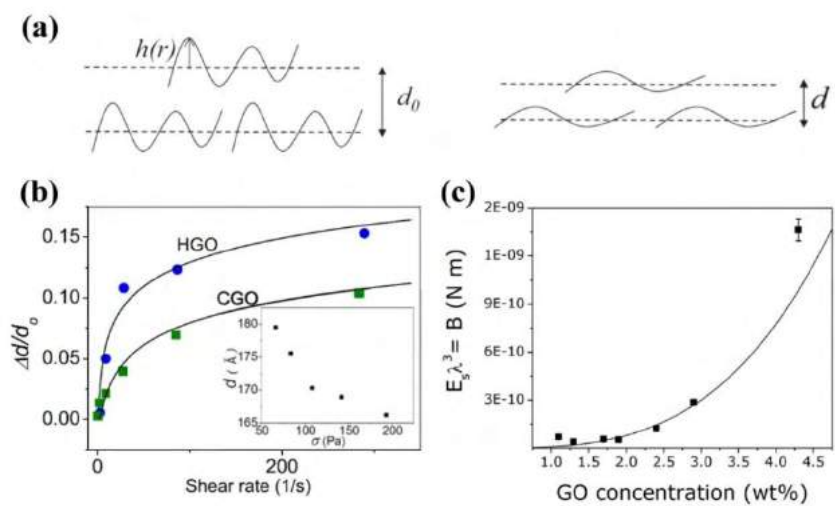

**Fig. 17** (a) Schematic of undulating GO flakes at rest (Left) and under shear (Right). (b) The relationship between the normalized spacing variation and the shear rate for the homemade GO (HGO), blue circles, and commercial GO (CGO), green squares, separately, as well as the fitting date using Eq. 2, black lines. Change of the smectic distance $d$ with the shear stress $\sigma$ in the HGO sample is shown in insert figure. Reproduced from Ref. [14]. (c) Bending stiffness as a function of GO concentration. Reproduced from Ref. [163].

Theoretically, Incze *et al*. [164] investigated the bending mechanical properties of GO with different coverage of epoxy groups by DFT calculations coupled with the continuum elasticity theory. The GO model is built on a (8 × 9) graphene supercell containing 288 carbon atoms with only epoxy groups located on one or both sides of the graphene surface. The number of oxygen atoms varies from 12 (coverage of 4.16%) to 36 (coverage of 12.5%). Due to existence of epoxy groups, the optimized GO models are corrugated, as shown in Fig. 18a. To calculate the elastic bending coefficient of GO, a displacement imposed on each atom along the $z$ axis, perpendicular to the basal $(x, y)$ plane, is supposed to be

$$\zeta = \zeta_0 \cos(2\pi x / L_x), \tag{3}$$

where $L_x$ and $L_y$ are the lateral dimensions of the GO structure. Particularly, for small values of the deformation amplitude $\zeta_0$, the curvature $\gamma$ and the in-plane deformation $\varepsilon_{xx}$ can be expressed as:

$$\gamma = d^2\zeta / dx^2, \tag{4}$$

and

$$\varepsilon_{xx} = \left(d\zeta / dx\right)^2. \tag{5}$$

Thus, the total variation of elastic energy of the deformed GO respect to the reference state at $\zeta_0 = 0$ is

$$\Delta E_c = 4\pi^4 \frac{L_y}{L_x^3} D\zeta_0^2 + 6\pi^4 \frac{L_y}{L_x^3} \Lambda\zeta_0^4, \tag{6}$$

where $D$ is the elastic bending coefficient and $\Lambda$ is the elastic extension coefficient. In Eq. 6, the first term is the cost in bending energy (curvature), proportional to $\zeta_0^2$; and the second term is the extensional energy proportional to $\zeta_0^4$. Based on these equations, the calculated elastic bending coefficient $D$ shows a linear increment as the oxygen concentration increases, as shown in Fig. 18b. Also, GO shows much larger $D$ than the pure graphene. For example, the $D$ of GO with an oxygen concentration of 12.5% is about 40 times larger than that of graphene (0.9 eV). This increase of $D$ can be attributed to the existence of epoxy groups, which bring corrugated structure and rigidify the C–C bond locally. Meanwhile, the elastic extension coefficient $\Lambda$ doesn't change with the oxygen concentration.

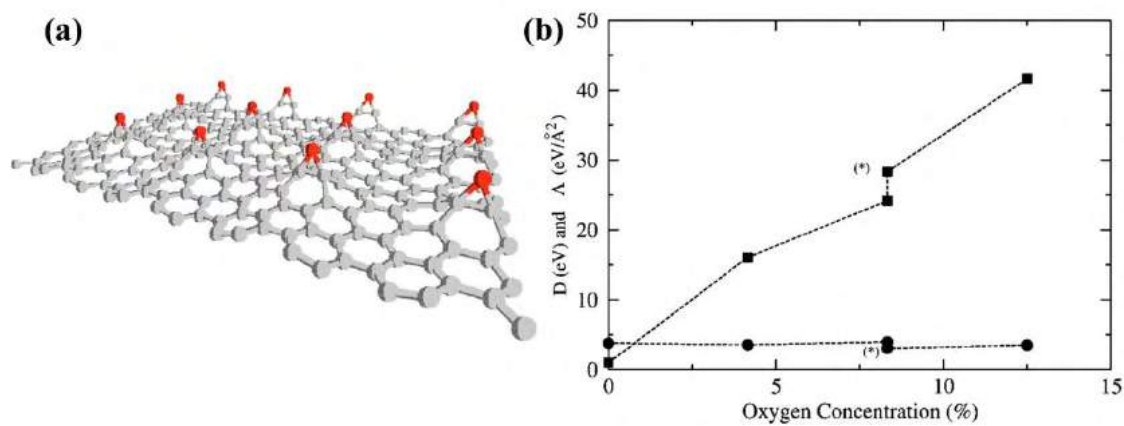

**Fig. 18** (a) The corrugated GO model with carbon in gray and oxygen in red; and (b) change of the bending coefficient $D$ (squares) and the extension coefficient $\Lambda$ (circles) with respect to the oxygen concentration. The two points at the same 8.3% concentration represent two configurations of adsorbed oxygen atoms, namely on one side or on the two sides (marked by a star) of the graphene plane. Reproduced from Ref. [164].

On the other hand, GO can be added into other materials to improve their compressive or bending mechanics. Using GO sheets as template skeletons and metal ions as catalysts and linkers, Wang *et al.* [165] fabricated the composite aerogels composed of resorcinol formaldehyde (RF) and GO. This RF-GO aerogel shows high compressibility, which can tolerate a compressive strain as high as 80% and quickly recover their original morphology after releasing the compression. Cheng *et al.* [166] added GO to polymer composites consisting of surfactant-wrapped/doped polyaniline (PANI) and divinylbenzene (DVB). It was found that even with only 0.45 wt% loading of GO, bending modulus of the composites exhibits at least 29.8% increment. As shown in Fig. 19a, for the PANI–DVB nanocomposites with 50% PANI and 50% DVB, the bending modulus increases as the content of GO increases and gradually converges to the highest value of 1422 MPa at the GO content of 0.45%, which is 29.8% larger than that the neat PANI–DVB composites. The bending strength also presents an overall increasing trend as the increase of GO content and achieves a high value of 46.6 MPa at the GO content of 0.6%. Moreover, the composite (50% PANI and 50% DVB) with GO shows larger stiffness than the composites without GO, as indicated by the load–deflection curves shown in Fig. 19b.

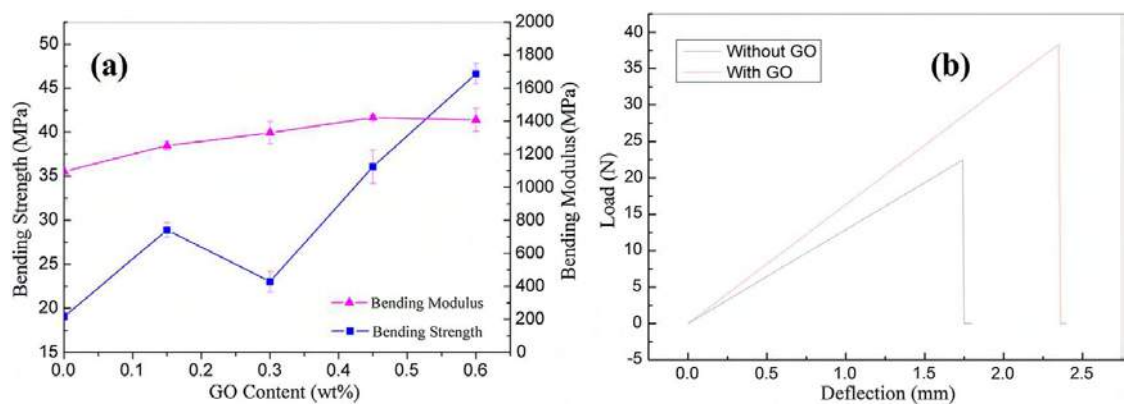

**Fig. 19** (a) Bending strength and modulus of the PANI–DVB nanocomposites with different GO content; (b) the load–deflection curves of the PANI–DVB nanocomposites with and without GO. Reproduced from Ref. [166].

In general, GO presents excellent pliability, which is also tunable depending on the structural details. The superior flexibility particularly makes GO promising in flexible devices. However, more theoretical works are required to understand the relationship between the flexibility and the GO structure details.

## 3 Electronic properties of GO and electric devices

### 3.1 Electronic band structure

The electronic band structure of GO is of great interest, since it is the key to understanding the electronic and optical properties of this material. In particular, the energy gap of GO is tunable by manipulating the chemical composition and atomic structure, making this material very attractive for semiconductor industry. So far, extensive efforts have been made to establish a relationship between the bandgap and chemical structure of GO.

On the experimental side, FETs using GO or RGO as the channel materials exhibit on/off ratios up to $10^3$ [167, 168], indicating bandgap opening in GO. Electron and photon spectroscopic techniques have also been utilized to directly measure the bandgap or optical gap of GO [169-173]. These results show that oxygen functionalization can open a bandgap up to a few eV in GO, and the gap increases with the oxidation level. Guo *et al.* [174] employed the femtosecond laser to reduce and pattern GO, and measured the bandgap of GO by diffuse reflectance spectra (Fig. 20a). By tuning the laser power, the oxygen coverage rate of GO drops from 0.62 to 0.17, and the bandgap decreases from 2.4 eV to 0.9 eV accordingly (Fig. 20b). The FETs based on GO channels show ON/OFF ratios of 6.5−56, as the bandgap is reduced from 2.1 eV to 1.7 eV. To corroborate these results, they built GO models with epoxy groups locally forming ordered pattern on one side of graphene. DFT calculations showed that the bandgap of these GO models increases with the O:C ratio, and attains 2.74 eV with an O:C ratio of 0.5 (Fig. 20c). The calculated density of states (DOS) suggests that the formation of bandgap in GO is due to the bonding of $p$-orbital electrons from oxygen species and graphene, which leads to the splitting of $p$-bonding and $p$-antibonding bands.

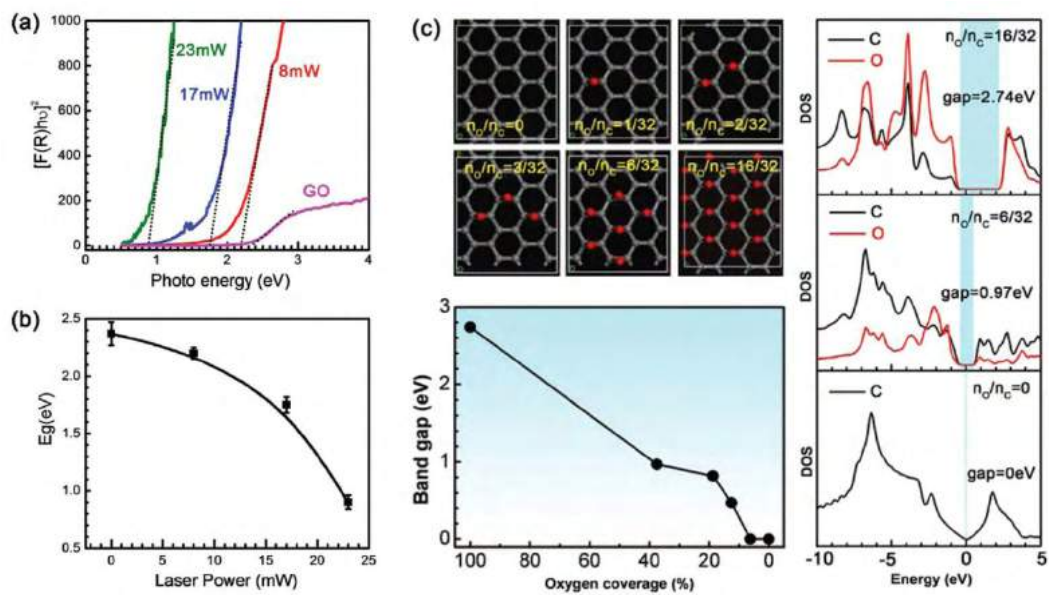

**Fig. 20** (a) Absorption spectra related to the Kubelka−Munk function of GO. (b) Bandgap ($E_g$) of laser reduced GO as a function of laser power. (c) DFT modeling of the bandgap of GO with various oxygen contents. Left top panel: model structures of GO including epoxy groups with increasing O:C ratios. Left bottom panel: calculated bandgap of the GO models as a function of oxygen coverage rate. Right panel: $p$-orbital density of states (DOS) of C and O atoms for selected GO models with oxygen coverage rates of 100%, 37.5% and 0%, respectively. Reproduced from Ref. [174].

On the theoretical side, numerous studies have been conducted to investigate the electronic band structures of different GO models [95, 96, 175, 176]. Ito *et al.* [177] focused on the GO systems containing epoxy groups only, in which the oxygen species uniformly distribute on one side of graphene. Their DFT calculations show that these GO models present bandgaps of 0.02, 0.22, 1.44, and 3.39 eV with the O:C ratios of 0.125, 0.167, 0.25, and 0.5, respectively (Fig. 21a). The width of π band becomes narrower as the oxidation content increases, that is, the transfer integral between carbon atoms in graphene decreases, indicating the localization of electronic states due to the oxygen adatoms (Fig. 21b-d). Yan *et al.* [16] investigated GO systems in ordered phases and at high oxidation levels. They found that the fully oxidized GO structures present bandgaps of 2.0–4.0 eV. For partially oxidized GO phases, the gap size varies significantly with the relative fraction of epoxy and hydroxyl groups. In particular, the gap vanishes for the intermediate ordered phases containing armchair graphene ribbons with $3n + 2$ ($n$ is an integer) rows of atoms or zigzag ribbons with even number rows of atoms.

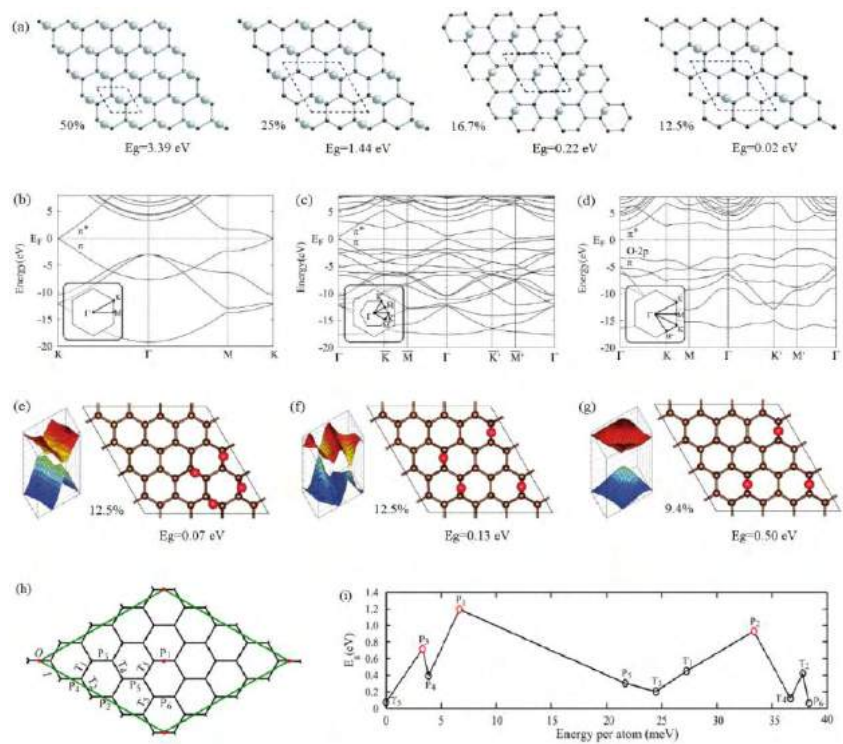

**Fig. 21** (a) GO models including epoxy groups in uniform ordered patterns on graphene, with O:C ratios of 50%, 25%, 16.7% and 12.5%. The numbers indicate the bandgap of each model calculated by DFT. The band structures of the GO models with the O:C ratio of 16.7% and 50% are shown in (c) and (d), respectively. The band structure of pristine graphene is shown in (b) for comparison. Reproduced from Ref. [177]. (e), (f), and (g) GO models including epoxy groups randomly distributed on graphene, with O:C ratios of 12.5%, 12.5% and 9.4%, respectively. The numbers indicate the bandgap of each model. The constant energy surfaces of conduction and valence bands are shown on the left of each model. Reproduced from Ref. [105]. (h) Adsorption positions for the second oxygen atom with respect to the first fixed oxygen atom (O), indicated by $P_n$ ($n = 1$ to 6) and $T_n$ ($n = 1$ to 5), respectively. (i) The bandgap of the configurations in (h) versus the average energy per oxygen atom. Reproduced from Ref. [178].

In addition to the degree of oxidization, the spatial distribution of oxygenated groups has a big impact on the electronic band structure of GO [105, 178] (Fig. 21e-g). Lian *et al.* [178] showed that a big bandgap can be opened in the highly reduced GO. For epoxy-functionalized graphene with O:C ratios of 0.063 and 0.094, the bandgap can attain as much as 1.19 eV and 1.58 eV, respectively, for particular patterns of oxygen species on the carbon layer (Fig. 21h and i). Analysis of DOS suggests that the arrangement of oxygenated groups affects the hybridization between the *p*-orbital of oxygen species and the frontier orbitals of graphene. Large bandgap is induced when oxygen atoms strongly interact with the energy bands of graphene near the Fermi level.

Using the structural models of GO described in Section 2.1, Liu *et al.* [17] explored the electronic properties of GO in disordered phases with different oxygen coverage and epoxy to hydroxyl ratios. As displayed in Fig. 22a and b, they found existence of midgap states in-between the conduction band minimum (CBM) and the valence band maximum (VBM) originated from $sp^3$ carbon species. The energy gap between CBM and VBM increases from 0.53 eV to 1.77 eV, as the oxygen coverage rate increases from 0.2 to 0.5, when the hydroxyl to epoxy ratio is fixed at 2.0. At a fixed coverage rate of 0.5, the energy gap changes from 1.46 eV to 2.45 eV, as the hydroxyl to epoxy ratio increases from 0.5 to 8.0. These midgap states may be induced by the monovalent oxygenated groups chemisorbed on the carbon layer in the disordered GO [98, 179]. For an isolated hydroxyl group on graphene, the electronic band structure shows a flat band at the Fermi level (Fig. 22c), leading to a small peak in the DOS. This is because the carbon atom bonded to the hydroxyl group is decoupled from the graphene π-electron system, resulting in a local imbalance between the numbers of carbon atoms in the two sublattices. For an epoxy group, the bonding with graphene is divalent, and thus the electronic band structure reserves the conical dispersion relation near the Fermi level. Due to the strong hybridization of *p*-orbitals of the oxygen atom and graphene, the Dirac point is shifted in **k** space and the Fermi velocity is significantly reduced compared to that of graphene. Studies show that these impurity states as scattering centers deteriorate the transport properties and limit the carrier mobility of GO/RGO [180-182]. Moreover, the electronic transport properties rely on the specific distribution of oxygen functional groups on the graphene plane. Compared to the same amount of randomly distributed oxygenated groups on graphene, clusters of oxygen functional groups weaken the electron scattering and suppress the contribution to the resistivity [183].

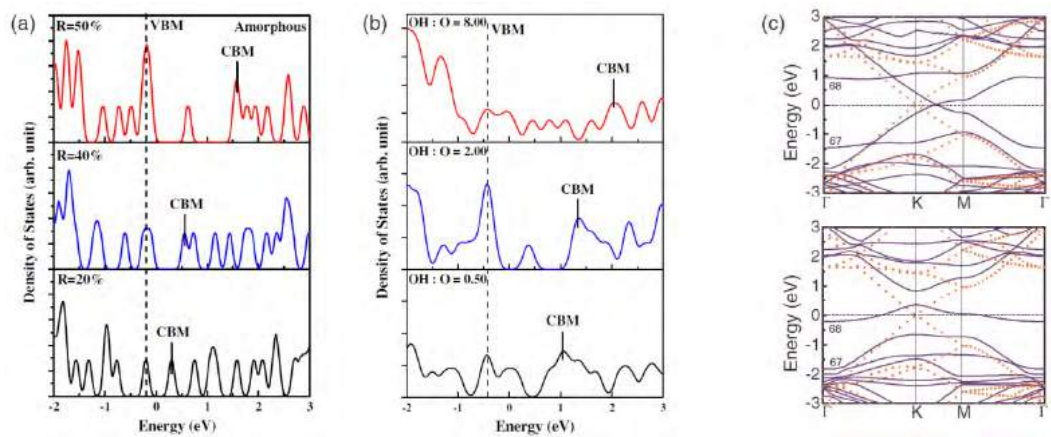

**Fig. 22** (a) DOS of amorphous GO models as a function of oxygen coverage rate. (b) DOS of amorphous GO with various relative fractions of hydroxyl and epoxy groups and with a fixed oxygen coverage rate of 50%. Reproduced from Ref. [17]. (c) DOS (solid lines) for a single epoxy (top) and hydroxyl (bottom) group on graphene. The dotted lines indicate the DOS of pristine graphene. Reproduced from Ref. [98].

To conclude, the electronic band structure of GO has a complicated dependence on the type, concentration and distribution of oxygen functional groups. By manipulating the chemical structure of GO, a variety of electronic properties can be achieved, which are in turn useful for many technological applications, such as film conductors [29, 88, 156, 184], FETs [8, 185, 186], and sensors [31-33, 187, 188], as we will discussed below.

## 3.2 Transport property

In GO, the chemisorption of oxygen functional groups disrupts the extended π conjugated network of the parent graphene, rendering GO an insulator. Hence, instead of common or highly oxidized GO, more attentions have been paid to RGO, which is a promising substitute to graphene for electronic devices. However, RGO still retains a considerable amount of oxygen residues and structural defects, which deteriorate its transport properties. The mechanism of electron conduction in RGO also differs from that of pristine graphene.

Gómez-Navarro *et al.* [189] first reported the electronic transport properties of chemically-derived RGO sheets on the silicon substrate. The GO monolayer has a differential electrical conductivity on the order of $10^{-3}$ S/m at a bias voltage of 10 V, close to the insulating behavior. By chemical reduction or hydrogen plasma treatment, the resulted RGO monolayer shows enhanced electrical conductivities of 0.05−2 S/m, which is still about three orders of magnitude lower than that of pristine graphene (Fig. 23a). As temperature decreases from 298 K to 4 K, the conductivity of RGO drops by over three orders of magnitude, compared to one order of magnitude decrease for graphene. The current of RGO exhibits a dependence on temperature as $\sim\exp(-1/T^{1/3})$ (Fig. 23b), in good agreement with the 2D variable range hopping mechanism for charge transport [190]. Particularly, after a Joule heating process at $\sim$2750 K in less than 1 minute, an extremely high conductivity of 3112 S/cm for the thin RGO films ($\sim$4 μm) can be achieved [191].

Similar to graphene [192], the field effect transport of RGO shows a nearly ambipolar dependence of resistivity on the gate voltage (Fig. 23c). The asymmetry between the effects of the positive and negative gate voltages may be caused by the asymmetric scattering of charged impurities in RGO. The room-temperature mobilities of RGO range from 2 to 200 cm$^2$ V$^{-1}$ s$^{-1}$ for holes, and from 0.5 to 30 cm$^2$ V$^{-1}$ s$^{-1}$ for electrons, respectively. These values are about two orders of magnitude lower than that of graphene (3000−10000 cm$^2$ V$^{-1}$ s$^{-1}$) due to the defective nature of RGO [193, 194]. At low bias voltages, bilayer RGO has a higher conductivity by over a factor of two than the monolayer RGO, since the conductivity of the first layer is limited by the interaction with the substrate. However, further increase of the number of layers does not affect the conductivity of RGO much (Fig. 23d).

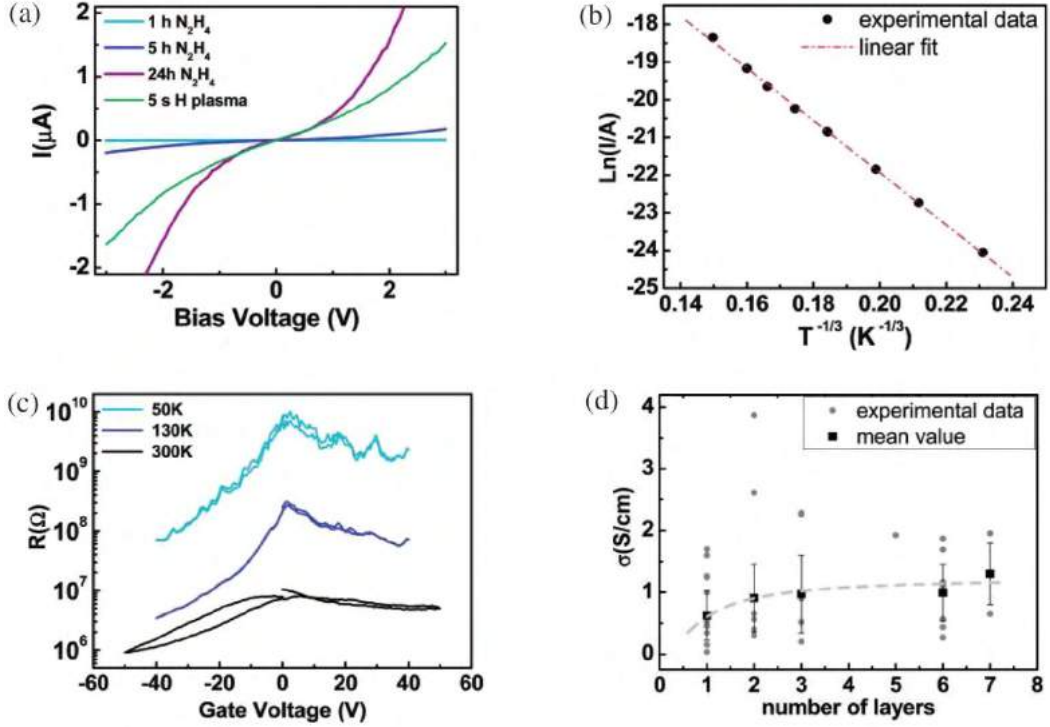

**Fig. 23** (a) Room-temperature current-voltage (I-V) curves of GO monolayers reduced by hydrazine for 1, 5 and 24 hours, and reduced by hydrogen plasma for 5 seconds, measured in ambient conditions. (b) Semilogarithmic plot of I versus $T^{-1/3}$ of a fully reduced GO monolayer, with a bias voltage of 0.2 V. (c) Resistance versus back gate voltage of a fully reduced GO monolayer at three different temperatures, with a bias voltage of 100 mV. (d) Mean volume conductivity as a function of the number of layers in a fully reduced GO film. Reproduced from Ref. [189].

The presence of oxygen residues and defects in RGO highly affects the charge transport mechanism of the system. Experimental TEM studies show that RGO contains intact graphene regions interspersed with $sp^3$ carbon and defective regions [94]. Such inhomogeneous disordered structure induces localized electronic states near the Fermi level. Hence, the electron conduction in RGO is dominated by phonon-assisted quantum tunneling between these localized states, rather than excitations of electrons from the valence band to the conduction band [195, 196]. In this so-called variable range hopping (VRH) model, the conductivity ($\sigma$) follows the characteristic temperature dependence of [197]

$$\sigma = \sigma_0 \exp[-(\frac{T_0}{T})^{\frac{1}{3}}].\tag{7}$$

The hopping parameter $T_0$ has an expression of

$$T_0 = \frac{3}{k_B N(E_F) L_1^2},$$ (8)

where $k_B$ is the Boltzmann constant, $N(E_F)$ is the electron density of state near the Fermi level, and $L_1$ is the localization length of the electron wavefunction involved in the hopping process. Actually, VRH model for charge transport in RGO has been supported by many experimental studies [198-200].

Kaiser *et al.* [201] investigated the electronic transport properties of RGO down to a very low temperature of 2 K. They found that at low bias voltages, the current of RGO follows the temperature dependence given by the VRH model. By applying a bias voltage of 2 V, the current attains a constant value below 100 K. Hence the electron conduction at low temperature and high bias voltages relies on tunneling of charge carriers between the localized states promoted by the strong electric field. In addition, they estimated the minimal energy barrier for the tunneling based on the potential drop yielded by field-driven conduction, which is on the order of 40 meV.

Eda *et al.* [167] explored the evolution of electronic transport in GO at various degrees of reduction. They found that the conductivity of GO increases by over five orders of magnitude upon reduction. For the lightly reduced GO, field effect transport shows distinct off state at a temperature of 78 K and on-off ratios greater than $10^3$, much larger than that of graphene nanoribbons [202] (Fig. 24a). The room-temperature mobilities of RGO are in the range of $10^{-3}$–$10^0$ cm$^2$ V$^{-1}$ s$^{-1}$, depending on the reduction degree (Fig. 24b). The highly reduced GO sample possesses mobility independent of the electric field, indicating the band-like charge transport in the system (Fig. 24c). The carrier density of RGO changes with temperature and exhibits two regimes of different slopes, which suggests that the density of states is non-monotonous (Fig. 24d). The bandgap is extracted from the low- and high-temperature regimes, which decreases as the reduction degree increases, and approaches zero in highly reduced GO (Fig. 24e). For lightly reduced GO, the apparent energy gap between the tail states of the valence and conduction bands is about 55 meV. Further analysis shows that, as reduction proceeds, the density of state near the Fermi level increases, while the localization length does not change much (Fig. 24e). These results suggest that further reduction of GO does not lead to delocalization of charge carriers, but increases the number of available hopping sites.

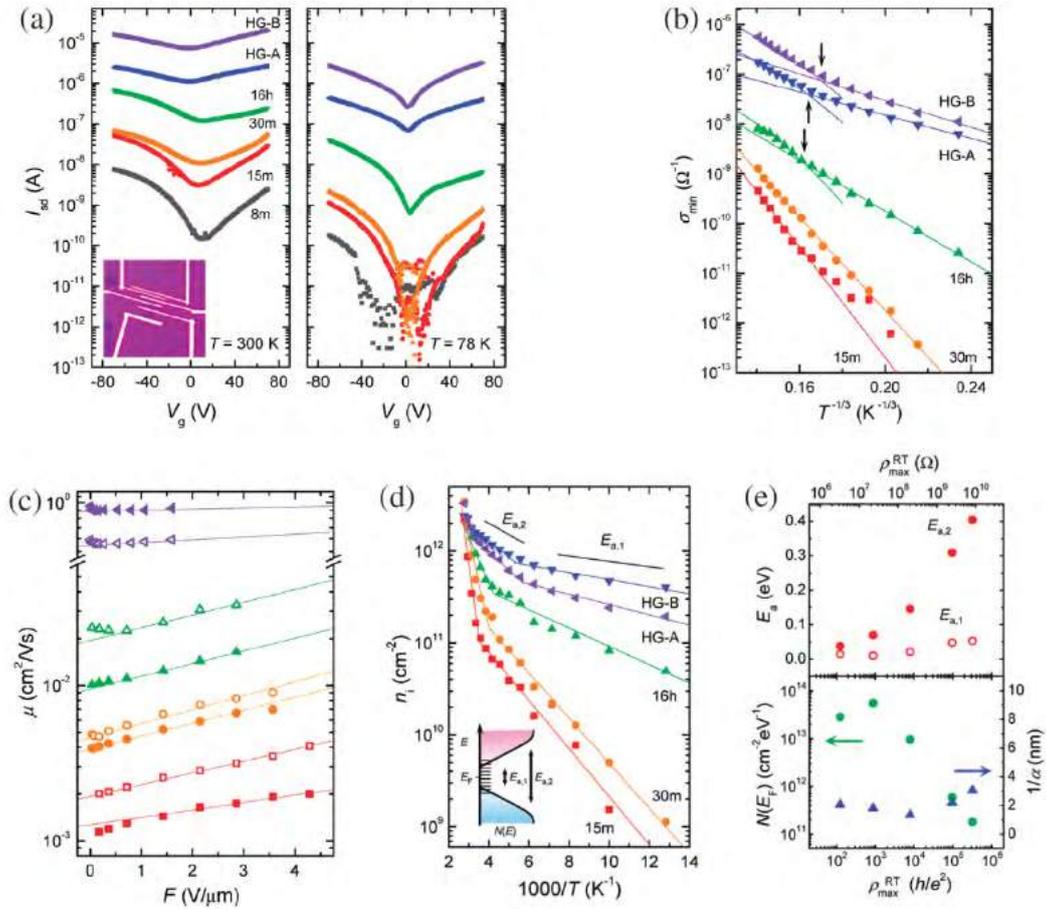

**Fig. 24** (a) Current versus gate voltage of RGO with different degrees of reduction measured at 300 K (left) and 78 K (right). The inset shows an optical micrograph of a typical RGO device. (b) Minimum conductivity versus $T^{-1/3}$ of RGO. (c) Field effect mobility of RGO as a function of average electric field F. Open and filled symbols represent hole and electron mobilities, respectively. (d) Arrhenius plot of intrinsic carrier mobility of RGO. Two regions are fitted with a line with different slopes. The inset is a schematic illustration of the energy band structure of RGO showing localized states near the Fermi energy and band tails. (e) Top panel: thermal activation energies as a function of the room-temperature maximum resistivity of RGO. Bottom panel: density of states at the Fermi level $N(E_F)$ and localization length $1/\alpha$ as function of the room-temperature maximum resistivity of RGO. Reproduced from Ref. [167].

In short, GO undergoes transition from an insulator to a semiconductor and finally to a semimetal upon gradual reduction. The electronic transport in RGO is governed by the hopping between localized states near the Fermi level. Although significantly being enhanced with regard to GO, the electrical conductance and charge carrier mobility of RGO are still much lower than those of pristine graphene, due to the presence of oxygen residues and defects.

### 3.3 Transparent and flexible electronic devices

Transparent film conductors are of great importance for optoelectronic devices, such as liquid crystal displays, photovoltaic cells, and organic light emitting diodes (LEDs). Currently, owing to the excellent electric conductivity and optical transparency, ITO and FTO are the two commercial standard materials for the transparent film conductors [30, 88]. For example, ITO has a transparency of >90% at a wavelength of 550 nm, sheet resistances of 10–30 $\Omega$/square and a favorable work function of 4.8 eV [203]. However, with the development of electronic technology, markets for the flexible transparent conductors grow rapidly, including touch screens, flexible displays, printable electronics, solid-state lighting, and thin film photovoltaics. These new markets put forward further requirements for the transparent film conductors with high flexibility, light weight, low cost, and large-scale production [156], which are beyond the capabilities of ITO and FTO. Therefore, GO, which satisfies nearly all the above requirements and is compatible with the large-scale manufacturing processes, is a promising candidate for the transparent film conductors of next generation [30].

### 3.3.1 Transparent conductors

Due to the atomically thin layered structure, GO is highly transparent in the visible spectrum. However, GO is typically insulating because $\pi$ bonds in graphene are destroyed by the oxygenated groups. Fortunately, one can tune the conductivity of GO by adjusting the oxidization degree. Thus, to restore the inherently high electric conductivity of graphene, GO should be reduced. Particularly, highly reduced GO can be semimetallic with electronic states similar to those of the disordered graphene [18]. Through thermal annealing, RGO films with sheet resistances as low as $10^2$–$10^3$ $\Omega$/square and optical transmittances above 80% for 550 nm light are obtained, indicating potential application of RGO for the transparent conductors [19]. Note that although the unreduced GO monolayer shows high transmittance of 96%, it presents low electric conductivity [204]. Generally, as the optical transmittance increases, the film conductivity of RGO decreases inversely, as displayed in Fig. 25a. Such conflict between the optical transmittance and the electric conductivity was also demonstrated by Eda *et al.* [205]. Afterwards, a range of transmittance and sheet resistance values have been reported for the RGO samples obtained from various methods, including oxidation, exfoliation, dispersion, and deposition procedures [8]. However, all the results suggest that the conflict between the optical transmittance and the electric conductivity is universal [18], as summarized in Fig. 25b. Therefore, to improve the performance of GO-based transparent conductors, an appropriate compromise of optical transmittance and electric conductivity should be chosen.

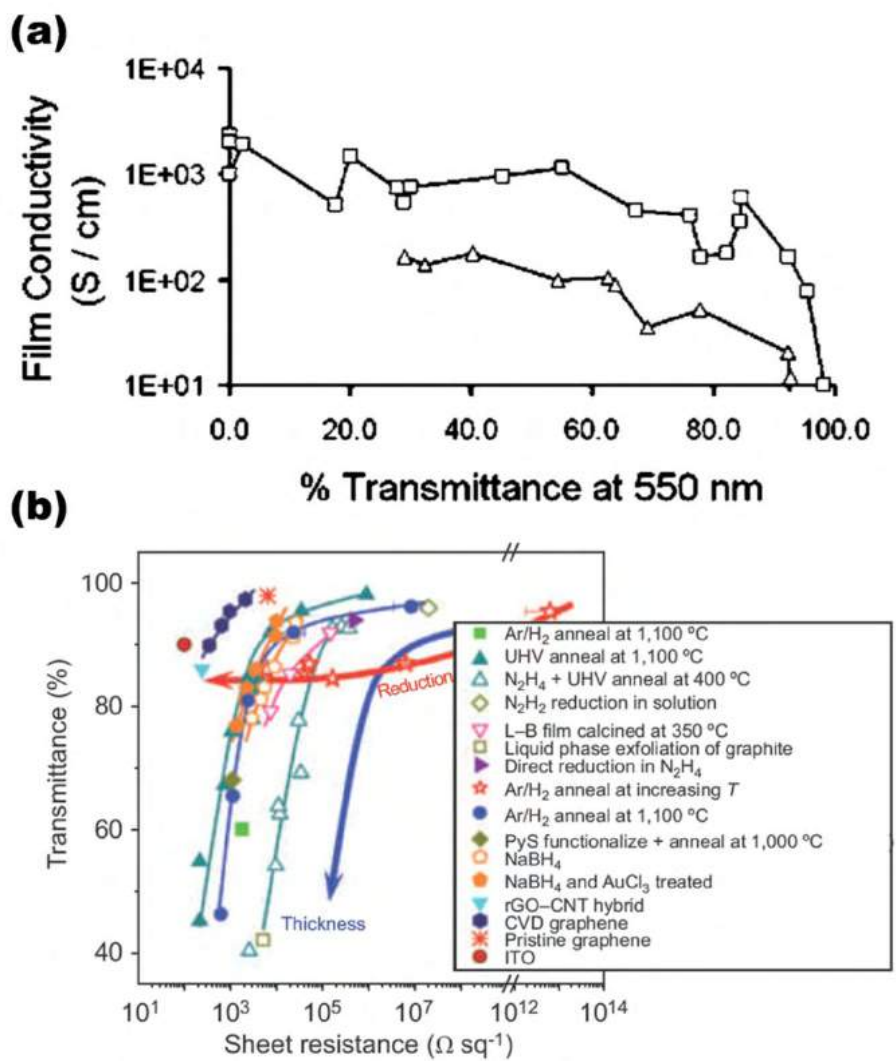

**Fig. 25** (a) Relationship between the film conductivity and the transmittance of RGO films. Reproduced from Ref. [19]. (b) Summarization of transmittance at wavelength of 550 nm as a function of sheet resistance of RGO films with different reduction and thickness reported previously. Reproduced from Ref. [18].

To improve the performance of GO-based transparent conductors, several approaches have been proposed to enhance the electric conductivity without impairing the optical transmittance. Generally, these approaches can be classified into three categories: (i) controlling the quality and morphology of RGO; (ii) doping; and (iii) formation of GO hybrids.

For the small-size RGO sheets, intersheet junctions will lead to a high contact resistance [30]. Therefore, a commonly used method is to fabricate large-area and high-quality RGO, which possesses enhanced electric conductivity but retains the optical transmittance. As demonstrated by Zhao *et al.* [184], the RGO film made from large-area sheets with an average area of ~7000 μm² exhibits a sheet resistance of 840 Ω/square at a transmittance of 78% at 550 nm wavelength, which is much lower than that of a RGO film (19.1 kΩ/square at a transmittance of 79%) made from small-area GO sheets of ~100–300 μm². For the large-area RGO membrane of $10 \times 10$ cm², its electric conductivity can be as high as $5.51 \times 10^5$ S/m [206]. Usually, for both as-prepared and reduced GO films, the electric conductivity consistently rises with increasing the average GO size [207], as demonstrated in Fig. 26a-d.

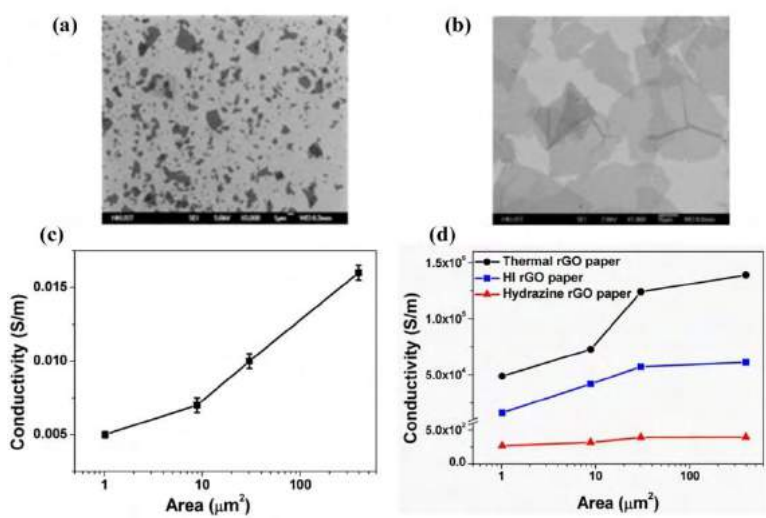

**Fig. 26** Typical SEM images of the (a) small ($1.07 \pm 1.7$ μm²) and (b) ultra large ($272.2 \pm 205.3$ μm²) GO papers. Electric conductivity of GO papers as a function of GO area (c) before and (d) after reduction. Reproduced from Ref. [207].

Additionally, high-quality RGO sheets with fewer defects can be obtained by improving the fabrication methods. For example, instead of the conventional sonication method, Jeong *et al.* [208] reported fabrication of large-area RGO with fewer defects through exfoliating GO into large-area nanosheets using a homogenizer via a shear-induced exfoliation mechanism and further reducing it by hydrazine in organic solvents. By applying shear stress in solution, high-quality RGO sheets can be generated without the need for post-annealing processes. As a result, the enhanced sheet resistance of the high-quality RGO sheets is found to be 2.2 k$\Omega$/square at a transmittance of 80% at 550 nm wavelength. Similarly, using the pressure-assisted reduction at relatively mild temperature, Shin *et al.* [209] reported high-quality RGO sheets with reduced sheet resistances to ~1 k$\Omega$/square at an optical transmittance of ~78% at 550 nm wavelength. Using a hydrohalic acid reducing method at low temperature based on the nucleophilic substitution reaction, high-quality RGO sheets without breaking their integrity and flexibility are achieved, which have a sheet resistance of 1.6 k$\Omega$/square and a transmittance of 85% at 550 nm wavelength [210].

Another way to enhance the performance of GO is chemical doping, which can increase the carrier concentration. Generally, there are two kinds of chemical doping: surface transfer doping and substitutional doping. The former can be achieved by electron exchange between RGO and dopants; the latter can be realized through heteroatom substitution. For the surface transfer doping, acid treatment is a simple and effective way. As mentioned above, GO can be formed by oxidizing graphene in acid solution, finally resulting in the oxygen-group functionalized graphene. As a consequence, oxygen atoms or ions binding with the dangling bonds of carbon in graphene can realize the charge doping [211]. For example, $HNO_3$ is a typical p-type dopant. Acid treatment of GO in a $HNO_3$ bath will reduce the sheet resistance by about 20–50% along with improved transmittance due to removal of the impurities originated from thermal annealing and etching of graphene film beneficial to electric conductivity [212].

In addition to the nitric acids, halogenating agents are also known to be useful in realizing charge transfer. In 2008, Eda *et al.* [205] proposed to enhance the electric conductivity of RGO through doping the films by dipping in thionyl chloride ($SOCl_2$). The obtained RGO thin films possess sheet resistances of ~$10^5$ $\Omega$/square at a transmittance of 65% for the 550 nm light because the $SOCl_2$ treatment would bring about p-type doping and shift the Fermi level into the valence band, as confirmed by the electronic structure calculations [213]. Similarly, doping of GO films in $SOBr_2$ solution will significantly improve the conductivity and transmittance with a sheet resistance of 1600 $\Omega$/square and a transmittance of 82% [212]. The enhanced conductivity of $SOBr_2$ doped GO films can be explained by the formation of –Br or –SOBr functional groups on the graphene basal layer, which show strong electronegativity and can act as electron acceptors to increase the hole density in GO, as shown in Fig. 27a-c.

Besides, Au ions can also act as electron-withdrawing groups. As reported by Jeong *et al.* [208], the fabricated high-quality RGO shows a sheet resistance of 2.2 kΩ/square at a transmittance of 80% at 550 nm wavelength. However, after doping with HAuCl₄, the sheet resistance of the high-quality RGO significantly decreases to 750 Ω/ square without degrading the transmittance.

On the other hand, substitutional doping can be achieved by hydrazine treatment. As reported by Hwang *et al.* [214], using hydrazine treatment and thermal reduction, N-doped RGO films are obtained and show a sheet resistance of as low as 300 Ω/square at a transmittance of 80% at 550 nm wavelength, where N atoms replace the original C atoms and form the C–N bonds. Also, the N-doped RGO with enhanced electric properties can be obtained through heating GO in a mixture of NH₃ and H₂ [159].

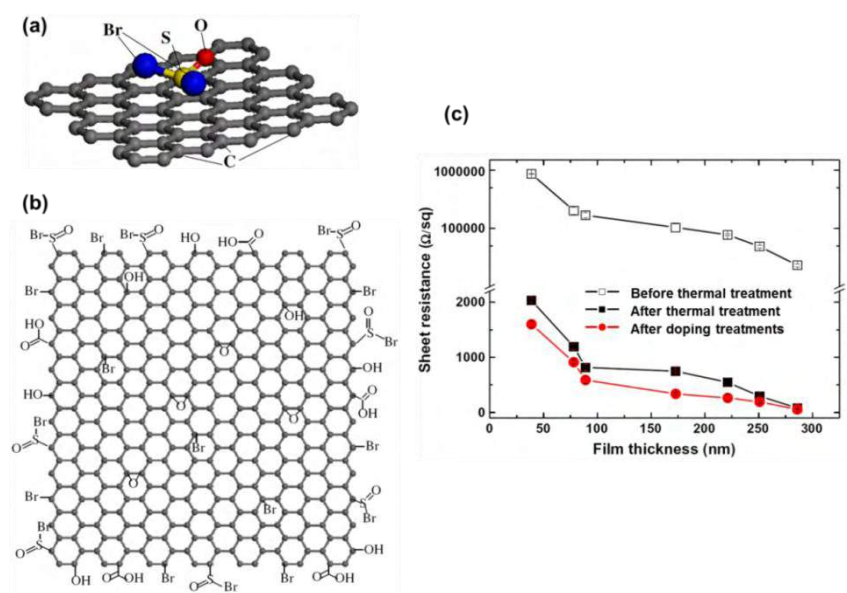

**Fig. 27** Schematic illustrations of SOBr₂ doped RGO: (a) atomistic structure of a SOBr₂ molecule absorbed on graphene surface, (b) final chemical structure of RGO sheet after doping treatments. (c) Comparison of sheet resistance at different treatment stages. Reproduced from Ref. [212].

Particularly, introducing short- and long-period wrinkles in the N-doped RGO films can further improve the performance. Mu *et al*. [215] demonstrated that due to the hierarchical wrinkles of the N-doped RGO layer, the transparent conductor can even show conductivity of 100–457 Ω/square and transmittance of 67–85% with high elasticity, which can be operated under large tensile strain of ~400% and bending deformation of ~180º.

Similar to the doping strategy, appropriately hybridizing GO with other materials is also an effective method to improve the performance of GO. In 2007, Watcharotone *et al.* [216] first reported the thin film transparent conductors based on RGO–silica hybrids, as illustrated in Fig. 28a. Although loading of RGO will reduce the optical transmittance of the silica, it can improve the electric conductivity, as shown in Fig. 28b. Meanwhile, the RGO–silica composites show consistently high transmittance (94% to 96%) in a wide wavelength range from 380 to 1000 nm even at the high RGO loading (11 wt.%), as shown in Fig. 28c. However, owing to the limitation of insulating matrix, the conductivity of RGO–silica composites is within 1 S/cm. To improve the electric conductivity, replacing the insulating silica by semiconducting or metallic materials is a good choice. It was found that for the RGO–carbon nanotube (CNT) hybrids, the sheet resistance is 240 Ω/square at an optical transmittance of 86% [217], as schematized in Fig. 28d. Moreover, depending on the hybrid RGO layer, the GO–single walled CNT thin films present remarkably low sheet resistance in the range of 180–560 Ω/square with an optical transmittance range of 77–86% at 550 nm wavelength [218]. In addition to the CNTs, RGO hybridized with metal nanowires (NWs) can also serve as high-performance transparent conductors [219]. For example, through dip coating by electrostatic force, silver nanowires (AgNWs) are successfully embedded into RGO films, which exhibits a sheet resistance of 150 Ω/square with a transmittance of 86% at 550 nm, comparable to the ITO [220]. Domingues *et al.* [221] demonstrated that the AgNWs hybridized with RGO deposited on Al foil present a sheet resistance of 74 ± 7 Ω/square with a transmittance of 89% at 550 nm. Meanwhile, using a dip-coating method, RGO–AgNW hybrid films show a sheet resistance as low as 27 Ω/square with a transmittance of 72% at 550 nm [222]. In addition, compared to copper (Cu) NW films, the RGO-Cu NW hybrid films exhibit lower sheet resistance at the same optical transmittance at 550 nm wavelength [223], as shown in Fig. 28e. It should be noticed that the higher the transmittance is, the larger the sheet resistance difference between the RGO–Cu NW hybrid films and Cu NW films. Besides, the RGO functionalized with 4-benzenediazonium sulfonate also exhibits an excellent optoelectrical property with a resistance of 1.1 kΩ for a transmittance of 89% at the wavelength of 550 nm [224].

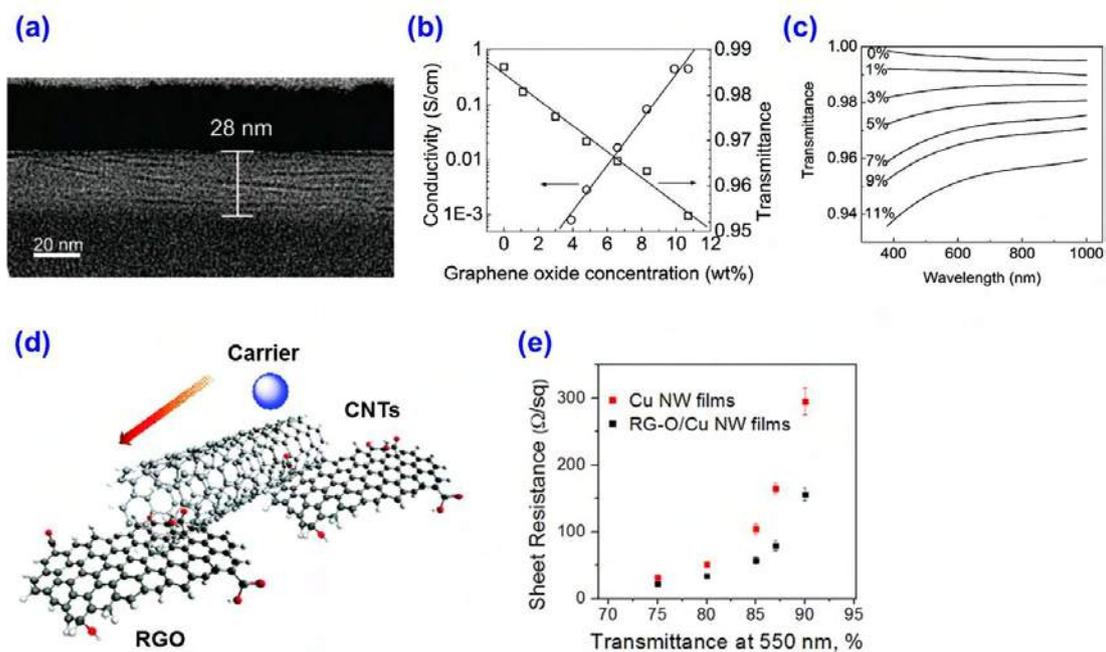

**Fig. 28** (a) TEM images of the cross sections of RGO–silica composite films with 11 wt. % of RGO after the high-temperature curing. From the bottom, the layers are glass substrate, composite film, Pt layer, and carbon layer, respectively. (b) Relationship between the bulk electric conductivity (circle) and light transmittance at 650 nm (square) of the RGO–silica films as a function of RGO concentration. (c) Optical transmittance for different wavelengths of the RGO–silica films with various RGO concentrations. Reproduced from Ref. [216]. (d) Schematic plot of RGO–CNT hybrid films. Reproduced from Ref. [217]. (e) Comparison of the transmittance and sheet resistance between the pure Cu NW films and RGO–Cu NW hybrid films. Reproduced from Ref. [223].

Generally speaking, GO/RGO materials are promising candidates for transparent conductors due to the high transparency and tunable electric conductivity. Through fabricating large-area and high-quality samples, doping or hybridizing, enhanced electric conductivity for GO/RGO-based transparent conductors can be achieved, making them useful in various fields, especially in the high definition display.

### 3.1.2 Flexible devices

Benefited from the functional groups and structural defect, GO possesses high flexibility, as discussed in Section 2.3. Considering the mechanical flexibility and electric conductivity, GO-based flexible electronic devices were widely studied [225]. Moreover, combined with the outstanding transparency, GO holds promise in constructing transparent and flexible devices. For example, He *et al.* [226] fabricated the all-RGO thin film flexible transistors. The transistor consists of solution-processed RGO electrodes and a micro-patterned RGO channel. Bending test shows that this transistor is very flexible. Moreover, the resistance only degrades by ~1% after the initial 200 bending cycles and then becomes very stable without obvious resistance change even up to 5000 cycles at the bend radius of 4 mm. Such all-RGO-based devices exhibit good performance with an ON/OFF ratio of 3.8 and a Dirac point of +0.4 V in the electrolyte-gating, which can be used to detect protein with good sensitivity and selectivity.

Liu *et al.* [227] also reported a nonvolatile memory device completely composed of RGO films. In this device, both the top and bottom electrodes are made up of highly reduced GO films obtained from high-temperature annealing of GO, whereas the active material is made up of lightly reduced GO obtained by low-temperature annealing and then light irradiation of GO, as illustrated in Fig. 29a and b. The fabricated diode shows electric bistability and nonvolatile memory effect with a current ON/OFF ratio of $10^2$. The memory device exhibits strong retention ability with a retention time as long as $10^3$ s under ambient conditions and its ON/OFF ratio shows no significant variation. In particular, this all-RGO-based memory exhibits excellent flexibility with desirable long life time under retention and bending tests, as shown in Fig. 29c. After repetitive bending and relaxing cycles of this device at a voltage of 2 V with a tensile strain of about 2.9%, there is no noticeable electrical degradation in both the ON and OFF states up to 1000 iterations (Fig. 29d), which indicates a high mechanical endurance of the fabricated device. The high flexibility of the device is mainly attributed to three factors: (i) the thermal insensitivity of lightly reduced GO at ambient conditions, (ii) the electrical stability of flexible highly reduced GO electrode, (iii) the compact contact between highly reduced GO and lightly reduced GO. The thermal insensitivity of lightly reduced GO is originated from the low-temperature annealing, which can partially remove the oxygen-containing groups on the carbon basal plane and make the RGO material less hydrophilic, resulting in enhanced film strength and better contact with the top and bottom electrodes. The electrical stability of RGO film can be ascribed to its graphene-like properties since graphene film can keep its resistance stable in both longitude and transverse directions under stretching of up to 11% [228]. Furthermore, the strong $\pi$–$\pi$ interaction between highly reduced GO and lightly reduced GO films enables their compact contact, giving rise to a stable electrical contact in the highly reduced GO/lightly reduced GO/highly reduced GO layers.

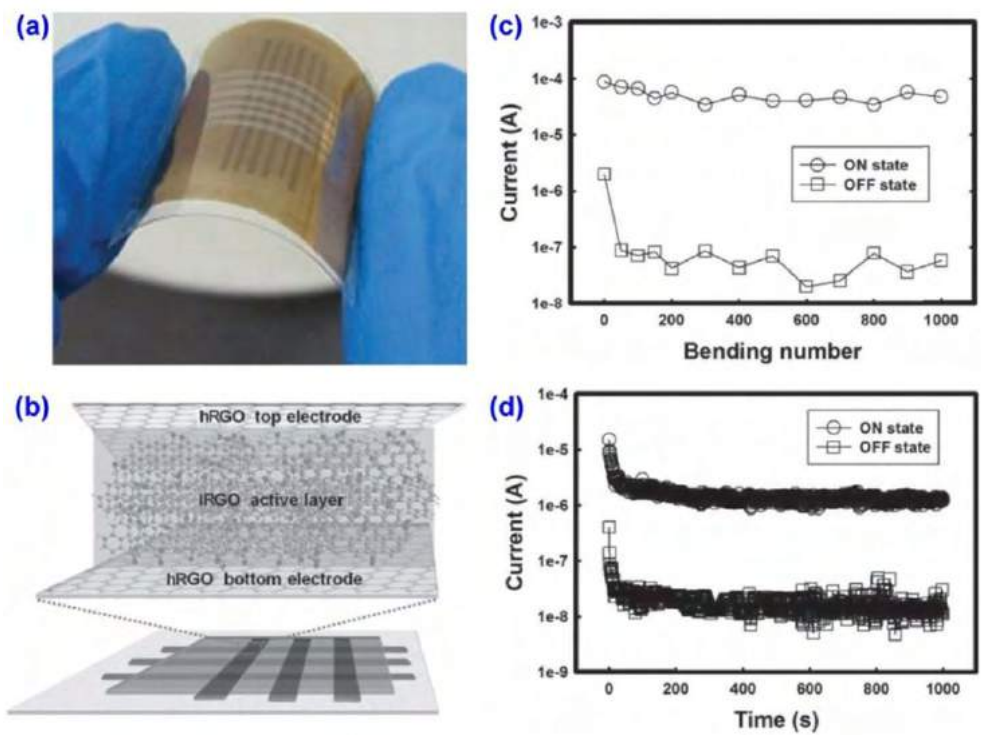

**Fig. 29** (a) Photograph of the flexible memory device with configuration of highly reduced GO / lightly reduced GO / highly reduced GO (hRGO/lRGO/hRGO) and (b) its schematic illustration. (c) Bending experiment of the all RGO-based memory device at a voltage of 2 V. (d) Retention test of the memory device at a reading voltage of 1 V in ambient conditions. Reproduced from Ref. [227].

In addition to the all-GO-based flexible devices, GO films are often deposited on flexible, bendable and transparent substrates for constructing flexible devices through various techniques and then reduced into RGO films. Generally, there are two kinds of approaches to fabricate GO films on substrates. One is solution-based approaches for the scalable fabrication of thin films, such as electrophoretic deposition (EPD), spin-coating, dip coating, Langmuir-Blodgett (L-B), and vacuum filtration [30, 229]. Then, RGO thin films are further obtained through reduction treatment of the GO films, such as hydrazine hydrate, hydriodic acid or thermal treatment. The other method is to directly deposit GO on the target substrate and then reduce it.

Usually, the GO-based devices fabricated by the EPD method possess high conductivity. For example, the GO films prepared by electrophoretic deposition on ITO exhibit high conductivity up to 150 S cm$^{-1}$ [230]. Especially, Wang *et al*. [206] reported a large-size flexible RGO membrane of $10 \times 10$ cm$^2$ prepared by the EPD method of heating-reduced RGO suspension, which reaches a much high electric conductivity of $5.51 \times 10^5$ S/m, and can be transferred to arbitrary substrates without damaging themselves.

Spin coating is a widely used technique to deposit uniform thin films with nanoscale thicknesses to flat substrates using sol-gel precursors. Using the spin coating method, Becerril et al. [19] spin-coated GO thin films on quartz and produced films with sheet resistances as low as $10^2$–$10^3$ $\Omega$/square with a transmittance of 80% for 550 nm light. Similarly, spin-coating GO thin films on other substrates, such as polyethylene terephthalate (PET) substrate [231-234], Si/SiO$_2$/Ni substrate [235], and Al-patterned substrate [236], can be made into flexible transparent devices with high performances. For example, the organic light-emitting diodes (OLED) with GO-based anode formed by spin-coating exhibit high stability in both mechanical and electrical performances during bending tests of at least 1000 cycles, as well as good conductivity and transparency comparable to the ITO electrode [237]. By spin-coating the GO dispersion onto the Au grid on a PET substrate, Qiu et al. [232] fabricated flexible electrochromic devices, exhibiting enhanced electrical performance and excellent flexural endurance, compared with the individual components. Despite the relatively large sheet resistance of RGO sheet, the sheet resistance of the RGO/Au grid hybrid film is even lower than that of the Au grid, as presented in Fig. 30a. Also, both the transmittance and sheet resistance decrease as the thickness of Au grid increases. Figure 30b compares the transmittance of the ITO film, PET substrate, Au grid, and RGO/Au grid hybrid films prepared with GO dispersion of various concentrations of 0.1, 0.2 and 2 g/L, indicating RGO/Au hybrid film is able to replace the ITO in flexible optoelectronics. The flexural endurance of RGO/Au hybrid film is depicted in Fig. 30c and d. As shown in Fig. 30c, the sheet resistance of RGO/Au film almost remains unperturbed by the bending up to a radius of 3mm. However, the sheet resistance of the ITO film starts to increase at the bending radius of 8 mm. The ITO film breaks at the bending radius of 3 mm with a dramatic sheet resistance increase. Figure 30d presents variation of the film sheet resistance under bending at the bending radius of 2.5 mm. It can be seen that after 500 bending cycles, the sheet resistance of RGO/Au hybrid film is only 6 times larger than that of its original state, but the sheet resistance of ITO increases by 67 times. Compared with Au grid alone, the flexible nature of RGO enable the RGO/Au grid hybrid film more stable behavior at very small bending radius (0.5 mm). Particularly, the RGO/Au film is still continuous even when cracks appear on the Au grid. On the contrary, the dramatic increase of the resistance demonstrates the brittle nature of ITO, which brings cracks under bending to further break the film. Furthermore, the sheet resistance of the obtained RGO/Au grid film remains the same even after one year, indicating the excellent chemical stability of the hybrid film.

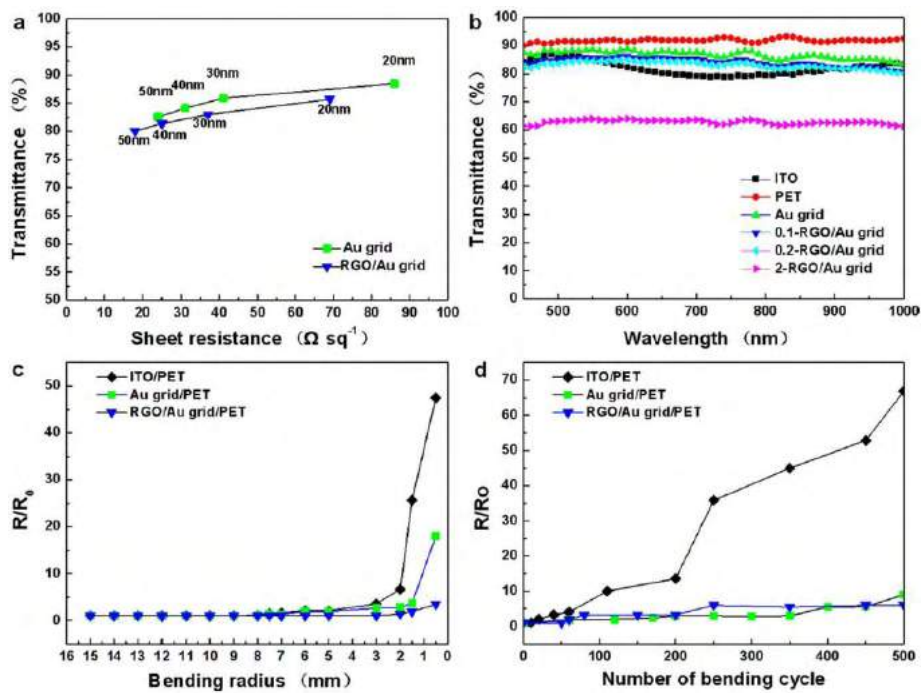

**Fig. 30** (a) Transmittance and sheet resistance of the RGO/Au grid hybrid and Au grid films with different thickness of Au. (b) Transmittance of the ITO film, PET substrate, Au grid, and RGO/Au grid hybrid films with different concentrations of GO dispersion. (c) The resistance change of the RGO/Au grid hybrid, Au grid and ITO films on PET substrates after one time bending and recovering at different bending radius. (d) The sheet resistance change of the RGO/Au grid hybrid, Au grid and ITO films on PET substrates under multiple bending times at the bending radius of 2.5 mm. Reproduced from Ref. [232].

As a popular alternative to spin coating, dip coating is able to fabricate thin films from sol-gel precursors, which is generally used for applying films onto flat or cylindrical substrates [238]. However, stacking of the GO sheets up to tens of layers in dip coating is often observed, as shown in Fig. 31a-c, indicating the existence of scrolled or folded structural defects that are detrimental to film transparency [239, 240]. To avoid sheet folding/wrinkling during deposition, a layer-by-layer sequential assembly of two oppositely (positively and negatively) charged RGO sheets is developed based on the dip coating method [241, 242]. This method takes the principle of electrostatic attraction between the two oppositely charged suspensions of RGO layers and provides highly controllable thin films with tunable thickness, transmittance and sheet resistance [241]. For example, using the layer-by-layer assembly approach, Hsiao *et al.* [242] deposited two oppositely charged suspensions of GO on the negatively charged electrospun waterborne polyurethane (WPU) fibers, and obtained the flexible and lightweight GO-based composites with enhanced electric conductivity of ~16.8 S/m and high electromagnetic interference shielding performance of ~34 dB over a frequency range from 8.2 to 12.4 GHz.

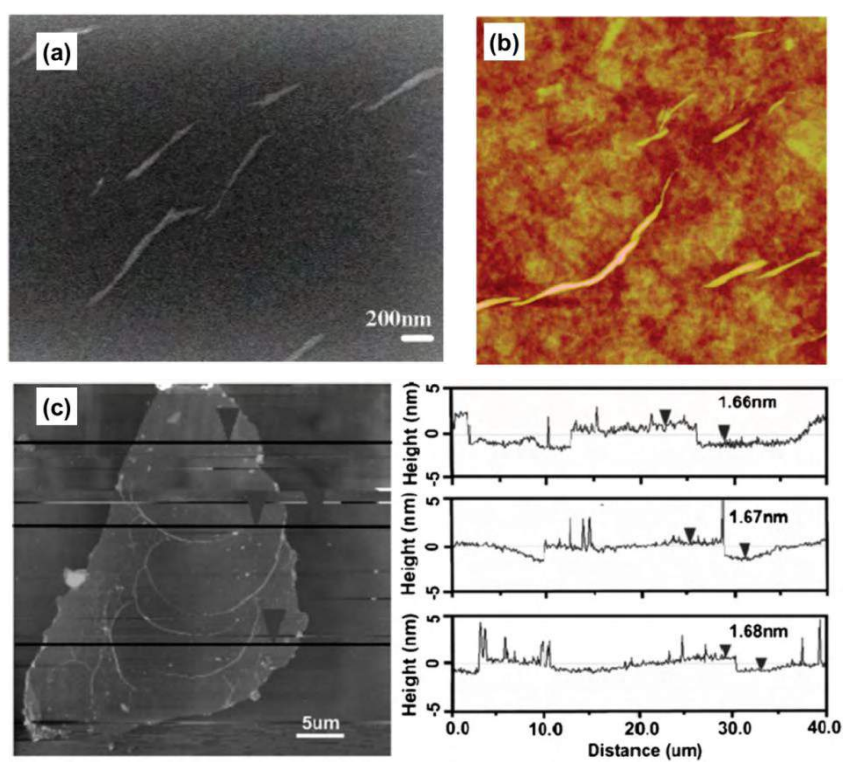

**Fig. 31** Folded and wrinkled morphology of the GO films fabricated by dip coating: (a) SEM image of GO films; (b) AFM height image (3.2 × 3.2 μm) (color scale: black to bright yellow, 30 nm). Reproduced from Ref. [239]. (c) AFM image of RGO monolayer with three height profiles measured at different locations, where the height profiles are taken along the black solid line in the AFM image. Reproduced from Ref. [240].

The L-B technique is a sophisticated approach to control the interfacial molecular orientation and packing density, which can effectively deposit GO sheets on solid supports. It is based on the layer-by-layer deposition of molecular-scale ordered domains formed at the air–water interface onto a solid substrate without intermediate transfer processes [30]. Using the L-B technique, Li et al. [243] fabricated large transparent conducting films from large amounts of GO sheets in organic solvents in a layer-by-layer manner. Moreover, the thickness of GO sheets is controllable during the transferring onto the quartz substrate, including the one-, two- and three-layer L–B GO films. Cote et al. [244] demonstrated that the GO monolayer could be uniformly and rapidly spread on water without any surfactant or stabilizing agent by the L-B approach, due to the strong electrostatic repulsion between the 2D confined layers. Such repulsion also prevents the single layers from overlapping during compression, leading to excellent reversibility of the monolayers. Then, GO monolayers are easily transferred to a silicon wafer with tunable densities continuously from dilute, close-packed to over-packed monolayers as the surface pressure increases, as illustrated in Fig. 32a– e. This was also identified by Zheng et al. [245] that the density and wrinkling degree of the GO monolayers can vary from dilute, close-packed flat GO to GO wrinkles and concentrated GO wrinkles by changing the L-B processing conditions. Besides, the film structures assembled by the L-B method highly depend on the size of GO sheets used. For the films produced from ultra-large GO sheets with a close-packed flat structure, a remarkable sheet resistance as low as ~500 Ω/square and an optical transmittance of 90% at 550 nm wavelength are obtained.

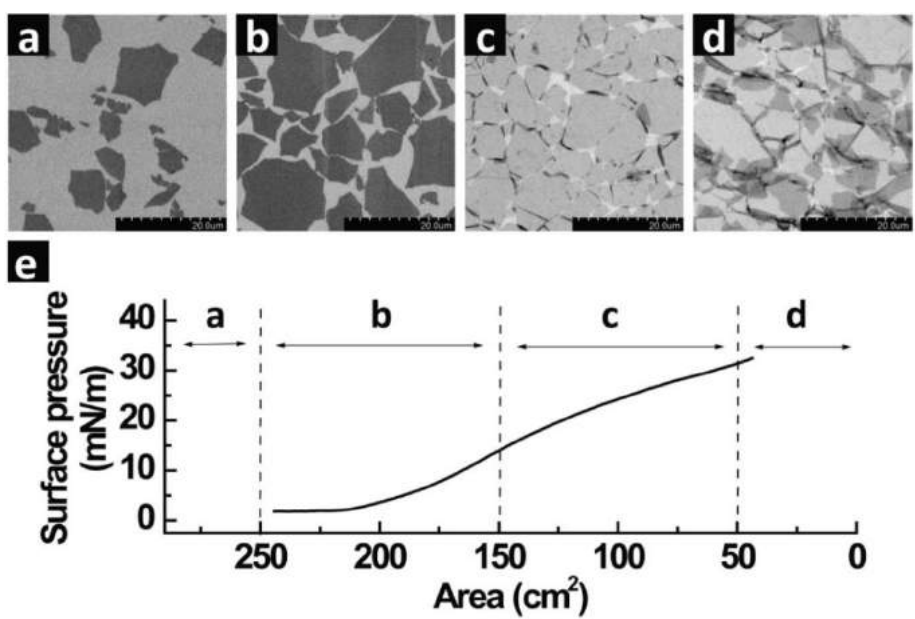

**Fig. 32** SEM images (scale bars of 20 µm) of the L-B assembly of GO monolayers on a silicon wafer at different regions of the isotherm: (a) dilute monolayer, (b) close-packed flat monolayer, (c) over-packed monolayer with sheets folded at interconnecting edges, (d) over packed monolayer with folded and partially overlapped sheets interlocking with each other. (e) Isothermal surface pressure/area plot showing the corresponding regions a–d at which the monolayers are collected. Reproduced from Ref. [244].

Vacuum filtration is a technique for separating a solid product from a solvent or liquid reaction mixture, which is most preferred for small batch in solution to quickly dry out small crystals. The solid is trapped by the filter and the liquid is drawn through the funnel into the flask below by a vacuum. Employing the vacuum filtration, a GO suspension was filtered through a commercial mixed cellulose ester membrane with an average pore size of 25 nm [15]. During filtering the suspension through the ester membrane, only the liquid is able to pass through the pores, while the GO sheets are blocked. This method allows good control over the film thickness by simply varying either the concentration of the GO in the suspension or the filtration volume. Afterwards, the GO flakes on the filter membrane can be transferred by placing the membrane with the film side down onto a substrate. Finally, a uniform GO thin film with good transparency and flexibility is obtained, as shown in Fig. 33a–c. Indeed, the transfer process is independent of the substrate since the van der Waals interactions provide sufficiently strong cohesive forces within the film and also between the GO sheets and the substrate to ensure a well adhered uniform film. The relationships between the sheet resistance and transmittance of the GO sheets and the filtration volume are presented in Fig. 33d and e, respectively. The lowest sheet resistance can be down to ~43 kΩ/square with a transmittance of ~90% at 550 nm wavelength. Employing vacuum filtration technique, Konios *et al.* [246] deposited uniform layers of GO on top of mixed cellulose ester (MCE) membrane and then transferred the GO films on glass or PET substrates for selective laser patterning. Through accurately controlling geometrical features of the RGO micromesh, transparency of the RGO micromesh can be readily changed from ~20% to up to ~85% with only a slight increase in the sheet resistance. As a result, the flexible organic photovoltaic (OPV) with the RGO micromesh as transparent electrodes can achieve a high power conversion efficiency up to 3.05%, the highest ever reported for flexible OPV devices incorporating solution-processed graphene-based electrodes. In addition, GO hybrid films with high conductivity and superior mechanical properties can be also produced by the vacuum filtration method, such as the poly(vinyl alcohol)/ GO composite films with a Young's modulus of 4.8 GPa and a flexible elongation of 36 ± 4% [247], polyimide/GO composites with a flexural strength increase of 22.5% [248], the GO/CNT composites with a sheet resistance of 240 Ω/square at a transmittance of 86% [217], and the GO/multi-walled CNT/epoxy composite materials with high storage modulus [249].

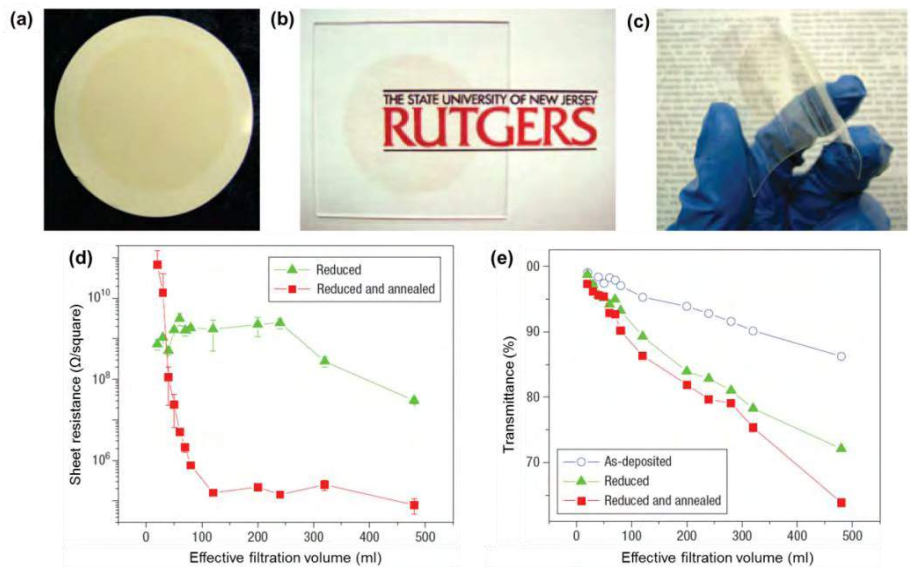

**Fig. 33** Photographs of GO thin films on (a) filtration membrane, (b) glass, and (c) plastic substrates. Sheet resistance (d) and transmittance (e) at a wavelength of 550 nm as a function of filtration volume for the GO thin films under different treatments. Reproduced from Ref. [15].

Besides, through drop coating GO solution on the PET films sputter coated with Au thin layers, Xue *et al*. [250] fabricated the PET-supported GO film and then exposed it to UV irradiation through a photomask with an interdigitated pattern for 24 hours. After removal of the photomask, the obtained GO–RGO patterns can be used as electrodes for flexible supercapacitors, achieving capacitances up to 141.2 F/g. Using overprints produced in inkjet technology, Rogala *et al*. [251] printed GO-based ink on PET substrates for flexible and transparent electronics. The obtained RGO overprints show a transmittance of ~84% at 550 nm wavelength and a sheet resistance of 150 kΩ/square. The significance of this inkjet printing with GO dispersion is fully compatible with standard industrial printers and polyester substrates, which allows to obtain fully continuous, conducting, and elastic overprints with low thicknesses, fine transparencies, and fine adhesion to the substrate.

However, since the RGO is difficult to be well dispersed, solution-based approaches for the scalable fabrication of thin films are quite limited. The other method is to directly deposit GO on the target substrate and then reduce it through hydrazine hydrate, hydriodic acid or thermal treatment.

Zhao *et al*. [184] fabricated large-area GO with the maximum area up to ~40000 $\mu m^2$ by a modified chemical exfoliation technique. To explore the application in flexible and transparent electronic devices, they further deposited large-area GO on PET substrates and reduced it by HI acid to improve the conductivity. Figure 34a and b compare the optical images of GO films assembled on PET substrates before and after HI acid reduction. It should be noticed that the RGO films maintain the integrity and flexibility of the original GO films. Moreover, compared with the almost transparent GO films, the RGO films change to light gray after HI acid reduction, indicating an appreciable improvement of electric conductivity. Besides, the sheet resistance of the RGO films significantly decreases with increasing GO sheet area at the same transmittance. As shown in Fig. 34c, the RGO-based transparent film conductors made from GO sample with area of ~7000 $\mu m^2$ has a sheet resistance of 840 Ω/square at a transmittance of 78% at 550 nm, much lower than those of made from GO sample II (ca. 1000–3000 $\mu m^2$ in area, 5.6 kΩ/square with a transparency of 77%) and from sample III (ca. 100–300 $\mu m^2$ in area, 19.1 kΩ/square with a transparency of 79%). This may be ascribed to the reduction in the number of intersheet tunneling barriers when increasing the average sheet area. Moon *et al*. [252] reported an efficient, one-pot vapour-phased reduction method of RGO paper from GO paper and thin film, using mixed reducing agent consisting of hydriodic acid with acetic acid (HI–AcOH), which allows mass production of highly qualified RGO. The RGO obtained by reducing GO prepared on PET substrates yields conductivity of ~9.44 S/cm with good flexibility.

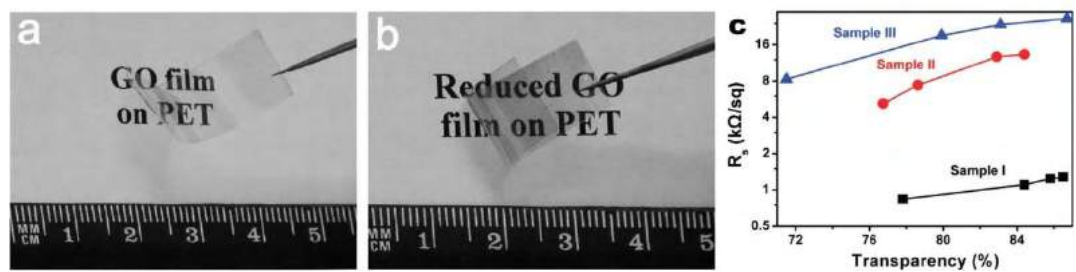

**Fig. 34** Optical images of GO films (a) before and (b) after HI acid reduction. (c) Sheet resistance

and transmittance at 550 nm of transparent film conductors assembled with sample I, sample II, and sample III GO sheets. Reproduced from Ref. [184].

Instead of using toxic chemical agents such as hydrazine and hydriodic, Ning *et al*. [253] used

a magnetron sputtered Sn metal layer to directly reduce preformed GO film on a transparent flexible

PET substrate at room temperature, which is an eco-friendly reducing process. It was found that increase of the Sn layer thickness will benefit the conductivity of GO film, as shown in Fig. 35a.

The conductivity achieves a preferable value at a Sn thickness of 8.6 nm, which gives the suitable

Sn thickness without waste. Also, increasing the thickness of GO film will benefit its conductivity but lower the transmittance, as shown in Fig. 35b. Generally, the fabricated RGO film can exhibit a sheet resistance of 6.7 kΩ/square at a transmittance of 75% at 550 nm and it is rather flexible and robust, as presented in Fig. 35c. Figure 35d further shows the application of flexible and transparent film in a circuit of a red LED. The RGO film can serve as a conductor and light the LED with an applied voltage of 3 V.

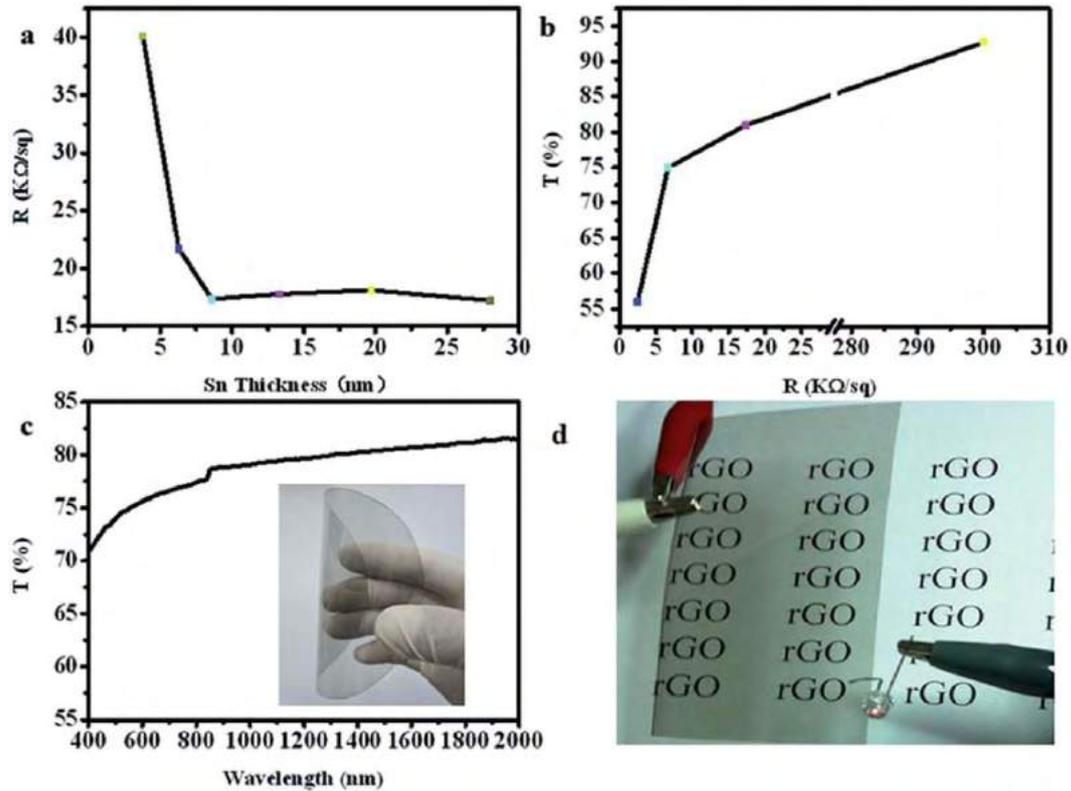

**Fig. 35** (a) Influence of the thickness of the sputtered Sn layer on the conductivity of the RGO films of 6.3 nm thick; (b) the transmittance at 550 nm versus sheet resistance for RGO films of different thicknesses; (c) optical transmittance of the RGO film with an average sheet resistance of 6.7 kΩ/square in the visible range (inset is the photo image of the flexible RGO film) and (d) photo image showing that the transparent conductive RGO film can be used in a circuit to light a red LED. Reproduced from Ref. [253].

To summarize, flexible electronic devices are considered as the next innovation in electronics. Considering the superior transparency, high conductivity, and outstanding flexibility, RGO is regarded as an attractive candidate material for the flexible electronic devices. So far, different kinds of all-GO-based flexible devices with satisfactory performances have been developed. Particularly, GO/RGO films are often deposited on flexible, bendable and transparent substrates to construct flexible electronic devices using various techniques, which greatly expand the applications of GO-based flexible devices.

### 3.4 Field-effect transistors (FETs)

It is well known that graphene possesses excellent field emission properties with extraordinary high carrier mobility (200000 $cm^2$ $V^{-1}$ $s^{-1}$ for perfect graphene), which is much higher than that of silicon and enables graphene promising applications in FETs [254, 255]. Due to the oxygen-containing functional groups, GO can present improved field emission performance compared with graphene. Indeed, exceptionally low threshold field emission from atomically thin edges of RGO was reported [256]. The average threshold field required to emit a current of 1 nA was found to be less than 0.1 V $\mu m^{-1}$. Such a low threshold field can be attributed to the combination of large enhancement factor and lower local work function at the edges of RGO. The edges provide an array of emission sites in the form of low work function C–O–C ether chains from which multiple electron beams are simultaneously emitted. Therefore, the enhanced field emission characteristics are ascribed to the termination of straight edges by cyclic ether, which constitutes the most stable form of oxygen in RGO. The excellent field emission properties of GO implies its promising prospect in FETs. Here, performances of GO-based FETs were discussed, including the carrier polarity, carrier mobility and ON/OFF current ratio.

Due to the unique band structure, the carriers in graphene are bipolar, that is, both electrons and holes can be continuously tuned by a gate electrical field [254]. Thus, the highly reduced GO also exhibits the graphene-like ambipolar field effect due to the graphene-like structural characteristics. For example, using a solution-based method, Eda *et al*. [15] were able to deposit RGO thin films with monolayer to several-layer thicknesses over large substrates, where the single-layer RGO films exhibit graphene-like ambipolar transistor feature. Reddy *et al*. [257] reported that the RGO-based FET exhibits ambipolar transfer characteristics with the gate-voltage of –1 to –20 V and from 44 to 57 V in the p- and n-type branches respectively, and the field-effect mobility for electron and hole being 8.98 × $10^{-6}$ $cm^2$ $V^{-1}$ $s^{-1}$ and 2.1 × $10^{-5}$ $cm^2$ $V^{-1}$ $s^{-1}$ respectively. Chang *et al*. [258] also demonstrated that the large-area few-layer RGO films exhibit ambipolar charge transport characteristics in a bottom-gate FET architecture, which can be used as a phototransistor. Yu *et al*. [259] showed that both the as-prepared RGO FET and FET with RGO modified by a bis-pyrenyl derivative, py-diIM-py, exhibit V-shaped ambipolar field effect behavior from p-type region to n-type region.

On the other hand, the bipolar feature of graphene makes the type and concentration of the carrier sensitive to doping. Therefore, due to abundant oxygenated groups, as-prepared GO and RGO typically show p-type conductivity. As reported by Jin *et al.* [260], GO with an optical band gap of 1.7 eV behaves as p-type semiconductor at room temperature in ambient condition. Moreover, the measured current versus the gate voltage ($I$–$V_g$) curves at a zero gate bias suggest that conductivity of the GO-based FETs is 0.034 S cm$^{-1}$ after subjecting to 5 min oxidation, indicating that GO could be a good candidate for semiconducting channel in the FET device. Karteri *et al.* [261] pointed out that both GO FET (Fig. 36a) and GO-based organic FET with PMMA and PVP (Fig. 36b) exhibited p-type enhancement mode operation with neat output and transfer behavior, as indicated by the output characteristics shown in Fig. 36c and d. The p-type semiconducting behavior in GO-based FETs can be mainly attributed to the polarization of entrapped oxygen groups in between the GO layers, which alter the sp$^2$ hybridized carbon atom in graphite into sp$^3$ hybridization, and further change the metallic behavior of graphite into a p-type semiconducting behavior. The p-type semiconducting behavior of GO and RGO were also observed in other literatures [262, 263].

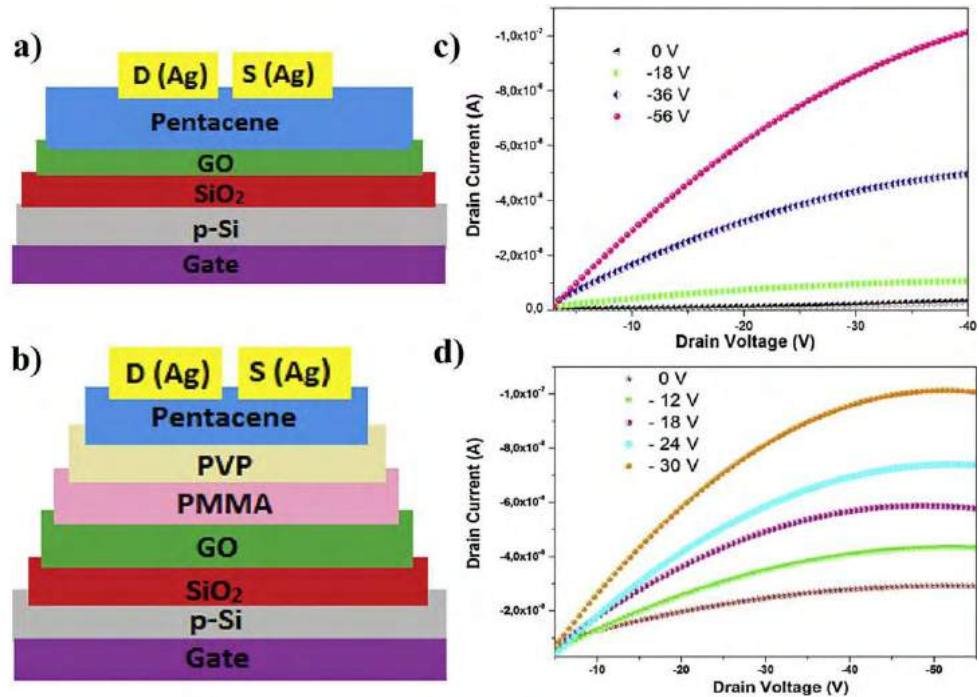

**Fig. 36** Schematic diagram of (a) the GO based FET and (b) the GO based organic TFT where D and S stand for drain and source electrodes, respectively; Output characteristics of (c) the GO based FET and (d) the GO based organic FET at various gate voltages. Reproduced from Ref. [261].

From the above statement, it can be concluded that the carrier polarity of GO is tunable. Indeed, plenty of efforts have been devoted to tuning the polarity of GO through controlling the oxidation degree or doping. Using chemically reduced GO sheets, Joung *et al.* [264] fabricated FETs with the chemically reduced GO sheets assembled between prefabricated gold source and drain electrodes using ac dielectrophoresis. When applying a backgate voltage, 60% of the devices show p-type FET behavior, while the remaining 40% show ambipolar behavior. After mild thermal annealing at 200 °C, the 60% of the original p-type RGO devices are transformed to ambipolar behavior and the original 40% ambipolar RGO FETs remain ambipolar with increased hole and electron mobilities. The maximum hole and electron mobilities of the devices are 4.0 and 1.5 $cm^2$ $V^{-1}$ $s^{-1}$, respectively. Figure 37a shows the current versus the gate voltage curves of the RGO FET, which directly demonstrates the change from p-type to ambipolar upon thermal annealing. Cai *et al.* [265] reported a FET based on a superlattice hybrid film of RGO coupled with photoactive titania ($Ti_{0.87}O_2^{0.52-}$) nanosheets with the polarity response changing from ambipolar to n-type unipolar. It was found that the electron carrier concentration increases as high as $7.6 \times 10^{13}$ $cm^{-2}$ after injecting the photoexcited electrons in $Ti_{0.87}O_2^{0.52-}$ into the RGO sheets.

Particularly, it is possible to covert the p-type GO into n-type behavior. Tu *et al.* [266] achieved remarkable conversion between n- and p-type RGO by changing the thermal annealing temperature. As shown in Fig. 37b, for temperatures of 300−450 °C and 800−1000 °C, the charge carriers in RGO are electrons, belonging to n-type. However, for temperatures at 250 °C and 450−800 °C, the charge carriers are holes, which are p-type. This phenomenon can be explained by the fact that different functional groups of RGO are predominant at different annealing temperatures. This is because the individual oxygen functional groups presented on RGO are determined by the annealing temperature. It was found that for temperatures of 300−450 °C and 800−1000 °C, the predominant oxygen functional groups are electron-donating groups, such as $sp^2$-bonded hydroxyl, ether and epoxy groups, leading to n-type RGO. While for temperatures at 250 °C and 450−800 °C, the predominant oxygen functional groups, such as carboxyl, carbonyl, and $sp^3$-bonded hydroxyl, ether, and epoxy groups, are electron withdrawing groups, which result in p-type RGO. This p–n–p double transformation tailored by thermal annealing temperature was also demonstrated by Wang and Chang during studying the gas sensing application of RGO [267]. They found that the as-prepared GO obtained by a modified Hummers' method is p-type. During thermal annealing, the RGO shows p-type at temperatures of 100 and 400 °C, but n-type at temperatures of 200 and 300 °C due to different predominant functional groups. Besides, the conversion of p to n-type RGO can be also achieved through laser annealing due to the reduction of GO framework where the C/O ratio can be as high as 23% after laser-assisted reduction [268].

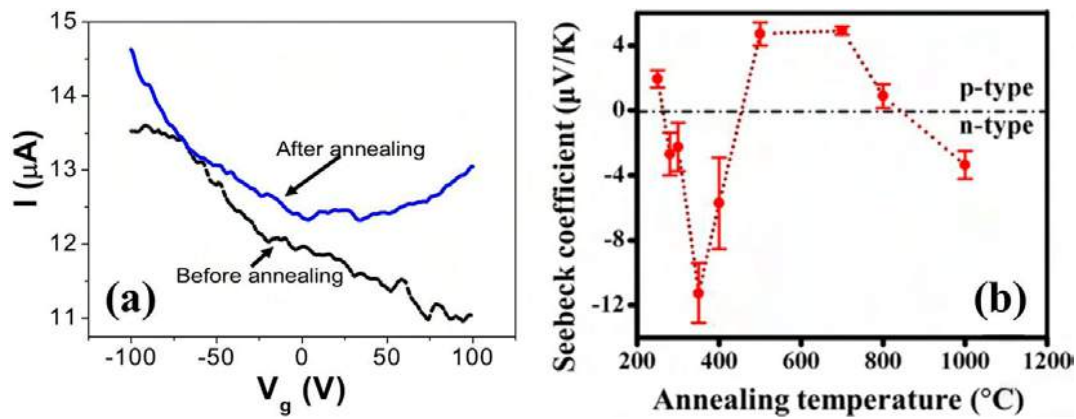

**Fig. 37** (a) I–V$_g$ characteristic of the RGO FET showing change from p-type to ambipolar upon thermal annealing. Reproduced from Ref. [264]. (b) Seebeck coefficient of RGO as a function of the annealing temperature. Reproduced from Ref. [266].

Besides, doping is another effective method to transform the p-type RGO to the n-type behavior. Through hydrazine treatment, Some *et al.* [269] demonstrated that the N-doped RGO can serve as an n-type semiconducting channel. The N-doped RGO can be achieved by reaction of as-prepared GO with hydrazine in an aqueous solution at an elevated temperature, followed by rapid thermal annealing. Moreover, enhanced N-doping effect and n-type behavior are observed through treating with as-prepared GO in a polar aprotic tetrahydrofuran (THF) solution. Afterwards, Li *et al.* [270] reported the fabrication of highly N-doped GO films and n-type GO FETs by simple irradiating the GO thin films in a NH$_3$ atmosphere. The measured electric field effect displays an n-type FET behavior with a charge neutral point (Dirac point) located at around –8 V and an electron mobility about 7.08 cm$^2$ V$^{-1}$ s$^{-1}$. The amino-like nitrogen was suggested to mainly contribute to this n-type behavior. Khan *et al*. [271] doped the RGO with tetra-methyl ammonium hydroxide (TMAH) and pointed out that the graphitic N leads to n-type doping, whereas oxidized N acts as electron withdrawing group and causes p-type doping. In addition to N doping, using the lithium and ethylenediamine (EDA) solvent for the reduction of GO, n-type RGO FETs can be also obtained [272]. The EDA, in the liquid phase at room temperature, is a strong base with di-amine terminal groups to accept the solvated electrons, and acts as a nucleophile in the substitution reaction, which can attain the N-doped RGO, as schematized in Fig. 38a. For the GO FET device, a constant source-drain bias at 0.5 V is applied with the gate voltage ranging from –40 V to 40 V. Under ambient conditions, all channels exhibit p-type doping behavior because oxygen can spontaneously adsorb onto the graphene surface. After removing oxygen and moisture at $10^{-4}$ Torr for 2 hours in a solution of Li–EDA system, the Li–RGO FET changes to n-type doping behavior with a typical Dirac point shift to –23.8 V, as shown in Fig. 38b. This suggests that the Li–EDA system can effectively reduce GO to RGO and induce n-type doping during reduction.

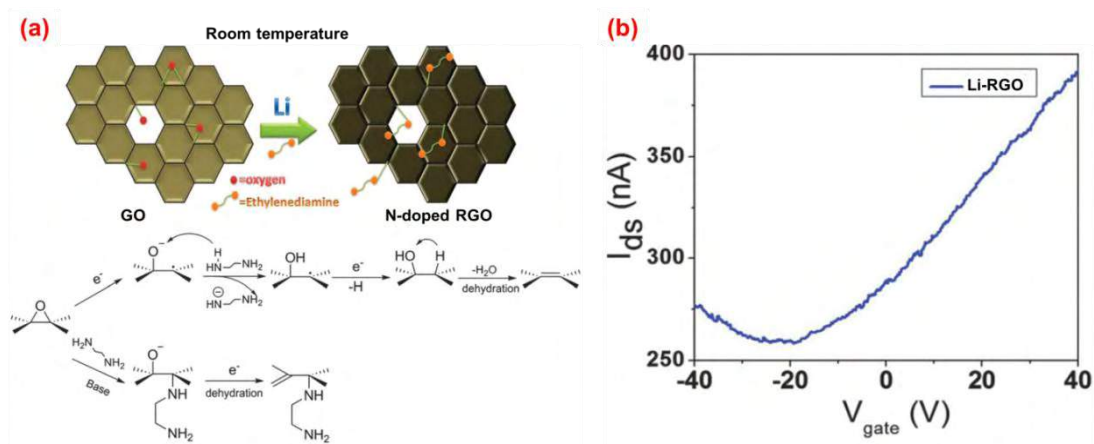

**Fig. 38** (a) Possible mechanism of the reduction of GO and nitrogen nucleophilic substitution-reduction on GO; (b) I–V characterization of a Li–RGO FET under vacuum conditions. Reproduced from Ref. [272].

In addition, through trapping the carriers, Zhou *et al.* [273] reported an approach to systematically control over polarity (from holes to electrons or vice versa) in GO-based transistors. For example, applying –40 V at the gate can induce an electron-rich FET, since the negatively shifted switch-on voltages indicate that more and more holes are trapped in the process and the device becomes electron rich eventually. On the contrary, a pre-applied +40 V gate pulse can induce a hole-rich device. Besides, Smirnov *et al.* [274] reported a method to change the carrier conduction of RGO by altering the relative humidity. It was found that in ambient conditions with a relative humidity of 7 %, the partially reduced GO films exhibit electronic conductivity. However, with increasing humidity, the electron conduction went down. In wet conditions with the relative humidity of 35 %, the RGO films show protonic conductivity.

Carrier mobility is another important parameter to character the performance of a FET. Eda *et*

*al.* [15] reported the RGO-based FETs consisting of one to five layers of RGO thin films. The mobilities of the FET devices in ambient conditions are found to be ~1 cm$^2$ V$^{-1}$ s$^{-1}$ for holes and ~0.2 cm$^2$ V$^{-1}$ s$^{-1}$ for electrons, respectively. Similarly, the RGO-composite thin-film FETs using RGO as the filler and polystyrene (PS) as the host, achieved the electron and hole mobilities of 0.2 and 0.7 cm$^2$ V$^{-1}$ s$^{-1}$, respectively [275]. Lim *et al.* [276] prepared the flexible organic FETs with inkjet-patterned RGO/poly(vinylalcohol) (PVA) composite electrodes on plastic substrates. The PVA is blended with GO to enhance exfoliation of the graphite sheets and to provide stable inkjet printing. The obtained FETs possess an enhanced field-effect mobility of 0.23 cm$^2$ V$^{-1}$ s$^{-1}$, which is significantly larger than that of FETs with Au (mobility of 0.015 cm$^2$ V$^{-1}$ s$^{-1}$) or polymer poly(3,4-ethylenedioxythiphene) (PEDOT) doped with poly(4-styrene sulfonate) (PSS) (PE DOT:PSS) (mobility of 0.035 cm$^2$ V$^{-1}$ s$^{-1}$) electrodes. Karteri *et al.* [261] fabricated the thin film FETs composed of RGO films, poly(methyl methacrylate) (PMMA) and PVP, exhibiting a mobility of 0.375 cm$^2$ V$^{-1}$ s$^{-1}$. A large carrier mobility up to 58 cm$^2$ V$^{-1}$ s$^{-1}$ at a carrier concentration of $1.2 \times 10^{12}$ cm$^{-2}$ was reported in the top-gated RGO FET [277]. The field-effect mobility in the range of 30–40 cm$^2$ V$^{-1}$ s$^{-1}$ for the GO–MoS$_2$ hybrid FETs was also reported [278].

Usually, the mobility of GO-based FETs depends on several factors, including the temperature, channel length, GO concentration, and external bending and illumination. Eda *et al.* [275] presented the dependence of the carrier mobility on the temperature in the RGO-based FETs, as shown in Fig. 39a. The carrier mobilities for both electrons and holes decrease rapidly with increasing temperature up to about 60 K and then increase gradually with further increasing the temperature. Furthermore, the carrier mobility of the RGO-based FETs also relies on the channel length. For the FET with the source-drain electrode separation ($L_C$) larger than the lateral size of the RGO ($L_S$), decrease of $L_C/L_S$ ratio will enhance the carrier mobility [275]. Then, Kobayashi *et al.* [279] systematically considered the FETs composed of 1–4 layers of randomly adsorbed RGO nanosheets with the lateral size $L_S$ of ~1 μm but different source-drain electrode separation $L_C$ ranging from 0.1 to 54.5 μm. As shown in Fig. 39b, the hole mobility ($\mu_{\text{hole}}$) mainly decreases with increasing the $L_C$. However, for the electron mobility ($\mu_{\text{elec}}$), it increases with increasing the $L_C$ in the range of $L_C < 2.5$ μm, but sharply decreases when $L_C$ is larger than ~2.5 μm, since the electron mobility is about one order of magnitude larger than the hole mobility. The dependence of total field-effect mobility ($\mu_{\text{FE}}$) on the source-drain electrode separation is the same as that of the electron mobility. Therefore, there is a critical $L_C$ for the RGO-based FETs, which is about 2.5 μm, close to the RGO lateral size $L_S$. When decreasing the $L_C$ to be smaller than 2.5 μm, the $\mu_{FE}$ exhibits a rapid increase, as shown in Fig. 39c. Moreover, from the variation of $\mu_{FE}$ in Fig. 39c, the variation of $\mu_{FE}$ reaches a peak at $L_C \approx L_S$. Thus, both decreasing and increasing $L_C$ from $L_S$ will reduce the variation in $\mu_{FE}$. Aleshin *et al*. [280] fabricated the organic FETs incorporating organic semiconductor poly(9-vinylcarbazole) (PVK) and GO flakes. It was found that the mobility PVK:GO hybrid organic FETs with Au–Al source-drain electrodes increases slightly from 3.8 to 6.7 $cm^2$ $V^{-1}$ $s^{-1}$ if concentration of the GO flakes increases from 3 up to 9.5 wt.%. Below the GO concentration of 3 wt.%, there is a sharp increase of the mobility from ~$10^{-9}$ $cm^2$ $V^{-1}$ $s^{-1}$ (~0.2 wt.% of GO) to 3.8 $cm^2$ $V^{-1}$ $s^{-1}$ (3 wt.% of GO). Wan *et al*. [233] studied the FETs composed of indium zinc oxide and GO hybrids on flexible graphene/PET substrates and found that bending will affect the field-effect mobility and threshold voltage. As shown in Fig. 39d, bending leads to a slightly increase of the mobility within ~$10^2$ bending times. Further increasing the bending times will gradually decrease the mobility. As a result, an estimated mean value of the field-effect mobility was ~15.5 $cm^2$ $V^{-1}$ $s^{-1}$ with a standard deviation of ~1.1 $cm^2$ $V^{-1}$ $s^{-1}$. Meanwhile, the threshold voltage shift was found to be less than 0.1 V even after 1000 times, indicating the good mechanical flexibility and stability of the GO hybrid FETs. In addition, it was suggested that the carrier mobility of the GO-based FET was related to the light illumination [281]. As presented in Table 2, the mobility increases with increasing the power of the incident light per area, from 0.3752 $cm^2$ $V^{-1}$ $s^{-1}$ under dark to 0.4557 $cm^2$ $V^{-1}$ $s^{-1}$ under 100 mW $cm^{-2}$ illumination. Besides, the threshold voltage also increases with the power of the incident light.

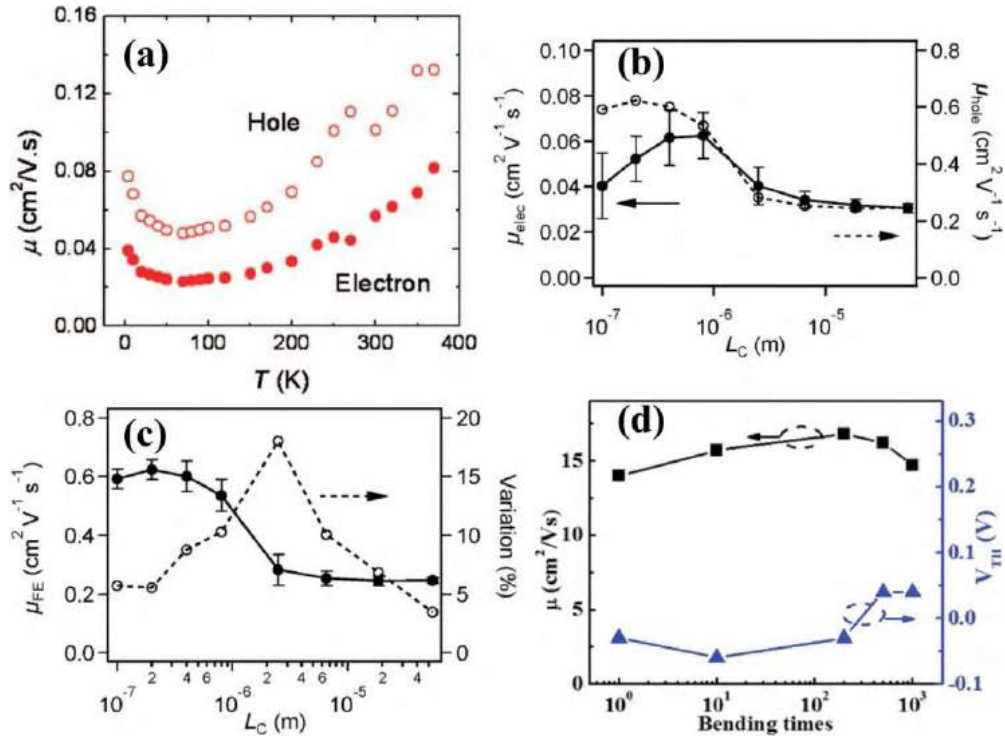

**Fig. 39** (a) The electron and hole mobilities ($\mu$) as a function of temperature. Reproduced from Ref. [275]. (b) $L_C$ dependences of field-effect electron ($\mu_{elec}$) and hole ($\mu_{hole}$) mobilities; and (c) $L_C$ dependences of the field-effect hole mobility $\mu_{FE}$ (solid circles) and their variations (open circles). Reproduced from Ref. [279]. (d) Field-effect mobility ($\mu$) and threshold voltage ($V_{TH}$) of the FETs measured after different bending cycle times. Reproduced from Ref. [233].

**Table 2** The values of mobility ($\mu$), threshold voltage ($V_{th}$), ON/OFF current ratio ($I_{ON}/I_{OFF}$) and sub-threshold swing (S) of the GO-based FET GO under dark and white light illumination, where S is a parameter controlling the voltage swing required to turn a transistor from "OFF" to "ON" state. Reproduced from Ref. [281].

| P (mW cm$^{-2}$) | $\mu$ ($\times 10^{-1}$ cm$^2$ V$^{-1}$ s$^{-1}$) | $V_{th}$ (V) | $I_{ON}/I_{OFF}$ ($10^2$) | S (V/decade) |
|---|---|---|---|---|
| Dark | 3.752 | −7.585 | 0.710 | 2.610 |
| 20 | 3.772 | 8.494 | 0.839 | 1.607 |
| 40 | 3.793 | 15.889 | 0.946 | 1.555 |
| 60 | 4.101 | 19.414 | 0.865 | 1.552 |
| 80 | 4.186 | 20.704 | 0.743 | 1.548 |

| 100 | 4.557 | 21.886 | 0.709 | 1.546 |

However, the field-emission devices based on pristine graphene have relatively low ON/OFF current ratios (usually in the range of 2–20) because of the finite minimum conductance of graphene at zero gate voltage [255]. Therefore, the ON/OFF ratios of FETs based on highly reduced GO are typically below 10 [282]. For example, Vasu et al. [277] reported an ON/OFF ratio of 4 for the top-gated FET made of RGO. Kim et al. [283] fabricated the RGO transistors operating below 4 V through two different methods, chemical reduction and solvothermal reduction. Both the RGO transistors have an ON/OFF ratio around 2. However, the RGO transistor based on chemical reduction shows higher hole and electron mobilities ($5.1 \pm 1.1$ and $3.3 \pm 1.5$ cm$^2$ V$^{-1}$ s$^{-1}$, respectively) than those ($1.8 \pm 0.7$ and $0.7 \pm 0.2$ cm$^2$ V$^{-1}$ s$^{-1}$, respectively) based on solvothermal reduction. Using GO as the gate dielectric, Eda et al. [284] fabricated the FET with an ON/OFF ratio between 2 and 3, and a field effect mobility of $\sim 10^3$ cm$^2$ V$^{-1}$ s$^{-1}$.

The ON/OFF current ratios can be enhanced by controlling the electronic gap in GO. It was found that the lightly reduced GO with a finite energy gap (10–50 meV) exhibits an ON/OFF current ratio more than $10^3$ at low temperature [167]. On the other hand, GO-based organic FETs usually exhibit high ON/OFF current ratios. Using GO sheets as charge-trapping layers, Kim et al. [285] reported a GO-based organic FET with an ON/OFF current ratio of $\sim 10^2$ for nonvolatile memory applications. As illustrated in Fig. 40a, the GO-based organic FET was fabricated on thermally grown 110 nm thick SiO$_2$ substrate using heavily doped n-type Si as the gate electrode. The GO nanosheets were prepared by a modified Hummer's method and then well dispersed on the SiO$_2$ substrate as charge-trapping layers. Poly(methyl methacrylate) (PMMA) was used as a charge-tunneling layer, which is dissolved in toluene at a concentration of 10 mg/ml, spin-coated at 3000 rpm for 40 s, and annealed on a hotplate at 120 °C for 15 min in a nitrogen-filled glove box. Finally, 50 nm thick gold source and drain electrodes are deposited with a shadow mask by thermal evaporation at a pressure of $\sim 10^{-7}$ Torr. Fig. 40b plots the transfer curves of the prepared GO-based organic FETs, which displays large gate bias dependent hysteresis with threshold voltage shifts over 20 V. After writing and erasing, the stored data are well maintained with an ON/OFF current ratio in the order of $10^2$ at $10^4$ s, as shown in Fig. 40c. Moreover, it was found that for the organic FETs with GO as the charge-trapping layers, the one with PMMA as charge tunneling layer exhibits higher mobility and ON/OFF ratio (0.134 cm$^2$ V$^{-1}$ s$^{-1}$ and 10$^6$, respectively) than those (0.084 cm$^2$ V$^{-1}$ s$^{-1}$ and 10$^5$, respectively) of the device with PVP as charge tunneling layer [286].

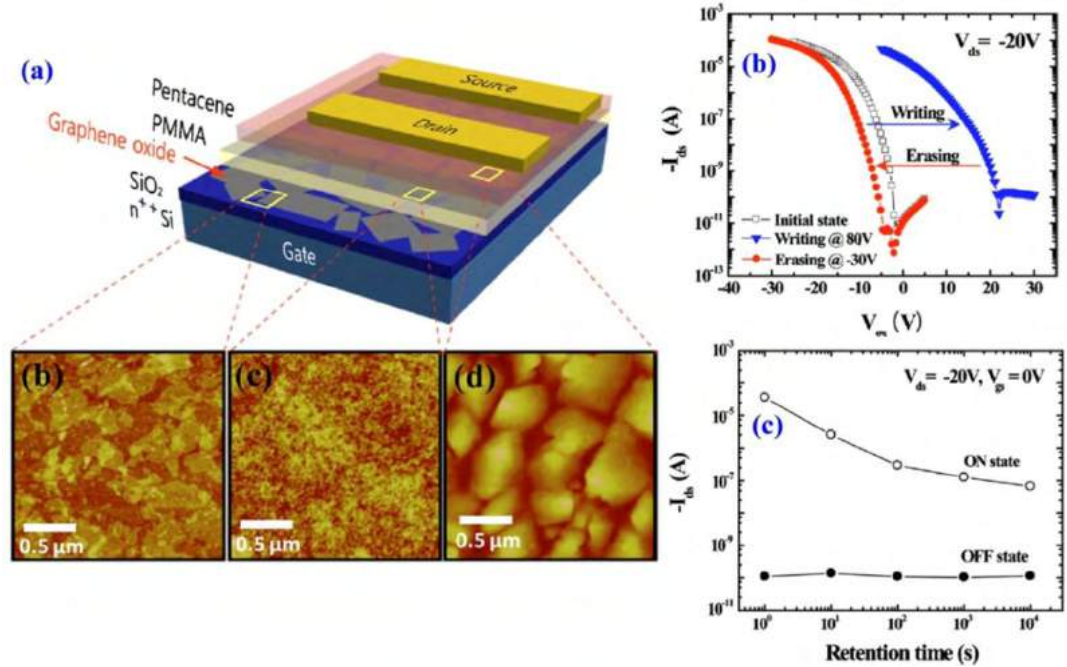

**Fig. 40** (a) Device schematics of GO nanosheets based GO-based organic FET with n$^{++}$ Si / SiO$_2$ / GO nanosheets/PMMA/pentacene/Au source-drain stacks. (b) Transfer curve characteristics (I$_{ds}$ vs V$_{gs}$) of GO-based organic FET, according to the different Writing (V$_{gs}$ = 80 V) / Erasing (V$_{gs}$ = −30 V) biases at drain bias (V$_{ds}$) of −20 V. (c) Retention characteristics of GO-based organic FET. After writing or erasing process, drain current (I$_{ds}$) is read for 10$^4$ s at drain bias (V$_{ds}$) of −20 V and gate bias (V$_{gs}$) of 0 V condition. Reproduced from Ref. [285].

Besides, the performance of organic FETs is closely related to the electrode materials and geometry. For example, the Au electrode covalently bonded with GO (GO-Au electrode) can overcome the poor contact quality between the metal electrodes and the organic semiconductors [287]. As shown in Table 3, performance of the organic FETs with different electrodes and semiconductors was compared, including the mobility, ON/OFF current ratio, threshold voltage and width-normalized contact resistance ($R_{con}W$). It can be seen that the organic FETs with covalently linked GO-Au electrodes usually have higher effective mobility and ON/OFF ratio than those with pure Au electrodes for different semiconductors, especially the Au bottom electrodes. Also, electrode geometry plays an important role on the performance of organic FETs. It was found that the organic FETs with ultrathin RGO electrodes show significantly improved charge injection efficiency compared to those with thick electrodes due to the narrow edges (~3 nm), which may facilitate the diffusion and assembly of organic semiconductors and thus form a uniform semiconductor film across the electrode/channel junction area [288]. As the RGO electrode edge minimizes from 20 to ~3 nm, the width-normalized contact resistance sharply decreases from ~120 to 49.8 kΩ·cm. As a result, the ultrathin RGO-based organic FETs attain the mobility, ON/OFF current ratio and threshold voltage of 0.5 cm$^2$ V$^{-1}$ s$^{-1}$, 1.4 × 10$^6$ and −10 V, respectively.

**Table 3** Electrical parameters of organic FETs with different electrodes and semiconductors. PDI-8CN2 and DNTT stand for N,N′-bis(n-octyl)-dicyanoperylene-3,4:9,10-bis(dicarboximide) and dinaphtho[2,3-b:2′,3′-f]thieno[3,2-b]thiophene, respectively. No field-effect property is found in Au bottom-contact DNTT devices. Reproduced from Ref. [287].

| Semiconductor | Electrode | Effective mobility (cm$^2$ V$^{-1}$ s$^{-1}$) | ON/OFF ratio | Threshold voltage (V) | $R_{con}W$ (kΩ cm) |
|---|---|---|---|---|---|
| Pentacene | Au-GO bottom | 0.070–0.350 | 10$^4$–10$^5$ | −10 to −32 | 16 |
| | Au bottom | 0.002–0.008 | 10$^2$–10$^3$ | −15 to −30 | 4168 |
| | Au top | 0.080–0.300 | 10$^3$–10$^5$ | −8 to −28 | 10.6 |
| PDI-8CN2 | Au-GO bottom | 0.01–0.075 | 10$^3$–10$^4$ | 0.2 to 4 | 62.8 |
| | Au bottom | 0.00023–0.002 | 10–10$^2$ | −20 to −30 | 191.5 |
| | Au top | 0.02–0.086 | 10$^3$–10$^4$ | −4 to 3 | 33.4 |
| DNTT | Au-GO bottom | 0.15–0.60 | 10$^5$–10$^6$ | −10 to −26 | 114.3 |
| | Au bottom | – | – | – | – |
| | Au top | 0.20–0.76 | 10$^5$–10$^7$ | −10 to −30 | 76.4 |

On the other hand, an important application of the GO-based FETs is to act as sensors owing to the inherent signal amplification capability and relatively high signal-to-noise ratio [33, 186, 289-291]. Compared with pristine graphene, the dangling bonds in GO/RGO sheets make them more suitable for sensing. When a molecule is adsorbed onto GO surface, the local change in the carrier concentration induces doping of the electrons, which can be monitored electrically in a transistor-like configuration [292].

The GO-based FETs are frequently used for biosensing, presenting distinct advantages including label-free, ultra-sensitivity, and real-time response. For example, Stine *et al.* [293] demonstrated the real-time and label-free detection of DNA hybridization using the RGO FET devices. As illustrated in Fig. 41a, in this RGO FET, an amino-silane film is formed prior to GO deposition to stabilize the RGO layer on the surface in aqueous solutions through electrostatic attraction between the positively charged amine groups and the negatively charged carboxyl groups in the GO. Then a flow cell is attached to the device to isolate the active RGO areas for sample exposure. The performance of a single isolated FET device for real-time, label-free DNA detection is presented in Fig. 41b, where the data are normalized through non-dimensionalization to account for variations between individual RGO sheets. By functionalizing with aminated single-stranded DNA and then being exposed to 1 μM solutions of the fully complementary target DNA and the fully non-complementary control DNA, the RGO FET exhibits distinct sensor behavior responding to both the complementary and non-complementary DNA exposure though a larger change in current. Thus, one pressing problem for the label free sensors is their response to non-specific binding. However, the reproducibility of RGO films can overcome this limitation since it allows usage of a second FET as reference. As shown in Fig. 41c, there are three data sets showing the sensor response after the introduction of solutions containing 1 μ M target DNA, 10 n M target DNA, and 1 μ M control DNA, respectively. The electrical signal is a differential change between this reference device and the device functionalized with the probe DNA. It was noticed that, when the device was switched from pure buffer to the two solutions containing the complementary target, distinct responses were observed. But introduction of the non-complementary strand shows no significant response and the signal returns to the baseline after the noise caused by switching the feed line. The reference device contributed to this increased specificity, which enables subtracting sources of interference common to both devices. Using peptide nucleic acid (PNA) as the capture probe of the RGO-based FET biosensor for targeting the complementary DNA sequence, a detection limit as low as 100 fM was achieved [294], which is one order of magnitude lower than that of the graphene-based FET biosensors based on DNA probes [295-297].

In addition to DNA sensing, He *et al.* [298] found that the conductance of RGO-based device changes with the concentration of dopamine molecule. At a gate voltage of –0.6 V, the conductance of RGO-PET FET device (RGO films on flexible PET substrate) increases with enhancing concentration of dopamine. However, at a gate voltage of +0.6 V, the conductance of RGO-PET device will decrease with increasing concentration of dopamine. This is because the RGO device has a Dirac voltage of +0.2 V. Above it, the device operates at n-type region; below it, the device operates at p-type region. After removing dopamine by thorough perfusion, the current returns to its initial level and the device can be reliably used for many times. Miao *et al.* [299] reported the highly sensitive GO-based FET biosensor for protein detection using thermally reduced GO sheets functionalized with AuNP–antibody conjugates. The probe antibody was labeled on the surface of RGO sheets through AuNPs for electrical detection of protein binding. It was found that binding of protein would significantly modify the resistance of the RGO sheet, which is referred to as the sensor response. The detection limit of this RGO-based FET biosensor is around 0.2 ng/mL, which is among the best of reported carbon nanomaterial based protein sensors and can be further optimized by tuning the sensor structure. Besides, GO-based FETs with high sensitivity to other biomaterials were also reported, including the glucose with a limit of detection of 1 nM and a rapid response within 1 s [263], GTP and ATP with a detection limit as low as 400 nM [259], and matrilysin (MMP-7) with a limit of detection of 10 ng/mL (400 pM) [300].

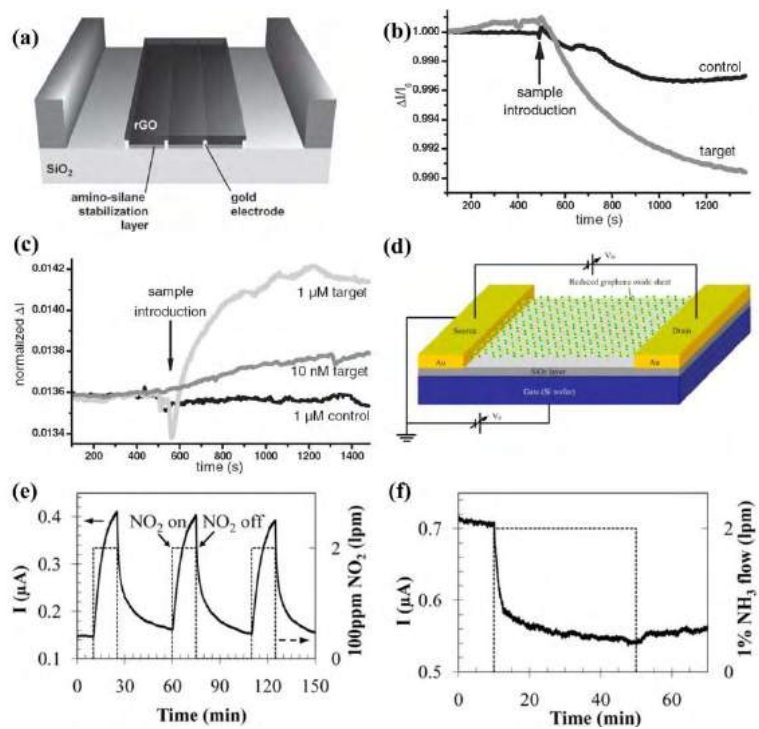

**Fig. 41** (a) RGO FET sensor schematic with an amino-silane film stabilizing the RGO layer. (b) Response of single FET devices after exposure to both target and control DNA solutions. (c)

Response of differential measurements using one of the FET devices as an internal reference where a definite change in the source-drain current is seen for the two samples containing the target, while no such response is seen for the control sample. Reproduced from Ref. [293]. (d) Schematic of the RGO FET device with a RGO sheet bridging the source and drain electrodes. (e) Repeatable response of the RGO FET to 100 ppm $NO_2$. (f) Dynamic sensing behavior of a RGO device (successively heated in Ar at 100, 200, and 300 ℃ for 1 hour each) for 1% $NH_3$ detection at room temperature. Reproduced from Ref. [301].

Another important application of the GO-based FET sensors is gas sensing. Lu *et al.* [301] investigated the RGO-based FET sensors for nitrogen dioxide ($NO_2$) and ammonia ($NH_3$) detection in the low concentration level, as schematized in Fig. 41d. FET with intrinsic GO sheets show no response to $NO_2$ or $NH_3$, indicating insignificant change in the electrical transport property of the non-reduced GO. However, after heating the GO at 100 ℃ for 1 hour followed by 1 hour heating at 200 ℃, or one-step heating at 200 ℃ for 2 hours, the partially reduced GO FET devices become highly responsive to $NO_2$ and $NH_3$, which may be attributed to vacancies or small holes of the graphene basal lattice as adsorption sites created during the thermal treatment. The observed sensitivity can be ascribed mainly to the electron transfer between the reduced GO and the adsorbed gaseous molecules of $NO_2$ or $NH_3$. First-principles calculation suggested that $NO_2$ is a strong oxidizer with electron-withdrawing power [302]. Thus, electron transferred from the RGO to the adsorbed $NO_2$ gives rise to an enriched hole concentration and further enhances electrical conduction in the RGO sheet [301]. Fig. 41e presents the repeatable response of RGO FET after successive multi-step heating at 100 and 200 ℃ for 1 hour. The sensing signal is highly dependent on the $NO_2$ concentration. The larger the $NO_2$ concentration is, the stronger the sensing signal will be. The lowest $NO_2$ concentration that can be detected by the GO sensor is 2 ppm. Moreover, compared with 200 ℃ annealing, annealing at higher temperature (300 ℃) will enhance the sensor sensitivity and shorten the response time, but lengthen the recovery time. The accelerated response is ascribed to the creation of more graphitic carbon atoms during annealing. The GO-based FET devices also respond to $NH_3$ after usual thermal treatments, as shown in Fig. 41f. However, the recovery process of the device after $NH_3$ sensing is extremely slow, taking more than 50 hours air flow. Moreover, the sensing behavior of $NH_3$ at room temperature of the RGO FETs can be modulated by the gate potential [303] . RGO under a positive gate potential (n-type conductance) exhibits an instantaneous response and fast recovery for $NH_3$ sensing, far superior to the performance in p-mode at zero or negative gate potential. Cheon *et al*. [304] studied the sensing abilities of GO-based organic FETs to different organic gas molecules using a high mobility copolymer, DPP-selenophene-vinylene-selenophene (DPP-SVS), doped with functional GO as an active layer. The polar functional groups in GO can serve as sensitizing centers for external polar organic gas molecules. But incorporation of GO will reduct the mobility of DPP-SVS. Therefore, an optimum GO concentration in the composite was found to be 1%. Then, it was found that for the polar organic molecule ethanol, acetone, and acetonitrile, each transistor parameter, threshold voltage responds independently to each gas molecule, which enables high selectivity of GO-based organic FETs for various gases. Particularly, the threshold voltage shift was more pronounced in the case of ethanol exposure (a shift of ~3.5 V) compared to that for acetone (~0.1 V) and acetonitrile (~0.5 V). The sensing mechanism of such GO-based organic FET gas sensors can be explained as follows. GO has a physical interface with the polymeric semiconductor and functional groups, enable adsorption of gas molecules. Upon adsorption of polar gas molecules, hole charge carriers in the semiconductor layer come across additional trapping centers, which will change the mobile charge carrier densities in the channel region of the FETs and further affect the FET parameters, including threshold voltage, charge carrier mobility, subthreshold swing, and ON/OFF ratio.

In addition, IR sensing [305, 306], pH sensing [307, 308] and ion sensing [262, 309] of GO-based FETs were also reported. Trung *et al*. [305] demonstrated the IR sensing ability of RGO FETs, which exhibited high sensitivity and reproducibility even with a temperature interval of 1 K. Then, high pH sensitivity of 57.5 mV/pH and excellent linearity of 0.9929 in a wide pH sensing range of 1–13 were achieved in the RGO-based FETs using an oxygen plasma treatment to increase the functional groups on the RGO films [308]. Besides, high sensitivity and excellent lower detection limit of 1 nM for real-time detection of $Hg^{2+}$ in an underwater environment were also obtained in the RGO-based FETs [262].

In brief, GO-based FET sensors offer low frequency noise, fast response time, and high sensitivity with a detection capability down to parts-per-billon (ppb) level or even to single molecule. Moreover, they are economical and massive to prepare, showing great potential for low cost commercial applications.

Benefited from the functional groups, GO/RGO sheets exhibit excellent field emission properties. Typically, the GO-based FETs are p-type under ambient conditions due to the p-type doping induced by oxygen-containing groups, but the highly reduced GO will present graphene-like ambipolar behavior. Fortunately, all of the carrier polarity, carrier mobility and ON/OFF current ratio of the GO-based FETs are tunable, making the GO FETs promising for future device applications.

## 4 Thermal properties and thermoelectric devices

### 4.1 Thermal property

The thermal properties of graphene and its derivatives have received increasing interest in recent years [310-312]. It is known that graphene possesses an extremely large thermal conductivity above 2000 W/mK at room temperature, which is very promising for heat removal in the electronic devices [22, 23]. On the other hand, a suppressed thermal conductivity is required in the thermoelectric materials [313]. Oxygen functionalization is an effective way to tailor the thermal transport properties of graphene for various applications [249, 314].

A few experiments showed that the thermal conductivity $\kappa$ of GO/RGO papers or thin films is significantly reduced with regard to graphene (1500~5000 W m$^{-1}$ K$^{-1}$ at room temperature [315]), i.e., below 10 W m$^{-1}$ K$^{-1}$ [316, 317], due to the phonon scattering at the junctions of GO nanoplatelets. On the theoretical sides, many efforts have been devoted to clarifying the relation between thermal conductivity and the chemical structure of GO, as well as to exploring the phonon scattering mechanism [24, 318-320]. The thermal transport of GO can be simulated by the non-equilibrium molecular dynamic (NEMD) method (Fig. 42a). Various algorithms have been proposed to impose a constant heat current flow J to the system by injecting/extracting or exchanging a kinetic energy between heat source and sink [321, 322]. The thermal conductivity of the system reaching the steady state is given by the Fourier's law: $\kappa = -J/\mathrm{Grad}(T)$, where Grad(T) is the gradient of temperature.

Zhang *et al*. [24] focused on the disordered GO at low oxidation levels. They employed the NEMD method and used an improved reactive empirical bond order potential of carbon/hydrogen/oxygen (REBO-CHO) for describing the interatomic interactions. At room temperature, pristine graphene shows a thermal conductivity of 3261 W/mK for a sample length L

= 10 µm. It was found that κ linearly increases with ln(L), indicating long phonon mean free path in graphene. As for GO, κ decreases rapidly with increasing oxygen content (Fig. 42b). For an O:C ratio of 0.04, κ drops to 4−10% of that of graphene at the same length. For an O:C ratio of 0.1 and L = 300 nm, κ attains 35 W m$^{-1}$ K$^{-1}$, 85 W m$^{-1}$ K$^{-1}$, and 29 W m$^{-1}$ K$^{-1}$ for GO with epoxy-, hydroxyl-, and ether-functionalization, respectively. Furthermore, the thermal conductivity of GO is independent of L when L > 100 nm, corresponding to diffusive phonon transport in GO. The reduction of κ is mainly caused by the oxygen defect scattering, which results in the decrease of phonon mean free path. The phonon group velocity and heat capacity of GO do not change much compared to graphene, since their phonon density of states show little difference. Mu *et al.* [318] reported similar results for GO consisting of epoxy groups (Fig. 42c). In particular, κ drops to 3% of that of graphene for GO with an O:C ratio of 0.2 and L = 20 nm, and κ shows little variance for L between 20 nm and 320 nm (Fig. 42d). For GO with O:C = 0.2, the diffusive limit of κ is down to 8.8 W m$^{-1}$ K$^{-1}$, which is even lower than the minimal κ of graphene at the amorphous limit (11.6 W m$^{-1}$ K$^{-1}$).

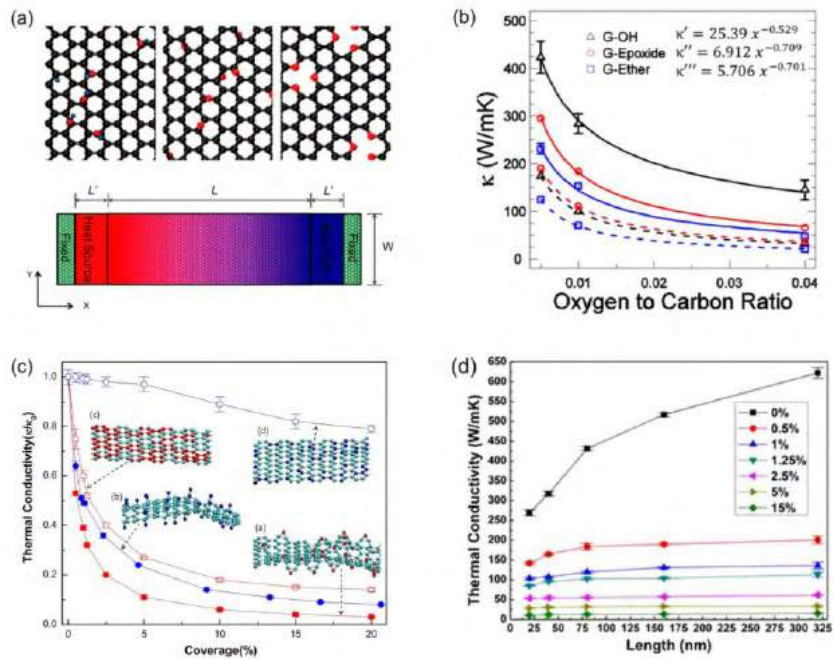

**Fig. 42** (a) Top panel: from left to right, GO models with hydroxyl, epoxy and ether functionalizations, respectively. Carbon, oxygen and hydrogen atoms are represented in black, red and blue colors, respectively. Bottom panel: schematic diagram of thermal conductivity calculation by NEMD simulations. (b) Thermal conductivity of GO with epoxy, hydroxyl and ether functionalization as a function of the O:C ratio at 300 K. The solid and dashed curve lines are fitted with the results from MD simulation and Klemens model, respectively. Reproduced from Ref. [24]. (c) Thermal conductivity of hydrogenated graphene (solid blue symbols), epoxy-functionalized graphene (solid red symbols), graphene with mass defects mimicking hydrogen (open blue symbol) and oxygen (open red symbols) atom weight, respectively, as a function of the oxygen/hydrogen coverage rate. Insets show the corresponding structures for the simulations. Carbon, oxygen and hydroxyl atoms are represented in cyan, red and blue colors, respectively. The sample length is 20 nm for all cases. (d) Thermal conductivities of epoxy functionalized graphene with different oxygen coverage rates as a function of the sample length. Reproduced from Ref. [318].

Lin *et al*. [319] considered GO models containing both epoxy and hydroxyl groups (the epoxy to hydroxyl ratio of 1:1). They used ReaxFF to describe the interactions between C, H and O species. They calculated $\kappa$ of GO in the ballistic regime, wherein $\kappa$ increases with the transport length below 50 nm (Fig. 43a). The relative thermal conductivity of GO to graphene $\kappa/\kappa_0$ decreases with increasing oxygen content (Fig. 43b). For L = 50 nm, $\kappa/\kappa_0$ drops to 20% and 10% for an O:C ratio of 0.05 and 0.2, respectively. At the same O:C ratio, $\kappa/\kappa_0$ decreases as the sample length increases. The intrinsic phonon mean free path $l_{ph-ph}$ and diffusive thermal conductively $\kappa_\infty$ are obtained by interpolating $1/\kappa \sim 1/l_{ph-ph} + 1/l_{ph-b}$ ($l_{ph-b}$ = L is the sample length), and both quantities decrease with increasing oxygen content (Fig. 43c). For GO with an O:C ratio of 0.05 and 0.2, $l_{ph-ph}$ drops to 3% and 1%, and $\kappa_\infty$ drops to 3% and 1.7% of that of graphene, respectively. The reduction of $\kappa_\infty$ of GO is mainly attributed to the oxygen defect scattering. Besides, the phonon density of states of C–C modes (G peak) decreases due to the presence of oxygenated groups, which suppresses the specific heat of this mode and also degrades the overall thermal conductivity of GO (Fig. 43d).

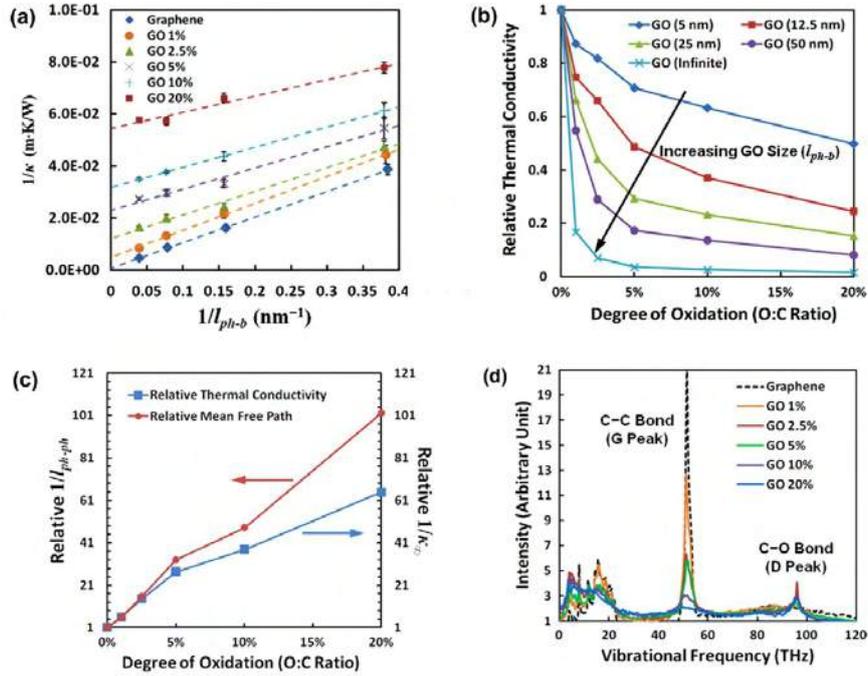

**Fig. 43** (a) 1/κ versus 1/l$_{ph-b}$ curves for graphene and GO with various oxygen contents. (b) The relative thermal conductivity of GO (normalized by the pristine graphene value) as a function of O:C ratio for different sample lengths. (c) Comparison between the intrinsic relative phonon mean free path 1/l$_{ph-ph}$ and the thermal conductivity 1/κ$_0$ (both normalized by the pristine graphene value) as a function of the O:C ratio. (d) The density of states of the in-plane phonon modes as a function of the phonon vibrational frequency for graphene and GO with various oxygen contents. Reproduced from Ref. [319].

Using reverse NEMD technique and COMPASS force field, Shen *et al*. [323] explored the effect of tensile strain on the thermal conductivity of GO containing both epoxy and hydroxyl groups in random distributions. Without strain, GO with L = 29 nm shows κ of about 8% and 4% of that of graphene, for an O:C ratio of 0.05 and 0.5, respectively. Applying a 10% uniaxial strain, κ of graphene drops by 28.4%, while κ of GO increases by 9−12%, depending on the oxidation level (Fig. 44a). The thermal conductivity shows opposite trends under external strain or stress for graphene and GO (Fig. 44b). Further analysis of phonon vibrational power spectra shows that both group velocity and phonon mean free path decrease with increasing stress for graphene, leading to the reduction of thermal conductance (Fig. 44c and d). As for GO, the impact of phonon softening is less prominent due to less elongation of C−C bond. The phonon mean free path of GO increases rapidly with stress, as the uniaxial strain reduces the vertical displacement of sp³ carbon species and flattens the structure (Fig. 44e). The reduced phonon scattering is more dominant, and hence the thermal conductivity of GO is enhanced under tensile strain.

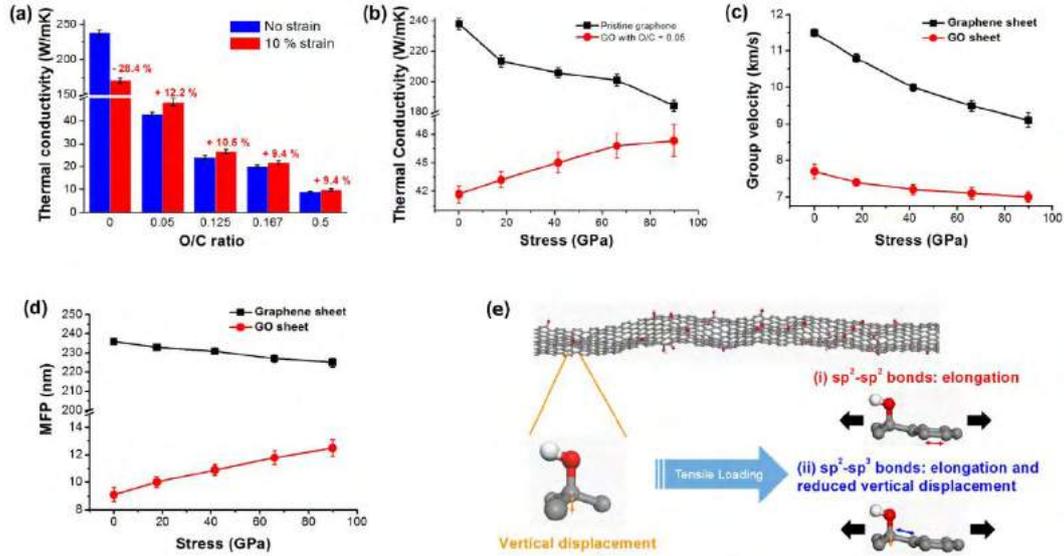

**Fig. 44** (a) Thermal conductivities of GO sheets as a function of the O:C ratio before and after applying a 10% strain. (b)-(d) Thermal conductivity (b), group velocity (c), and phonon mean free path (d) of graphene and GO with an O:C ratio of 0.05 as a function of tensile stress, respectively. (e) Schematic of vertical displacement of sp³ carbon atoms due to functionalization and the changes in bond structures after tensile loading. Reproduced from Ref. [323].

## 4.2 GO-based thermoelectric devices

A thermoelectric device converts heat to electrical energy and vice versa via the Peltier–Seebeck effect, and the efficiency is characterized by a figure of merit ZT defined as [324]:

$$ZT = S^2\sigma T/\kappa, \tag{9}$$

where S is the Seebeck coefficient (or thermopower), $\sigma$ is the electrical conductivity, T is the absolute temperature, and $\kappa$ is the thermal conductivity from electrons and phonons. Materials with ZT ~ 1 are usually regarded as good thermoelectric materials, while devices with ZT > 3 are competitive to the conventional energy conversion techniques [325].

The difficulty to achieve high thermoelectric efficiency relies on the delicate compromise between S, $\sigma$ and $\kappa$ [326]. A material with high $\sigma$ is usually a good thermal conductor with high $\kappa$. A bandgap is required to obtain a large S, which however is detrimental to the electrical conductance [327]. To balance all these factors, narrow-bandgap semiconductors are expected to be candidate thermoelectric materials with high efficiency [328, 329]. As a semimetal, graphene has a low ZT value of about 0.01 at room temperature due to the small S and large $\kappa$ [330, 331]. Oxygen functionalization of graphene not only opens a bandgap, but also introduces phonon scattering centers and reduces the thermal conductance, and therefore providing a chance for improving the ZT value. So far, a few experimental studies have been carried out to investigate the performance of GO-based thermoelectric devices [21, 332-334].

Tian *et al*. [21] fabricated a macroscopic scale thermal rectifier using RGO papers, as schemed in Fig. 45a. It was found that the thermal rectification closely depends on the asymmetric shape. For the 60º triangular shaped RGO thermal rectifier, it has a rectification coefficient of 1.28 and the rectification coefficient increases with the increasing of the angle, as shown in Fig. 45b. While, the two-rectangular shaped sample has a rectification coefficient of 1.10, which can be enhanced by increasing the width ratio. Moreover, the rectification factor of the GO-based thermal rectifier is also related to the concentration of functional groups [335], as shown in Fig. 45c.

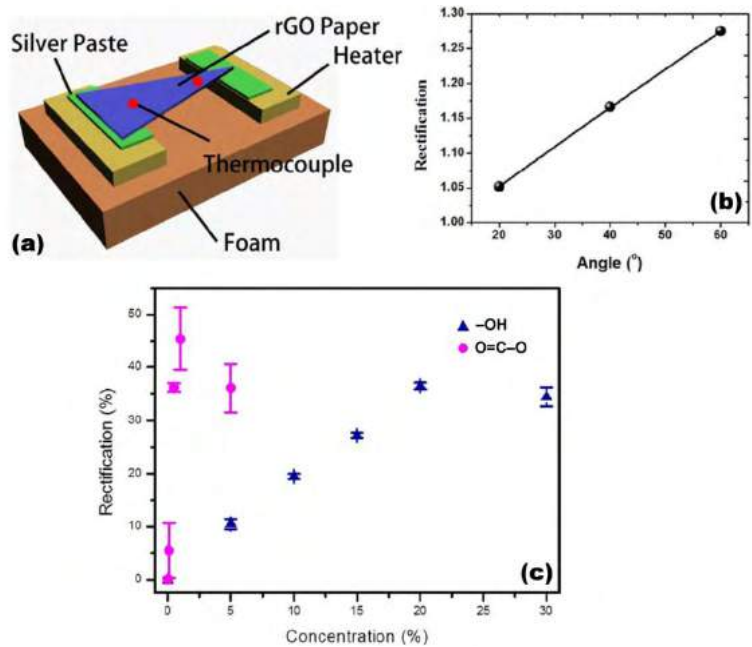

**Fig. 45** (a) Schematic view of the RGO-based thermal rectifier. (b) Relationship between the rectification coefficient and the shape angle at a temperature difference of 40 K. Reproduced from Ref. [21]. (c) Rectification factor as a function of the concentration of functional groups for the GO-based thermal rectifier. Reproduced from Ref. [335].

Generally speaking, to ensure outstanding device performance, high thermopower values of GO-based thermoelectric devices should be reached. Several approaches have been proposed to improve the thermopower. For example, Xiao *et al*. [330] reported enhanced thermopower of few-layer graphene films with oxygen plasma treatment. Before oxidation, graphene films show thermopower of 40–80 µV/K and electrical conductivity of 4–5×$10^4$ S/m in the temperature range of 300–575 K. After exposed in oxygen plasma for 15 seconds, the thermopower of the obtained GO films is significantly enhanced to about 700 µV/K at T = 575 K, and the electrical conductivity remains on the order of $10^4$ S/m (Fig. 46a and b). As a result, the power factor $S^2\sigma$ of GO attains as much as 4.5×$10^{-3}$ W/mK$^2$, which is about 15 times larger than that of the pristine graphene film (Fig. 46c). In contrast, the chemically reduced GO made by the Hummers method shows thermopower below 20 µV/K and electrical conductivity of 2–9×$10^3$ S/m, and its power factor is only 0.8–3.6×$10^{-6}$ W m$^{-1}$ K$^{-1}$ (Fig. 46d). The enhancement of thermopower for GO made by oxygen plasma treatment may be attributed to the bandgap opening. For RGO samples, the remaining oxygen content may be insufficient to open a gap. In addition, a considerable amount of defects may be generated in RGO during the aggressive synthesis process, which deteriorates electronic transport properties and leads to very poor thermoelectric performance of RGO.

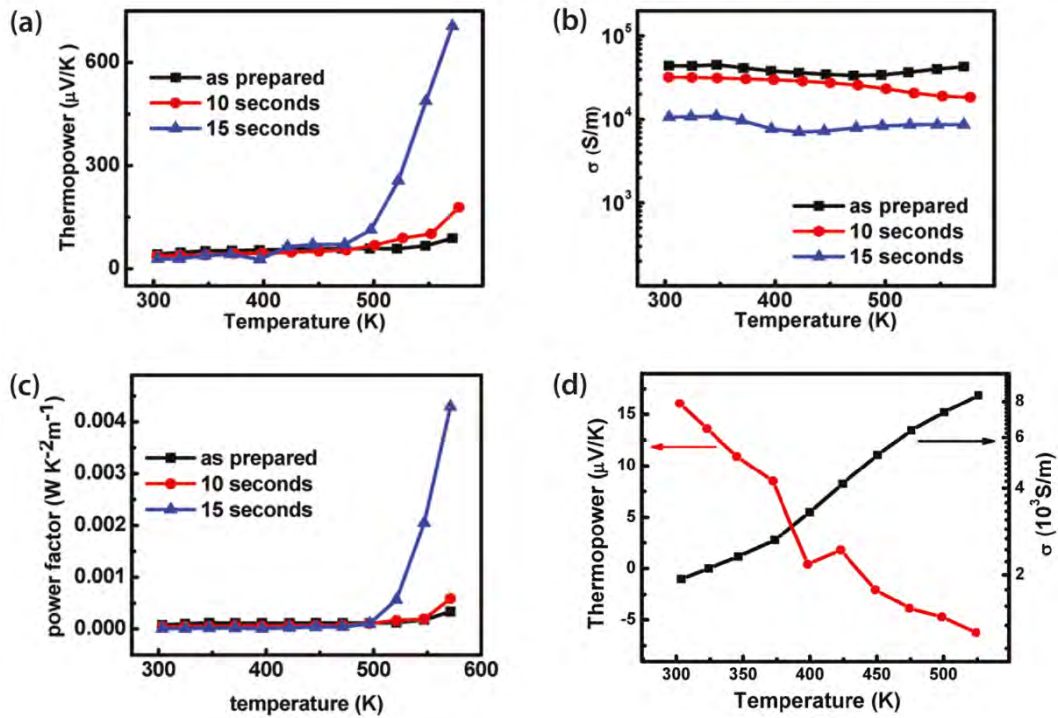

**Fig. 46** Temperature dependent of thermopower (a), electrical conductivity (b), and power factor (c) for few-layer graphene films with oxygen plasma treatments. (d) A comparison of thermopower and electrical conductivity of RGO films as a function of temperature. Reproduced from Ref. [330].

Choi *et al*. [316] suggested that the thermoelectric properties of GO depend on the oxidation degree. They made GO papers by Hummers oxidation of graphite flake and controlled the oxygen content via chemical reduction. As the oxidation level (characterized by the ratio of total area of O1s and C1s XPS spectra) decreases from 0.85 to 0.32, the electrical conductivity of GO increases from 0.14 S/m to 880 S/m, and the thermopower reduces from 60 μV/K to 11 μV/K at room temperature (Fig. 47a). The power factor reaches the highest value of $10^{-7}$ W/mK$^2$ at the oxidation level of 0.32 (Fig. 47b), consistent with the result reported by Xiao *et al*. [330]. The thermal conductivity of RGO with oxidation level of 0.32 is found to be 0.3 W/mK, which is about four orders of magnitude lower than that of graphene, due to the phonon scattering at the junctions of RGO nanoplatelets. However, the highest ZT value achieved is only $10^{-4}$, because of the low power factor of RGO samples.

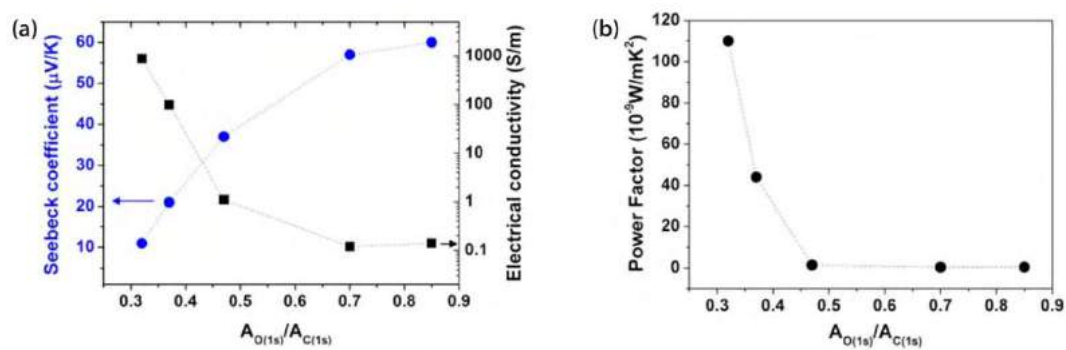

**Figure 47** Seebeck coefficient, electrical conductivity (a), and power factor (b) of RGO bucky papers as a function of the ratio of O 1s and C 1s XPS area. Reproduced from Ref. [316].

On the theoretical side, Zhou *et al*. [336] patterned GO nanoroads (GONRDs) functionalized by epoxies on graphene monolayer and investigated their ballistic thermoelectric transport properties using DFT calculations and the nonequilibrium Green's function method. They showed that GONRDs are all semiconductors with band gap tunable by the road width, edge orientation, and the structure of GO matrix, which remarkably enhances the Seebeck coefficient. The power factor of GONRDs increases to 4–22 times and lattice thermal conductance is reduced to 15–22% of the values of graphene, achieving a figure of merit up to 0.75. These theoretical results offer a strategy for designing GO based thermoelectric devices of high efficiency.

Although few GO-based thermoelectric devices with high thermoelectric efficiency have been obtained, several studies show the wide-range tunability of the electronic and thermal properties of GO, which shed light on the potential of GO-based materials for thermoelectric devices. More efforts from both experimental and theoretical sides are necessary to unravel the correlations of the three key parameters, i.e., S, σ, and κ, with the chemical structure of GO, in order to achieve the best thermoelectric performance of this material.

## 5. Nonlinear optics

The heterogeneous atomic and electronic structures enable GO abundant optical properties, such as tunable optical transmittance, intrinsic fluorescence and nonlinear optical properties [18, 35, 337-339]. Generally, GO is transparent with high optical transmittance in the visible spectrum region due to its atomically thin nature [18]. The optical transmittance depends on both thickness and oxygen content of GO. As the thickness of GO increased from 6 nm to 41 nm, the transmittance was found to drop from 90% to 20% at the 550 nm wavelength [19]. Meanwhile, the transmittance of GO also decreases with increasing oxidation level. It was found that for the GO films with 1–5 nm thickness, the optical transmittance at a wavelength of 550 nm drops from 95% to 60% after chemical reduction and thermal annealing treatments [15, 205], which was attributed to the restoration of $sp^2$ hybridization in carbon atoms. The transparent feature of GO stimulates its applications in display industry [88]. On the other hand, GO exhibits intrinsic fluorescence in the near-infrared (NIR), visible and ultraviolet regions [340-342], which is strikingly different from graphene since graphene has no energy gap [343]. The fluorescence nature makes GO promising in various fields, especially the biosensing [35, 37, 344, 345].

Another attractive optical property is the nonlinear optics of GO. The nonlinear optics, promoted by laser technology, mainly deals with various new optical effects and phenomena arising from the interactions of intense coherent radiation with matter. Since the $sp^2$ fraction of carbon domains in GO can be tailored through careful and controllable removal of oxygen-containing groups, rich nonlinear optical properties were observed in GO, including saturable absorbers, two-photon absorption, the four-wave mixing effect and optical limiting.

### 5.1 Saturable absorbers

Saturable absorption is a phenomenon that the absorption of light decreases with increasing light intensity. At high enough incident intensity, atoms in the ground state of a saturable absorber material are excited into an upper energy state. If the rate for the atoms decaying back to the ground state is lower than the depleting rate of ground state, the absorption will subsequently saturate [339]. Employing femtosecond pump-probe and Z-scan techniques, Zhao *et al.* [346] demonstrated that few-layered GO films show a fast energy relaxation of hot carriers and strong saturable absorption. From the transient differential transmission spectra presented in Figs. 48a and b, it can be noticed that the decay time $\tau$ of few-layered GO is about 0.17 ps, which is much faster than that of GO suspensions ($\tau = 0.37$ ps). The saturable absorption was measured by the Z-scan technique. The peak in Z-scan curves indicates that saturable absorption exists for both few-layered GO and GO suspensions due to photo-induced bleaching, as shown in Fig. 48c. Linear absorption coefficients of GO suspensions and few-layered GO at 800 nm are 0.62 cm$^{-1}$ and 10.6 cm$^{-1}$, respectively. Thus, the few-layered GO has stronger saturable absorption than GO suspensions. The fast carrier relaxation and large saturable absorption of few-layered GO were suggested to arise from the large fraction of sp$^2$ carbon atoms inside the few-layered GO sheet together with oxidation mainly existed at the edge areas.

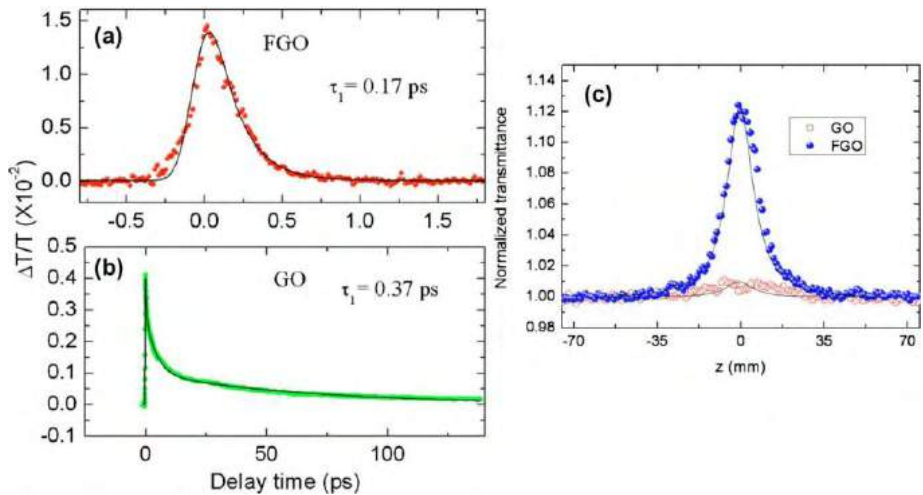

**Fig. 48** Transient differential transmission spectra of the few-layered GO (FGO) (a) and GO (b) suspensions as a function of probe delay in degenerate pump-probe experiments at 800 nm. The solid lines are fittings; and (c) Open-aperture Z-scan curves of FGO (filled circles) and GO (open circles). The solid lines are theoretical fittings. Reproduced from Ref. [346].

The large absorption, ultrafast carrier dynamics and Pauli blocking make GO promising in saturable absorbers. Compared with semiconductor saturable absorber mirrors and single-walled carbon nanotubes (SWNTs), GO-based saturable absorbers present good broadband tunability and large operating wavelength ranges, which are useful for mode-locking and Q-switching [347].

### 5.1.1 Mode-locking

Mode-locking is an optical technique by which a laser can be utilized to produce light pulses of extremely short duration, on the order of picoseconds or femtoseconds. Sobon *et al.* [348] comprehensively studied GO and RGO based saturable absorbers for mode-locking of Er-doped fiber lasers, as illustrated in Fig. 49a. Table 4 summarizes the laser parameters with GO and RGO based absorbers. The laser performance of the GO and RGO based saturable absorbers has no significant difference. Both the saturable absorbers provide stable and mode-locked operation with a soliton pulse duration of 390 fs over 9 nm of full width at half maximum (FWHM) bandwidth at 1560 nm center wavelength. Meanwhile, both GO and RGO based saturable absorbers show low non-saturable loss around 15%, as shown in Fig. 49b. However, the RGO-based absorber has a relatively high modulation depth than that of the GO-based absorber, 21% and 18%, respectively. Considering of the cost of production, GO seems to be a better saturable absorber than RGO. So far, passive mode-locking operations presented at 1 μm [349, 350], 1.5 μm [348, 351-355], and 2 μm [356, 357] spectral ranges were reported for the GO-based saturable absorbers.

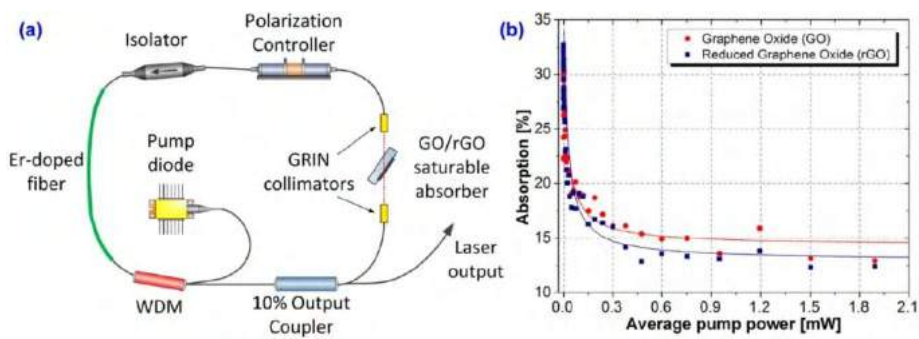

**Fig. 49** (a) Experimental setup of the mode-locked laser; and (b) power-dependent transmission of the GO and RGO. Reproduced from Ref. [348].

**Table 4** Summary of the laser parameters with GO and RGO based absorber. FWHM is the full width at half maximum; TBP is the time-bandwidth products; RF is radio frequency; and SNR is signal to noise ratio. Reproduced from Ref. [348].

| Parameter | Value | |
|---|---|---|
| | GO | RGO |
| FWHM bandwidth | 9.3 nm | 9.2 nm |
| Pulse duration | 390 fs | 390 fs |
| TBP | 0.448 | 0.442 |
| Pulse Energy | 33.7 pJ | 29.8 pJ |
| Peak Power | 86.4 W | 76.4 W |
| Soliton Order N | 0.79 | 0.75 |
| RF SNR | 60 dB | 60 dB |
| Pump power | 92 mW | 82 mW |
| Output power | 1.96 mW | 1.68 mW |
| Center wavelength | 1558 nm | 1559 nm |

On the other hand, the performance of GO-based saturable absorbers can be tuned by varying the number of GO layers. Boguslawski *et al.* [354] studied GO paper based saturable absorbers in Er- and Tm-doped fiber lasers. It was found that stable mode-locked operation provides 613 fs soliton pulses centered at 1565.9 nm in Er-doped fiber lasers, and 1.36 ps soliton pulses centered at 1961.6 nm in Tm-doped fiber lasers, respectively. In particular, by increasing the number of GO paper layers in the Er-doped lasers, the FWHM of the optical spectrum can be scaled from 1.7 to 4.2 nm, corresponding to 1.5 ps and 613 fs pulse durations, as shown in Fig. 50a and b. Li *et al.* [355] also demonstrated that parameters such as transmittance, saturation intensity, and modulation depth, can be controlled by changing the piece number of the GO-based saturable absorber pieces in Er-doped fiber lasers, as listed in Table 5. For the saturable absorber with four GO-PVA pieces, the pulse width is about 953.8 fs. Besides, it was found that the number of saturable absorber pieces influences the noise characteristics of mode-locking. By optimizing the number of saturable absorber pieces, relaxation oscillation can be completely suppressed and the phase noise can be reduced by more than 8 dB at 10 kHz in a laser mode locked by GO-PVA saturable absorbers.

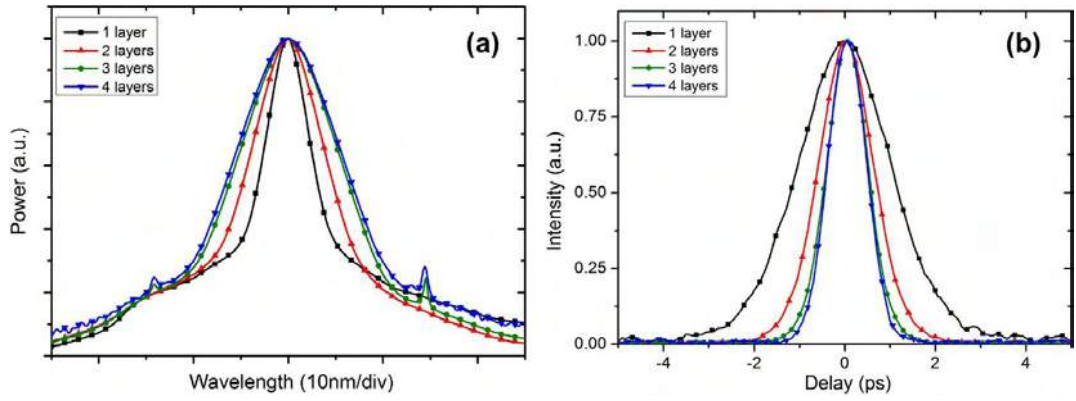

**Fig. 50** (a) Comparison of the FWHM optical spectra, and (b) autocorrelation traces of output pulses for Er-doped fiber laser with different layers of GO paper as a saturable absorber. Reproduced from Ref. [354].

**Table 5** Summary of the nonlinear transmittances with different pieces GO-PVA saturable absorbers at a measurement wavelength of 1560 nm and the output performances of the mode-locked fiber ring lasers based on GO-PVA saturable absorbers. Reproduced from Ref. [355].

| Parameter | Number of GO-PVA saturable absorber | | | |
|---|---|---|---|---|
| | 1 | 2 | 3 | 4 |
| Starting transmission (%) | 89.91 | 77.78 | 67.95 | 55.39 |
| Saturated transmission (%) | 90.49 | 79.03 | 68.78 | 56.65 |
| Nonsaturable loss (%) | 9.51 | 20.97 | 31.22 | 43.35 |
| Modulation depth (%) | 0.58 | 1.25 | 1.83 | 1.96 |
| Saturation intensity (MW/cm$^2$) | 0.205 | 1.786 | 1.820 | 2.254 |
| Pump current (mA) | 156 | 168 | 184 | 203 |
| Output power ($\mu$W) | 299 | 212 | 277 | 289.7 |
| Spectral bandwidth (nm) | 1.6 | 2.26 | 2.9 | 2.79 |
| Center wavelength (nm) | 1559 | 1558.9 | 1559 | 1558 |
| RF carrier power (dBm) | −28.1 | −26.2 | −32.5 | −33 |
| Phase noise@1 kHz (dBc/Hz) | −102 | −110.8 | −106.6 | −100 |
| Phase noise@10 kHz (dBc/Hz) | −113.1 | −120.7 | −120.6 | −119.4 |

In addition, the nonlinearity of the GO-based saturable absorbers can lead to polarization switching in a mode-locked fiber lasers [358]. It was found that in a passively mode-locked fiber laser with RGO as a saturable absorber, the laser moves from the self-pulsing mode to the rectangular-pulse mode during increasing the pump power, as shown in Fig. 51a. Also, varying the pump power will produce different polarization switching characteristics. At low pump powers generally within 150 mW, the laser output consists of weak but sharp pulses, indicating a weak polarization switching effect, as shown in Fig. 51b. While, at a high enough pump power such as 252 mW, the sharp pulses change into rectangular-shaped pulses (Fig. 51b), and the switching frequency is twice of the fundamental cavity round-trip frequency. Beyond a certain high pump power, the polarization switching becomes intense and the switching frequency is equal to the fundamental cavity round-trip frequency.

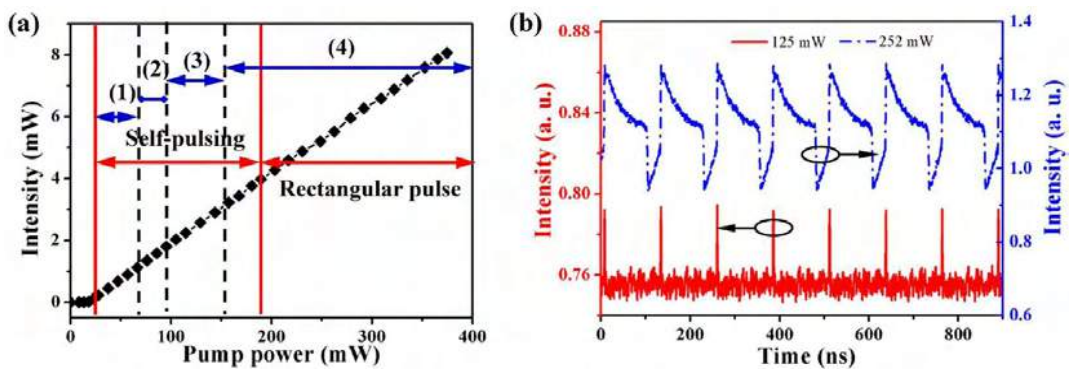

**Fig. 51** (a) Change from the self-pulsing mode to the rectangular pulse mode with increasing pump power, where "(1)", "(2)", "(3)", and "(4)" specify the regions with different polarization switching characteristics. (b) Pulse trains measured at 125 and 252 mW, respectively. Reproduced from Ref. [358].

### 5.1.2 Q-switching

Q-switching is also a technique used to pulsed output beam. It allows producing light pulses with extremely high peak power on the order of gigawatt. Compared with mode-locking, Q-switching leads to much lower pulse repetition rates, much higher pulse energies, and much longer pulse durations. Wang *et al.* [359] demonstrated that GO can be used as a highly efficient saturable absorber for the Q-switched Nd:GdVO$_4$ laser operating at a wavelength of 1064 nm. They fabricated a sandwich structure composed of a piece of GO/PVA absorber, a piece of round quartz and an output coupler mirror by the vertical evaporation method and obtained 104 ns pulses and 1.22 W average output power with the maximum pulse energy at 2 μJ and a slope efficiency of 17%. As shown in Fig. 52a, the average output power increases linearly with the incident pump power with a corresponding slope efficiency of 17%. Within the incident pump power of 10 W, an average output power of 1.22 W was obtained. Further increasing the incident pump power will decrease the output power and damage the absorber. Therefore, at the pump power of 10 W, the maximum pulse energy of 2 μJ was achieved from the Fig. 52a. Meanwhile, from the Q-switched pulse profile under the incident pump power of 10 W, the minimum pulse width of 104 ns was obtained [360], as shown in Fig. 52b.

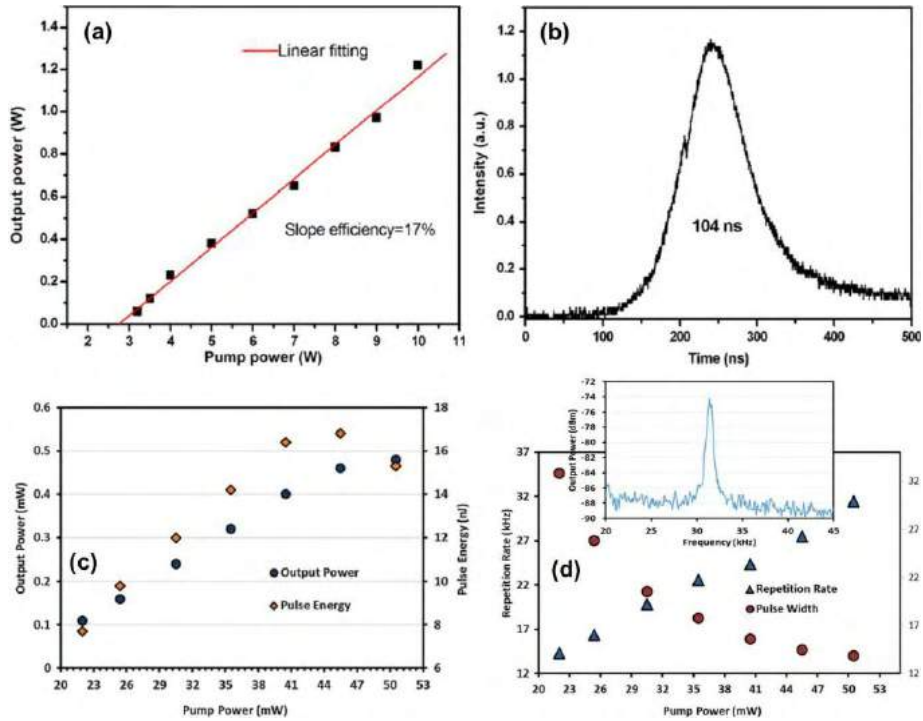

**Fig. 52** (a) Average output power versus incident pump power for Q-switching operation. (b) Q-switched pulse profile under the incident pump power of 10 W. Reproduced from Ref. [359]. (c) Average output power and pulse energy as functions of the pump power at a wavelength of 980 nm. (d) Repetition rate and pulse width as functions of the pump power at a wavelength of 980 nm. Inset shows the RF spectrum of the Q-switched laser at the pump power of 50.5 mW. Reproduced from Ref. [360].

Usually, there is a pump threshold for the Q-switched laser. Sobon *et al.* [361] reported a passively Q-switched Erbium-doped fiber laser based on RGO saturable absorber. When the pump power was below 120 mW, the laser operated in continuous-wave regime. Above the threshold of 120 mW, Q-switched operation was observed. Depending on the pump power, the output pulse energy varied between 90 nJ and 125 nJ. At a pump power of 175 mW, the maximum average output power of 14.6 mW and pulse of 1.85 µs were obtained. Then using a GO-deposited tapered fiber device as a saturable absorber, Liu *et al.* [362] fabricated the high-energy Q-switched double-clad thulium-doped fiber laser operating at a wavelength of 2 μm. The passively Q-switched fiber laser started continuous-wave operation at the pump power of 4.8 W and transited to Q-switched mode when the pump power increased to 5.1 W. Afterwards, it became robust when the pump power reached 5.2 W. Further increasing the launched pump power to beyond 6.4 W, the Q-switched operation started to become unstable. At a pump power of 6.27 W, the average power of 302 mW and the corresponding pulse energy of 6.71 μJ were achieved with a slope efficiency of ~21.8% and a pulse width of ~3.8 μs. Besides, for the Q-switched erbium-doped fiber lasers using GO paper as a saturable absorber, a pump threshold of 22 mW was obtained, at which a stable and self-starting Q-switched operation began to be observed [360]. However, when the pump power exceeds 50.5 mW, the Q-switching pulse disappears, indicating a damage threshold. The relationships between the average output power/pulse energy and the pump power are presented in Fig. 52c. It can be noticed that as the pump power increased from 22 to 50.5 mW, the average output power increased from 0.11 to 0.48 mW, and the pulse energy varied within 7.7 to 16.8 nJ. At the pump power of 45.5 mW, the maximum pulse energy of 16.8 nJ is obtained. Meanwhile, the pulse repetition rate changed from 14.3 to 31.5 kHz, whereas the corresponding pulse width decreased from 32.8 to 13.8 μs, as shown in Fig. 52d. As a result, the GO-based Q-switched lasers can operate at a maximum repetition rate of 31 kHz, with pulse duration of 7.04 μs and pulse energy of 2.8 nJ. Similarly, using a GO wafer as the saturable absorber, Lin *et al*. [363] reported a passively Q-switched 1066 nm c-cut Nd:YVO$_4$ laser, which achieves the maximum average output power of 514 mW at the pump power of 3 W, corresponding to an optical-to-optical conversion efficiency of 17.1%. It was found that the Q-switched laser has a low pump threshold of 1.5 W. During increasing the pump power from 1.5 W to 3 W, the pulse repetition rate changed widely from 62 kHz to 386 kHz. At the maximum repetition rate of 386 kHz, the minimum pulse width of 246 ns and pulse energy of 1.33 μJ were obtained.

In addition to the GO-based Q-switched fiber lasers operated at a single wavelength, such as 1.06 μm [359, 363], 1.5μm [360, 364] and 2 μm [362], dual-wavelength and multi-wavelength Q-switched lasers were also reported. For example, Song *et al*. [365] reported a passively Q-switched dual-wavelength Nd:GYSGG laser operating at 1057.28 and 1060.65 nm with GO as the saturable absorber, as shown in Fig. 53a. The maximum dual-wavelength average output power of 521 mW was achieved under the absorbed pump power of 5.4 W, corresponding to the optical-to-optical conversion and slope efficiency of 9.8% and 21%, respectively. The minimum pulse width was 115 ns with a pulse repetition rate of 338 kHz. Particularly, Zhao *et al*. [366] realized five-wavelength operation and pulse output in a GO-based Q-switched all fiber laser system. As presented in Fig. 53b, at a launched pump power of 240.6 mW, the output spectra were centered at 1548.004 nm, 1548.876 nm, 1549.928 nm, 1551.04 nm, and 1551.968 nm, with 3-dB bandwidth of 0.196 nm, 0.196 nm, 0.192 nm, 0.180 nm, and 0.184 nm, respectively. During the five-wavelength Q-switched operation, the minimum pulse duration of 2.72 μs, the maximum repetition rate of 72.25 kHz, and the maximum pulse energy of 229.74 nJ could be generated with a slope efficiency of ~4.9%.

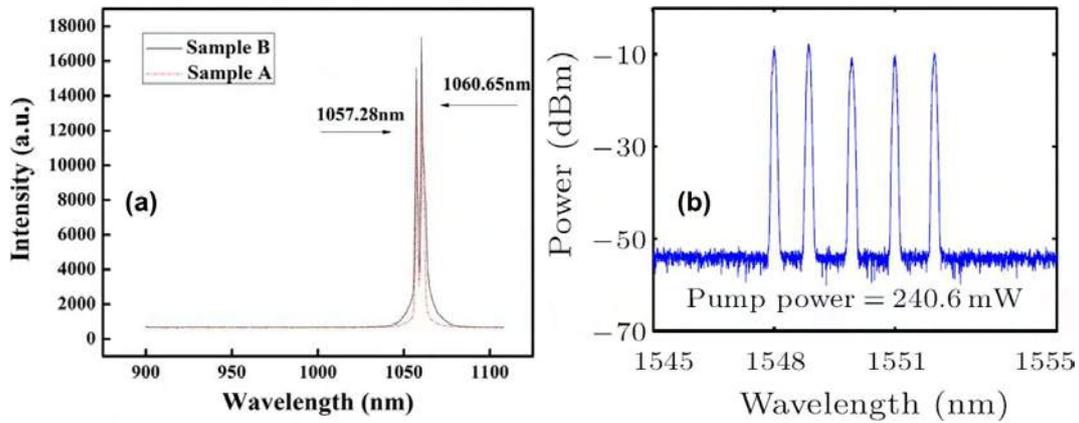

**Fig. 53** (a) Output spectrum of passively Q-switched Nd:GYSGG laser with GO saturable absorber. Samples A and B indicate alcohol-dispersed GO solution in 6 and 8 ml of alcohol, respectively. Reproduced from Ref. [365]. (b) Output spectra at the pump power of 240.6 mW of the erbium-doped ring-cavity Q-switched fiber laser with a GO-based saturable absorber. Reproduced from Ref. [366].

## 5.2 Reverse saturable absorption

In addition to the saturable absorption, the other nonlinear absorption is reverse saturable absorption, which indicates that the excited state absorption is larger than the ground state absorption. Materials with reverse saturable absorption will be less transmissive when excited, which is opposite to the case of saturable absorption and thus called reverse saturable absorption [339].

Jiang et al. [367] studied the nonlinear optical properties of GO thin films for femtosecond laser pulses at 800 and 400 nm. As shown in Fig. 54a, at a low excitation fluence of 5 mJ/cm$^2$ at 800 nm, the normalized transmittance increased as the sample moved into the beam focus at z = 0, indicating an optically induced transparency or saturable absorption behavior. However, as the excitation fluence increased, reverse saturable absorption was observed instead, with decreased normalized transmittance. Besides, the GO films showed systematical reverse saturable absorption even at very low excitation fluence when 400 nm laser pulses were used, as presented in Fig. 54b, indicating the existence of stronger nonlinear absorption at 400 nm than that of 800 nm. Moreover, reverse saturable absorption of the GO thin films relies on the film thickness. As shown in Fig. 54c and d, at a certain input fluence, the normalized transmittance decreases as the film thickness increases for the laser pulses both at 800 and 400 nm.

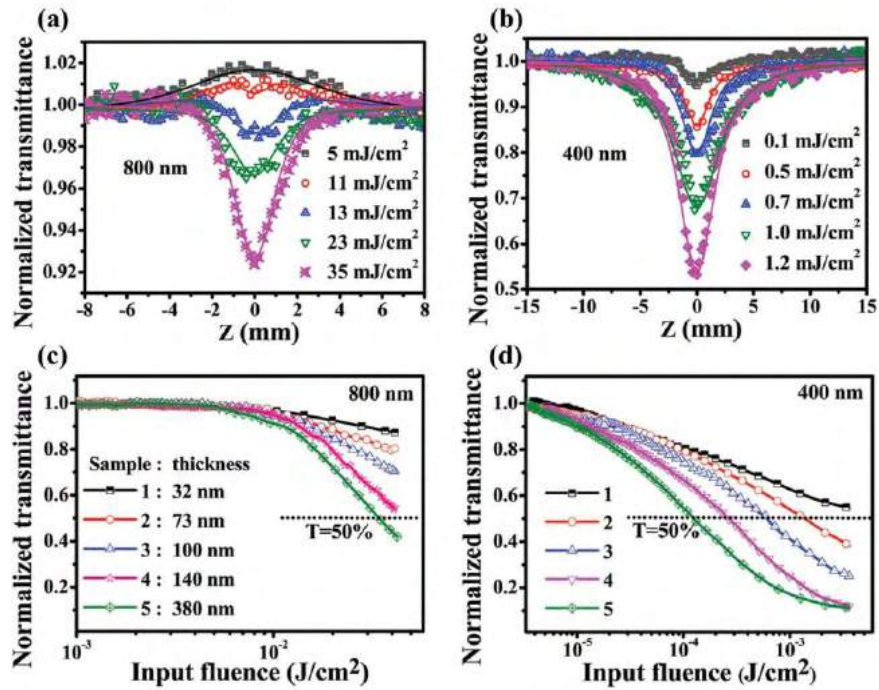

**Fig. 54** Open aperture Z-scan results of a GO film at 800 (a) and 400 nm (b) under different input fluences. Transmittance as a function of input fluence for GO films of different thickness to femtosecond laser pulses at 800 (c) and 400 nm (d). Reproduced from Ref. [367].

Such transition from saturable absorption to reverse saturable absorption by increasing the pulse intensity was also reported by other groups [368-372]. Particularly, Cheng et al. [371] carried out simulation on a GO mode-locked Yb-doped fiber laser. The mode-locked fiber ring laser was modeled by the scalar generalized nonlinear Schrodinger equation. It consisted of a piece of Yb-doped fiber with a length of 1 m, a piece of single mode fiber with a length of 65 m, a Gaussian shaped filer with bandwidth of 8 nm, an output coupler with output ratio of 80%, and a saturable absorber, as illustrated in Fig. 55a. The corresponding repetition rate is 2.99 MHz, with the same group velocity dispersion and nonlinear coefficient of 0.025 $ps^2$/m and 3 $W^{-1}km^{-1}$ for all the fiber in the ring, respectively. As a result, the GO-based saturable absorber shows a saturable absorption property within the range of power 0~10 W, where transmissivity increases as power increases; but reverse saturable absorption is observed in the range of 10~20 W, in which transmissivity decreases as power increases, as presented in Fig. 55b. Besides, the pulse-intensity-dependent transition from saturable absorption to reverse saturable absorption was also found in GO hybrids [373, 374].

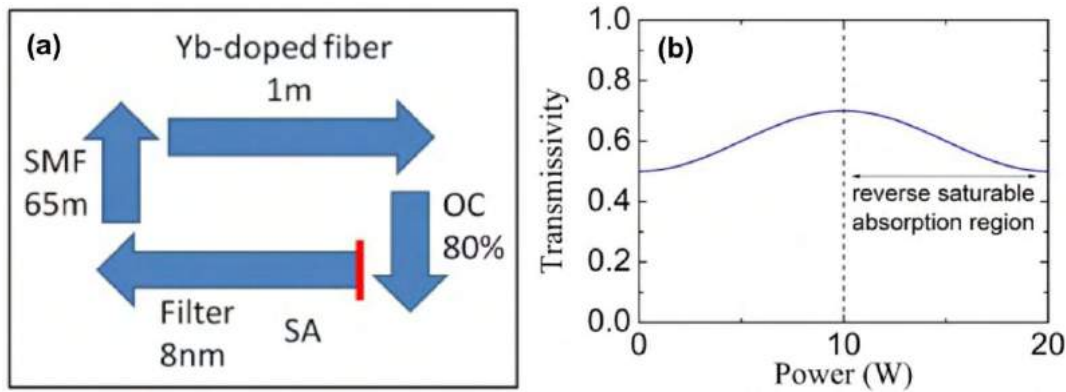

**Fig. 55** (a) Schematic of the Yb-doped fiber laser, where SMF indicates single mode fiber, OC is output coupler, and SA means saturable absorber, respectively. Red line is the start and end point for every round trip. (b) Transmissivity as a function of the pump power. Reproduced from Ref. [371].

In addition to the pulse intensity, other factors also play roles on the reverse saturable absorption of GO, such as GO concentration [368, 370, 372], types of solvents [369, 375], degree of oxidation [376] and chemical functionalization [373, 377-379]. Roy and Yadav [370] demonstrated that increase in concentration resulted in reverse saturable absorption to saturable absorption transition for GO thin films with femtosecond laser pulses at 800 nm. As shown in Fig. 56a, the GO thin films show reverse saturable absorption at low concentration, but switch to saturable absorption at high concentration. The transition takes place at 11.8 mM on increase in concentration. Liaros *et al*. [369] studied the nonlinear optical properties of few-layered GO dispersed in various organic solvents, such as N,N-Dimethylformamide (DMF), Tetrahydrofuran (THF), N-Methyl-2-Pyrrolidone (NMP), Chlorobenzene (CB), 1,2-Dichlorobenzene (o-DCB), and 1-Chloronaphthalene (CN). It was found that among the different solvents, use of heavy solvents like CN and o-DCB for the GO dispersions resulted in significantly large reverse saturable absorption. Besides, the $sp^2$ fraction and $sp^2$ clustering are crucial parameters for the reverse saturable absorption of GO [376]. As shown in Fig. 56b and c, increase of the degree of oxidation in GO will degrade the reverse saturable absorption both at 532 and 1062 nm in over a wide range from ps to ns laser pulses, where $GO_1$, $GO_2$ and $GO_3$ are GO samples with the fraction of $sp^2$ carbons of 49.8%, 17.0%, and 5.2%, respectively.

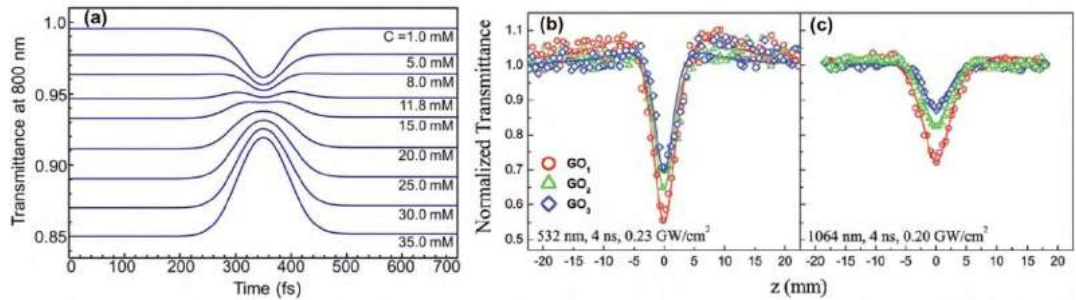

**Fig. 56** (a) Variation of transmittance with time for 100 fs pump pulse excitation for different concentration values at a pulse intensity of $5 \times 10^{10}$ W/cm$^2$. Reproduced from Ref. [370]. Open-aperture Z-scans of the $GO_1$, $GO_2$ and $GO_3$ dispersions under 4 ns at 532 nm (b) and 1064 nm (c) laser excitation. The solid lines correspond to the best fits of the experimental data. Reproduced from Ref. [376].

Another method to improve the reverse saturable absorption is chemical functionalization. For example, Zhu *et al*. [378] fabricated a soluble GO covalently functionalized with zinc phthalocyanine (PcZn) or GO–PcZn by an amidation reaction. It was found that at the same level of linear transmission, GO–PcZn exhibited better nonlinear optical response than both GO and PcZn alone at 532 and 1064 nm, as shown in Fig. 57a and b, which plot the normalized transmission and the corresponding scattering as functions of input energy density. Reverse saturable absorption was believed to contribute the enhanced nonlinear optical response in GO–PcZn at 532 nm. Similarly, the RGO–ZnO hybrid also showed enhanced reverse saturable absorption due to possessing great extent of π conjugation [380]. It was found that enhancement in the nonlinear absorption coefficient of the RGO–ZnO hybrid could be 15 times of that for the bare ZnO and 2.8 times larger compared to the bare GO, respectively. The enhanced reverse saturable absorption was also reported in the hybrid multilayers containing GO, α-CoPc(QnSO$_3$Na)$_4$ (CoPc), and polydiallydimethyldiallylammonium chloride (PDDA), fabricated by electrostatic self-assembled layer by layer technique [381]. Through measuring the third-order nonlinear properties of the film by Z-scan measurements at 532 nm with 4 ns pulses, it was found that at the input intensity of 1.16 μJ, the 25 bilayer CoPc/GO film exhibits strong reverse saturable absorption effect with a nonlinear absorption coefficient value of $1.6 \times 10^{-5}$ mW$^{-1}$, which is 5 and 3 orders of magnitude larger than that of CoPc solution sample and GO solution sample, and 1.5 times larger than that of the 35 bilayer CoPc/PDDA film, respectively. Besides, it was suggested that among various GO composites where GO was covalently functionalized with H$_2$, Cu, Zn, Sn, VO porphyrin, the metal free GO–H$_2$ porphyrin composite shows significant nonlinear absorption behaviour as well as the highest fluorescence quenching behavior in comparison to the other GO–porphyrin composites [379].

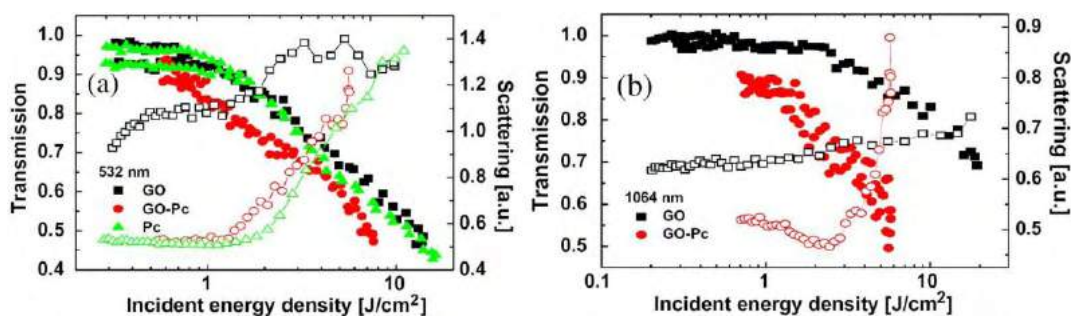

**Fig. 57** (a) Nonlinear transmission (solid symbols) and scattering (open symbols) of the GO–PcZn and GO dispersions and PcZn solutions at 532 nm (a) and 1064 nm (b). Reproduced from Ref. [378].

### 5.3 Two-photon absorption

Usually, the reverse saturable absorption can be achieved by multi-photon absorption, with two-photon absorption being dominated [382]. Two-photon absorption is the simultaneous absorption of two photons of identical or different frequencies in order to excite a molecule from one state (usually the ground state) to a higher energy electronic state. The energy difference between the involved lower and upper states of the molecule is equal to the sum of the energies of the two photons.

Compared with saturable absorption, two-photon absorption dominates at high intensities. Liu *et al*. [383] studied the nonlinear optical properties of GO at 532 nm in nanosecond and picosecond regimes. It was found that two-photon absorption dominates nonlinear absorption process of GO in the case of picosecond pulses at high intensities, while saturable absorption dominates at low intensities. As shown in Fig. 58a, at low input intensity of $2.1 \times 10^9$ W/cm$^2$, the normalized transmission curve shows a symmetrical peak with respect to the focus (z = 0), indicating that saturable absorption dominates the nonlinear absorption mechanism. However, when increasing the input intensity to $4.5 \times 10^9$ W/cm$^2$, a valley within the peak appears at the focus, as shown in Fig. 58b, indicating appearance of two-photon absorption. Further increasing the input intensities will lead to a deeper valley but keep the two-photon absorption, as shown in Fig. 58c and d. Further investigation by Liu *et al*. [384] demonstrated that the saturable absorption in GO is governed by sp$^2$-hybridized domains, whereas the two-photon absorption arises from the introduction of the sp$^3$-hybridized domains caused by oxidation.

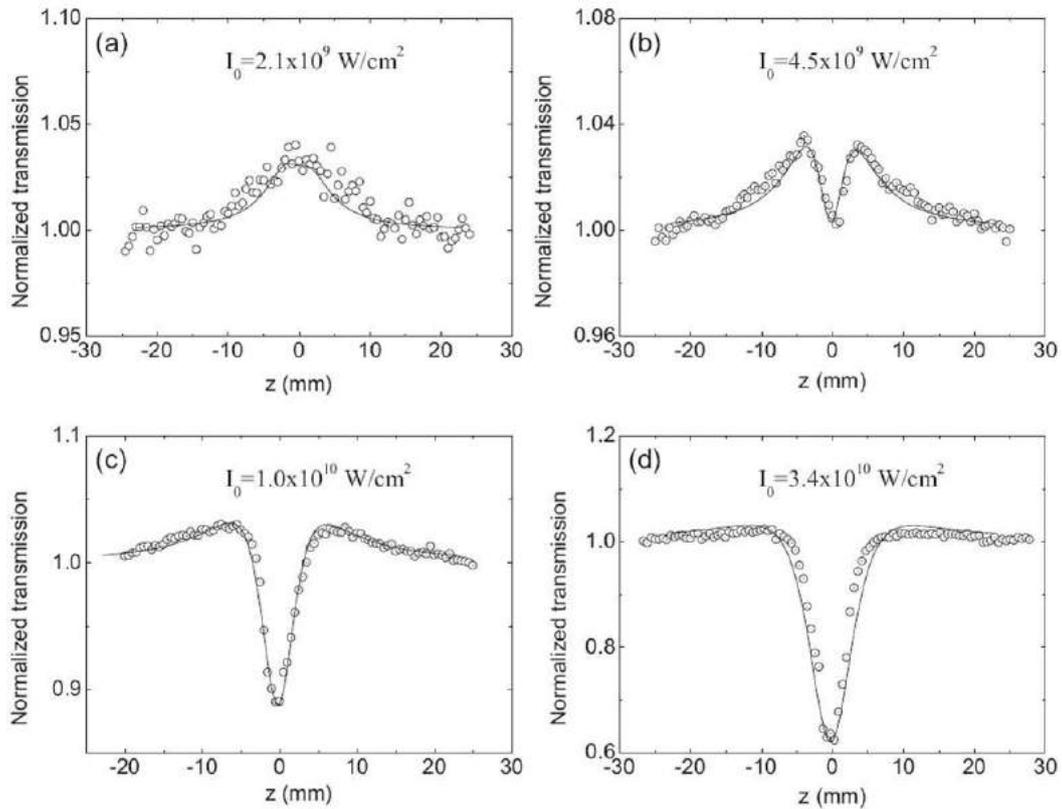

**Fig. 58** Open aperture picosecond Z-scan results of GO in 0.5 mg/ml DMF (*N, N*-Dimethylformamide) at different input intensities: (a) $2.1 \times 10^9$ W/cm$^2$, (b) $4.5 \times 10^9$ W/cm$^2$, (c) $1.0 \times 10^{10}$ W/cm$^2$, and (d) $3.4 \times 10^{10}$ W/cm$^2$. Solid lines are obtained from theoretical fits. Reproduced

from Ref. [383].

Therefore, the nonlinear optical response of GO sheets can be tailored by manipulation of the degree and location of oxidation [385, 386]. For example, Shi *et al.* [385] studied nonlinear optical properties of both GO and RGO by single beam Z-scan measurement at 532 nm in the picosecond region. It was found that saturable absorption and two-photon absorption strongly depend on the intensity of the pump pulse. Saturable absorption occurs at lower pump pulse intensity, with a saturation intensity of ~1.5 $GW/cm^2$, whereas two-photon absorption dominates at higher intensities, larger than 5.7 $GW/cm^2$. Particularly, varying the concentrations of $sp^2$ domains and $sp^2$ clusters in RGO via chemical reduction will lead to changes of the two-photon absorption coefficient and saturation intensity. During increasing the concentration of $sp^2$ C–C bonds from 48.31% to 61.75% by reduction, the two-photon absorption coefficient increases from 1.44 to 4.60 cm/GW and the saturation intensity changes from 1.5 to 2 .5 $GW/cm^2$. Similarly, Bhattachraya *et al.* [386] pointed out that, for the RGO measured by open aperture Z-scan technique, saturation dominates at low intensity about 127 $GW/cm^2$, while two-photon absorption becomes prominent at higher intensities from 217 to 302 $GW/cm^2$. For the GO samples with different concentration of $sp^2$ C–C bonds, the values of two-photon absorption coefficient and the saturation intensity increase with increasing extent of reduction, as shown in Table 6, where the concentrations of $sp^2$ carbons are 49.27% for GO, 57.76% for RGO1, 60.66% for RGO2 and 90.54% for RGO3, respectively.

**Table 6** Nonlinear optical absorption parameters derived from fitting of open-aperture Z-scan data. Samples GO, RGO1, RGO2, and RGO3 have different concentrations of $sp^2$ carbons of 49.27%, 57.76%, 60.66% and 90.54%, respectively. $I_{sat}$ and $\beta_{eff}$ represent the saturation intensity and the two-photon absorption coefficient, respectively. Reproduced from Ref. [386].

| Input pulse energy (μJ) | GO | | RGO1 | | RGO2 | | RGO3 | |
|---|---|---|---|---|---|---|---|---|
| | $I_{sat}$ (GW/cm²) | $\beta_{eff}$ (cm/GW) | $I_{sat}$ (GW/cm²) | $\beta_{eff}$ (cm/GW) | $I_{sat}$ (GW/cm²) | $\beta_{eff}$ (cm/GW) | $I_{sat}$ (GW/cm²) | $\beta_{eff}$ (cm/GW) |
| 1.5 | 57 | 0.0022 | 94 | 0.0047 | 137 | 0.0072 | 194 | 0.0128 |

| | | | | | | | |
|---|---|---|---|---|---|---|---|
| 2.5 | 57 | 0.0037 | 94 | 0.0062 | 137 | 0.0093 | 194 | 0.0143 |
| 3 | 57 | 0.0043 | 94 | 0.0064 | 137 | 0.0094 | 194 | 0.0136 |
| 3.5 | 57 | 0.0044 | 94 | 0.0065 | 137 | 0.0095 | 194 | 0.0131 |

## 5.4 Optical limiting

An important application of reverse saturable absorption or two-photon absorption is optical limiting. Reverse saturable absorption takes place on a laser irradiation of nano-second or longer timescale, owing to the relatively long excited state lifetimes of materials, while optical limiting materials based on two-photon absorption can work under the laser irradiation of pico-second or shorter timescale [387]. Optical limiting can achieve high transmittance for low incident light intensity, and low transmittance for high incident intensity, which is of great importance in protecting human eyes and sensitive instruments from laser-induced damage. Moreover, for real life applications, a practical optical limiter has to satisfy particular requirements such as low optical limiting threshold, broadband and fast response, high linear transmittance and physical/chemical stability [369].

Previously, Feng *et al.* [388] systematically investigated the nonlinear optical and optical limiting properties of a series of graphene-based materials at 532 and 1064 nm in a nanosecond regime, including the GO nanosheets, graphene nanosheets, GO nanoribbons, and graphene nanoribbons, as well as multi-walled CNTs (MWCNTs) for comparison. They found that reduced graphene samples exhibit stronger nonlinear optical and optical limiting responses than their GO precursors because of the increased crystallinity and conjugation, as presented in Table 7. At 532 nm, the optical limiting performances follow the sequence: graphene nanosheets > graphene nanoribbons > GO nanosheets > MWCNTs > GO nanoribbons, whereas the scattering signals follow another sequence: GO nanoribbons > graphene nanosheets > graphene nanoribbons > MWCNTs > GO nanosheets. However, the result at 1064 nm is quite different, that is, the optical limiting performances and scattering signals follow a consistent sequence: graphene nanoribbons > GO nanoribbons > graphene nanosheets > MWCNTs > GO nanosheets, indicating the predominant contribution of nonlinear scattering at this wavelength.

**Table 7** The optical limiting threshold ($F_{th}$, J/cm$^2$) at which the transmittance falls to 50% of the normalized linear transmittance, and nonlinear absorption coefficient ($\alpha_2$, $\times 10^{-13}$ cm/w, input energy: 250 µJ/pulse) for dispersions of the as-synthesized graphene families at 532 and 1064 nm. Those of MWCNTs are also included for comparison. Reproduced from Ref. [388].

| Samples | 532 nm | | 1064 nm | |
|---|---|---|---|---|
| | $F_{th}$ | $\alpha_2$ | $F_{th}$ | $\alpha_2$ |
| GO nanosheets | > 3 | 0.29 | – | 0.12 |
| Graphene nanosheets | 0.5 | 1.36 | 6.3 | 1.16 |
| GO nanoribbons | 1.0 | 1.09 | 4.0 | 1.40 |
| Graphene nanoribbons | 0.7 | 1.26 | 3.4 | 1.60 |
| MWCNTs | 1.4 | 1.05 | > 10 | 1.11 |

While, the optical limiting threshold of GO, which is defined as the input fluence at which the transmittance falls to 50% of the normalized linear transmittance, can be significantly degraded by hybridization, leading to a better optical limiting response than that of the benchmark material, i.e. C$_{60}$. For example, Balapanuru et al. [389] designed a PNP$^+$GO$^-$ complex formed by positively charged dye, 4-(1-pyrenylvinyl)-N-butylpyridinium bromide (PNPB), and negatively charged GO. They pointed out that the optical limiting threshold value of PNP$^+$GO$^-$ is 1.55 J/cm$^2$ at 532 nm. This optical limiting threshold was lower than that of the benchmark optical liming material C$_{60}$ (3.0 J/cm$^2$) measured on the same system [390]. Meanwhile, at 1064 nm, the PNP$^+$GO$^-$ solution showed a limiting threshold of 8.10 J/cm$^2$, which was also better than that of carbon nanotubes (10 J/cm$^2$) measured on the same system. Lim et al. [375] reported that the alkyl-functionalized sub-stoichiometric GO (sub-GOx) in chlorobenzene solvent exhibited strong optical limitation and achieved an optical limiting threshold of 10 mJ/cm$^2$ for a linear transmittance of 70% for nanosecond pulses, which was 5–10 times lower than those of C$_{60}$ in toluene and SWCNTs in tetrahydrofuran, as shown in Fig. 59a. Compared with fully oxidized stoichiometric GO, the sub-GOx has a significant fraction of sp$^2$ carbon atoms retained in the basal plane, providing its excellent optical limiting response. Zhang et al. [391] investigated the optical limiting properties of the GO hybrid material coordinated with Fe$_3$O$_4$ nanoparticles, GO–Fe$_3$O$_4$, using the Z-scan technique at 532 nm in the nanosecond and picosecond regimes. Compared with $C_{60}$ in toluene at different concentrations, GO–$Fe_3O_4$ exhibits a weaker optical limiting effect than $C_{60}$ at high concentration, but shows a stronger optical limiting effect than $C_{60}$ at low concentration in the high input fluence region. Note that lower concentration leads to higher transmittance. With the linear transmittance of 49%, GO–$Fe_3O_4$ exhibited better optical limiting effect compared with GO, but weaker than $C_{60}$. In detail, at the input fluence of 20 J/$cm^2$, the optical limiting thresholds were 2.82, 10.19, and 0.41 J/$cm^2$ for GO–$Fe_3O_4$, GO, and $C_{60}$, respectively. Thus, $C_{60}$ has the lowest optical limiting threshold, indicating the best optical limiting effect of $C_{60}$ at high concentration. However, at the same input fluence of 20 J/$cm^2$ with the linear transmittance of 87%, the optical limiting thresholds were 3.70, 10.38, and 8.58 J/$cm^2$ for GO–$Fe_3O_4$, GO, and $C_{60}$, respectively, as shown in Fig. 59b. Therefore, compared with GO and $C_{60}$, GO–$Fe_3O_4$ shows the best optical limiting effect at low concentration.

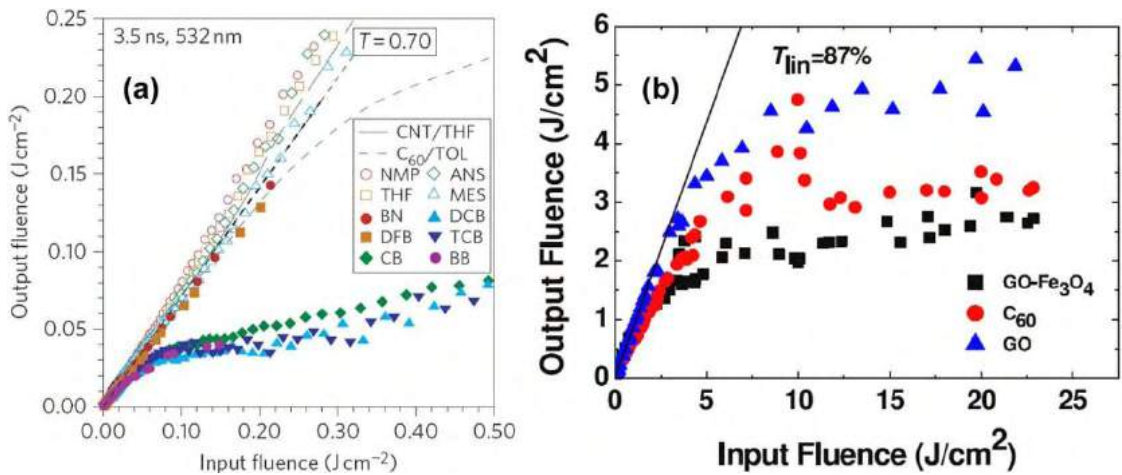

**Fig. 59** (a) Plot of output versus input fluence for sub-GOx dispersed in different solvents compared with $C_{60}$ in TOL and single-walled CNT in THF, all in cells with 1.0 mm path length. NMP, N-methylpyrrolidinone; THF, tetrahydrofuran; ANS, anisole; MES, mesitylene; DFB, 1,3-difluorobenzene; BN, benzonitrile; CB, chlorobenzene; BB, bromobenzene; DCB, 1,2-dichlorobenzene; TCB, 1,2,4-trichlorobenzene; TOL, toluene. The linear transmittance $T$ is 0.70. Reproduced from Ref. [375]. (b) Output fluence versus input fluence for GO–$Fe_3O_4$, GO and $C_{60}$ with the same linear transmittance of 87% with nanosecond pulses. Reproduced from Ref. [391].

Besides, the optical limiting threshold of a GO hybrid is usually lower than that of its component alone [392-397]. The enhanced optical limiting performance in GO hybrids can be ascribed to a combination of optical limiting mechanisms, including nonlinear optical absorption and scattering, as well as photo-induced electron or energy transfer in GO-organic hybrids [18]. For instance, GO–Er hybrids with GO being covalently functionalized with $NaYF_4:Yb^{3+}/Er^{3+}$ nanoparticles also presented lower optical limiting threshold of 134 mW than that of the GO (380 mW) [392]. Similarly, the PANI–GO hybrid exhibited an optical limiting threshold of 0.8 $GW/cm^2$, much lower that of its component, i.e., 3 $GW/cm^2$ for GO and 2 $GW/cm^2$ for PANI [393]. Song *et al.* [394] examined the nonlinear optical properties of RGO hybrid material covalently functionalized with zinc phthalocyanine (ZnPc). As shown in Fig. 60a, at the same level of linear transmittance of 80%, both RGO–ZnPc and GO–ZnPc hybrids exhibited lower optical-limiting threshold values, i.e., 0.98 $J/cm^2$ and 1.74 $J/cm^2$, in comparison with the GO (2.12 $J/cm^2$) and ZnPc (1.97 $J/cm^2$). It was believed that a combination of different nonlinear optical absorption mechanisms contributes to the superior optical limiting performance of RGO–ZnPc hybrid, including two-photon absorption originating from the $sp^3$ domains, saturable absorption from the $sp^2$ carbon clusters and excited state absorption from numerous localized $sp^2$ configurations in RGO moiety, reverse saturable absorption arising from ZnPc moiety and the contribution of efficient photo-induced electron transfer or energy transfer process between ZnPc and RGO.

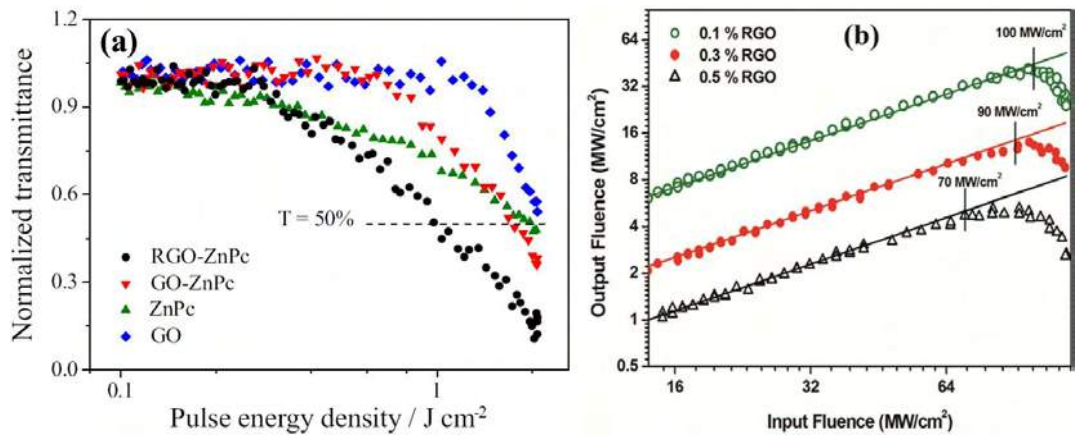

**Fig. 60** (a) The optical limiting of RGO–ZnPc, GO–ZnPc, ZnPc and GO excited at 532 nm with 4 ns pulses. Reproduced from Ref. [394]. (b) Output laser intensity as a function of input laser showing optical limiting properties of RGO-PVA nanocomposites at different mass ratios of RGO at 532 nm. Reproduced from Ref. [398].

In addition, other factors, such as hybrid matrix and concentration also affect the optical limiting thresholds of GO hybrids. It was found that for both GO nanosheets and GO nanoribbons, when prepared into Ormosil hybrid glasses by a facile sol–gel process, optical limiting thresholds as low as ~0.03 J/cm$^2$ can be achieved, surpassing those of corresponding suspensions by a factor of 5–20, as summarized in Table 8 [399]. For GO–APTES hybrids (GO functionalized with $NH_2(CH_2)_3Si(OC_2H_5)_3$, APTES), the optical limiting threshold decreases as the concentration of GO–APTES increases [400]. Similarly, Wang *et al.* [401] demonstrated that for the CMC–GO composite films obtained by doping GO into a carboxymethyl cellulose (CMC) matrix, as the CMC:GO mass ratio dropped from 100:0.5 to 100:1 and 100:1.5, the optical limiting threshold of CMC–GO composite films decreased from 0.120 to 0.056 and 0.030 J/cm$^2$ accordingly, indicating an enhanced optical limiting performance during increasing the GO concentration. This enhanced optical limiting performance due to increasing the GO concentration was also found in GO–PVA composite sheets [398, 402]. As shown in Fig. 60b, as the mass ratio of RGO in GO–PVA composite increased from 0.1 to 0.5 wt.%, the optical limiting threshold dropped from 100 to 70 MW/cm$^2$.

**Table 8** The optical limiting threshold ($F_{th}$) and nonlinear absorption coefficient ($\beta$, input energy of 200 mJ) for the samples. GONS: graphene oxide nanosheet, and GONR: graphene oxide nanoribbon. Reproduced from Ref. [399].

| Samples | $\beta$ ($\times 10^{-9}$ cm/W) | 1 $F_{th}$ (J/cm$^2$) |
| --- | --- | --- |
| GONS–$H_2O$ | 0.74 | 0.76 |
| GONR–$H_2O$ | 1.1 | 0.15 |
| GONS–Ormosil | 4.4 | 0.035 |
| GONR–Ormosil | 4.5 | 0.030 |

In addition to the low optical limiting threshold, GO-based optical limiting materials also present broadband and fast response. Lim *et al.* [375] demonstrated that the sub-GOx in chlorobenzene exhibits strong and optical limiting. The wavelength dependence of the optical limiting response in chlorobenzene was plotted in Fig. 61a. It can be seen that a broadband over 450–1064 nm wavelength range was confirmed, but with a gradual roll off above 700 nm. Similarly, Perumbilavil *et al.* [403] also pointed out that RGO exhibited broadband ultrafast optical nonlinearity in a spectral region of 400–700 nm. Jiang *et al.* [367] investigated the nonlinear optical properties of GO thin films on glass and plastic substrates. The as-prepared GO films exhibit excellent broadband optical limiting behaviors over a wide range of wavelength, as shown in Fig. 61b. Moreover, the optical limiting properties of GO films can be tuned by the extent of reduction. It can be noticed that the transmittance gradually decreases with increasing exposure time to hydrazine, which is attributed to formation of larger $sp^2$ conjugated network upon chemical reduction. Liaros *et al.* [404] suggested that depending on the solvent used, few layered GO dispersions show broadband optical limiting action under infrared laser wavelengths from 1064 nm up to 1800 nm. Besides, GO–PVA composite sheets displayed highly efficient broadband optical limiting activities for femtosecond laser pulses at 400, 800, and 1400 nm with very low limiting thresholds [402].

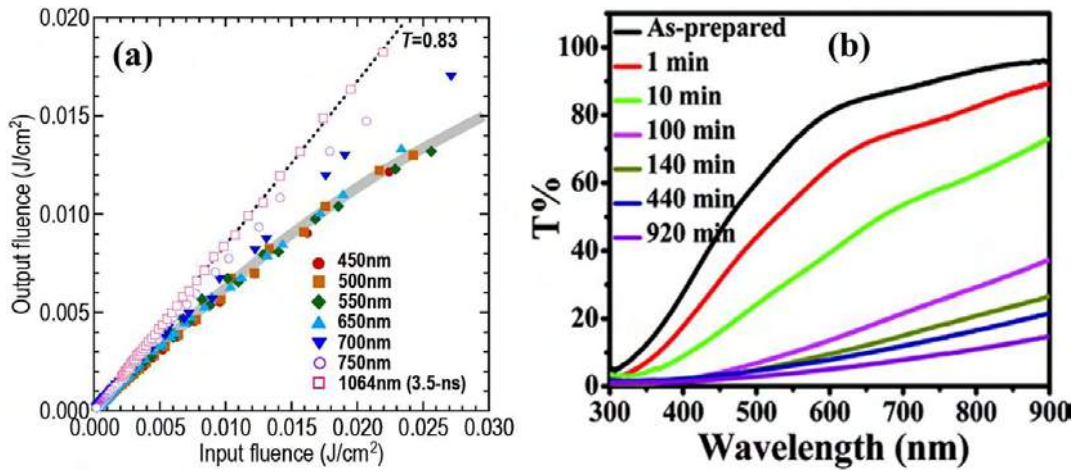

**Fig. 61** (a) Wavelength dependent output versus input influence characteristics of functionalized sub-GOx dispersed in CB, measured using a 7 ns tunable laser from 450 nm to 750 nm by Z-scan technique. Data for 1064 nm comes from Nd:YAG laser pulses (3.5 ns) also by Z-scan. The linear transmittance *T* at 1064 nm is 0.83, and at 450–550 nm is ~0.70. The grey line is a guide to the eye for the 450–650 nm data. Reproduced from Ref. [375]. (b) Transmittance measurements of GO after different periods of exposure time. Reproduced from Ref. [367].

One important application of GO-based optical limiters is for protecting human eyes. Yung *et al.* [405] investigated the optical filtering effect of RGO for eye protection. As shown in Fig. 62, the optical transmittance within the predominant reaching-earth solar spectrum ranging from 350 to 800 nm was measured by the laser treatment. It can be noticed that during increasing the photon energy, the absorption of the RGO film on PET increased dramatically from 17% at 800 nm to 96% at 350 nm, indicating an eye protection effect at high photon energy. Generally, the transmittance spectrum can be divided into three sections. For low energy visible regions with wavelength ranging from 500 nm to 700 nm, the transmittance rate increased from 40% to 75%, revealing transparent features. While, the absorption rates were all above 55% at the high energy visible blue light region from 400 to 500 nm. Particularly, for the ultraviolet region from 350 to 400 nm, the absorption rate is completely over 85%.

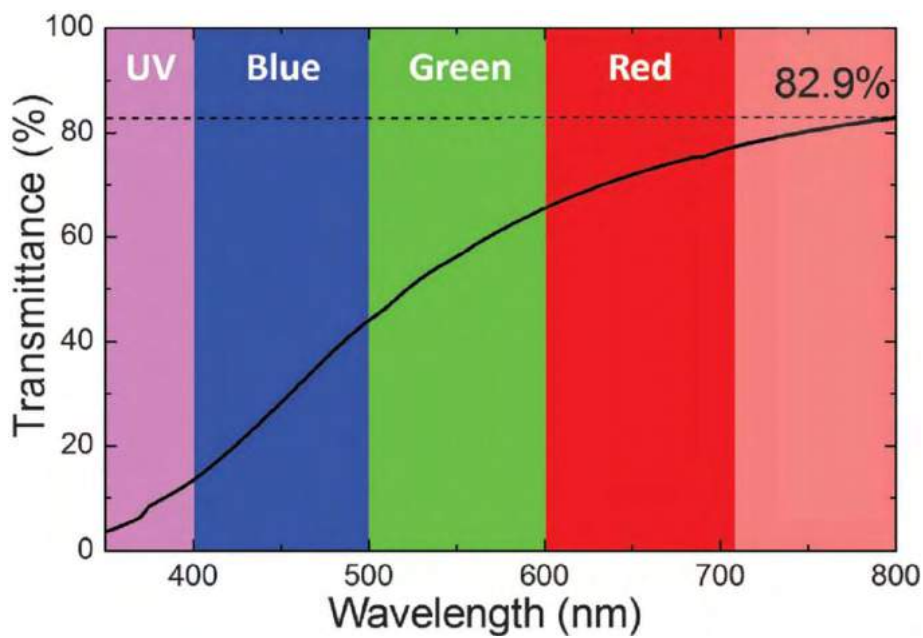

**Fig. 62** UV-Vis transmittance spectrum of the RGO after the laser treatment. Reproduced from Ref. [405].

In summary, GO related materials and composites show superior nonlinear optical properties due to the combined effects of $sp^2$ and $sp^3$ carbons. Moreover, the nonlinear optical properties of GO can be modulated by several factors, including thickness, concentration, oxidation extent, and hybridization. The low optical limiting threshold and broadband response enable GO-based materials promising in optical limiters, which can be used for eye protection.

## 6 Summary and outlook

In this article, the physical properties and related device applications of graphene oxide have been reviewed. Generally speaking, GO is a non-stoichiometric and amorphous material with monolayer or multilayer structure. Microscopic and spectroscopic measurements demonstrate that GO preserves the honeycomb lattice of graphene, whereas the carbon sheet is distorted due to the chemisorbed oxygenated groups, mainly epoxy and hydroxyl species, and their concentrations and spatial distributions depend on the synthesis process. Generally, GO samples with the O:C ratios of 0~0.5 can be achieved in experiment, and the oxidized regions with disordered and ordered structures have both been observed. On the theoretical side, low-energy ordered crystalline phases of GO have been proposed, in which epoxy and hydroxyl groups are closely arranged in chain-like structures with $sp^2$ carbon regions in-between. Amorphous GO with highly oxidized domains is nearly as stable as the ordered phases. Partially oxidized GO is thermodynamically unstable in air, however, the agglomeration of oxygenated groups can enhance the kinetic stability of the system, making GO a kind of metastable materials.

GO possesses many interesting physical properties owing to its unique atomic and chemical structure. Most importantly, the electronic properties of GO vary from insulating to semiconducting, and ultimately back to semi-metallic, depending on the oxidation degree. Therefore, GO and RGO can be used for dielectric layers, channels in FETs, and transparent conductive electrodes, and so on. The presence of oxygen functional groups can not only open a bandgap in GO and alter the absolute positions of energy levels (e.g., work function), but also significantly reduce the thermal conductivity of the system, making GO a promising candidate for thermoelectric materials. GO inherits the superior mechanical strength of graphene, and its mechanical properties can be further modulated by either controlling the concentration and species of oxygenated groups or intercalating molecules and metal ions into the gallery of multilayer GO layers. Hence GO can be used as GO papers, nanomechanical resonators, and flexible electronic devices. Moreover, GO shows intrinsic fluorescence and nonlinear optical properties; thus it holds great potentials for photodetection, photoimaging, optical limiters, and saturable absorbers. The main parameters to characterize the physical properties of GO are summarized in Table 9.

**Table 9** Summary of main physical parameters of GO materials, including the in-plane Young's modulus E, intrinsic strength $\tau_c$, fracture strain $\varepsilon_c$, perpendicular Young's modulus $E_\perp$, bending modulus $D$, thermal conductivity $\kappa$ at room temperature (RT), optical transmittance $T$ at 550 nm wavelength, electronic bandgap $E_g$, electrical conductivity $\sigma$, and optical limiting threshold $F_{th}$ at 532 nm laser. The $h$ is the thickness of GO material, and RT is room temperature.

| Sample | GO paper | GO multilayer | GO monolayer |
|---|---|---|---|
| E (GPa) | 6–42 [13, 147] | 34–77 [12, 145] | 247–542 [11, 139] |
| $\tau_c$ (GPa) | 0.015–0.133 [147] | 4–5 [12] | 21–54 [11, 139] |
| $\varepsilon_c$ (%) | 0.1–0.8 [147] | 8–15 [12] | 10–20 [11, 139] |
| $E_\perp$ (GPa) | 3.6–35 [134, 157] | 19–22 [157] | — |
| $D$ ($k_B$T) | — | — | 0.5–2 [14] |
| $\kappa$ at RT (W/mK) | 0.30 [316] | 0.14–2.87 ($h$ = 75 nm) [317] | 8.8–42 [318, 323] |
| $E_g$ (eV) | 0–3.5 depending on the oxidization degree [168, 170, 173, 174] | | |
| $\sigma$ (S/m) | $10^{-3}$–$10^3$ depending on the oxidization degree [8, 173, 189, 316] | | |
| $T$ at 550 nm (%) | <50 (h > 100 nm) [212] | 20–80 (h < 100 nm) [19, 212] | 85–96 [7, 204] |
| $F_{th}$ (J/cm$^2$) at 532 nm | ~10 [391, 396] | 2.12–~3 [388, 394] | 0.96 [395] |

Despite of the comprehensive research efforts on GO and the inspiring results that have been achieved, so far there are still several critical issues to be solved before realizing the wide applications of GO. First, the atomic structure of GO has to be precisely controlled in order to tailor its physical properties, not only the type and concentration, but also the distribution of oxygen functional groups on the carbon layers. The latter is a nontrivial task. Mild annealing treatment of GO induces phase segregation, forming oxidized domains embedded in the pristine graphene regions. AFM and thermochemical nanolithography technique can be used to locally reduce GO and hence create specific nanopatterns on GO. The atomic-resolved STM tip can even manipulate the positions of individual oxygen atoms on GO surfaces. These experimental techniques have to be well developed in order to effectively pattern the desired GO structures. Meanwhile, further theoretical works are necessary to elucidate the response of the GO structure to the external stimuli, such as temperature, pressure, humidity, light, and chemical species and to provide guidance for experimental fabrication and structural control of this material.

Second, large-scale production of high-quality GO samples at low cost is crucial for the technological applications of this material in the future. The conventional strategy of synthesizing GO is based on Hummers oxidation of graphite and then exfoliation of graphite oxide. The harsh treatment during the synthesis process unavoidably results in defects and holes in GO, which in turn affect the physical properties of GO and are detrimental to the performance of GO-based devices. Direct oxidation of epitaxial graphene produces GO samples with well-structured oxide layers and limited defect density; however, the preparation of epitaxial graphene is of high cost. Improvement in the synthesis method has to be made to diminish the damage to the lattice structure of GO. Alternatively, post-synthesis treatment that can heal the defective structures of as-prepared GO is also helpful for improving the quality of GO materials.

Third, precisely tailoring the electronic band structure is of key importance for the electronic, optoelectronic and thermoelectric applications of GO. For instance, to use GO as the channel materials in FETs, a bandgap is desired to achieve large ON/OFF ratios. This can be achieved by controlling the oxidation degree, which is, however, often accompanied by the loss of carrier mobility. Therefore, it is necessary to design peculiar band structure for GO having suitable bandgap and maintaining high carrier mobility simultaneously. For the thermoelectric applications of GO, the power factor has to be enhanced to maximize the ZT value. But there is a conflict between the electrical conductivity and the Seebeck coefficient — the former increases with carrier density, and the latter decreases. Further studies are necessary to understand the relations between these two parameters to attain an optimum power factor of GO.

Fourth, the optical properties of GO have not been well understood from the fundamental point of view, such as the origin of fluorescence in GO, the key factors for affecting the nonlinear optical properties and optical limiting behavior of GO, and the relationship between atomic structures, chemical composition and optical properties of GO. In this regard, elaborate experiments of GO with well controlled structure and oxidation degree, along with atomistic theoretical simulations with realistic structural models are desirable to provide vital insights into these issues.

Finally, the commercial and industrial applications of GO-based materials are still in the starting stage with many remaining difficulties and challenges. Although numerous promising applications have been proposed, the basic physical mechanisms underlying these applications are still unclear. In particular, there are nearly infinite combinations of GO/RGO with other nanomaterials in principles, making the GO/RGO based composites a largely unexplored area with many opportunities. Understanding the interaction mechanism between GO/RGO with the nanomaterials in those nanoscale hybrids from the atomic point of view is imperative. Deeper insights into the intrinsically synergistic effects of GO/RGO composites are critical for designing various GO-based devices and improving their overall performances.

Considering the outstanding performance of the GO-based materials in many aspects, we encourage more researches from fundamental science to explore the basic physical properties of GO, as well as more efforts in expanding the physics device applications of GO utilizing these properties. With the currently increasing attentions on GO/RGO and the related materials, graphene oxide, a material has been studied for over one and half centuries, has a bright future in the next-generation electronic, optical, and optoelectronic devices.

**Acknowledgments**

This work was supported by the National Natural Science Foundation of China (11604039, 11974068 and 11504040), the Fundamental Research Funds for the Central Universities of China (DUT18LK07, DUT16RC(4)66 and DUT17RC(4)52), and the Supercomputing Center of Dalian University of Technology. The authors thank Prof. S.B. Zhang (Rensselaer Polytechnic Institute, USA), Prof. A. Bongiorno (Georgia Institute of Technology, USA), Prof. L. Wang (Suzhou University, China), Prof. X. Jiang and Prof. Y. Su (Dalian University of Technology, China) for collaboration on graphene oxides and the related projects.